\documentclass[iop,appendixfloats]{emulateapj}

\newbox\grsign \setbox\grsign=\hbox{$>$}
\newdimen\grdimen \grdimen=\ht\grsign
\newbox\laxbox \newbox\gaxbox
\setbox\gaxbox=\hbox{\raise.5ex\hbox{$>$}\llap
     {\lower.5ex\hbox{$\sim$}}}\ht1=\grdimen\dp1=0pt
\setbox\laxbox=\hbox{\raise.5ex\hbox{$<$}\llap
     {\lower.5ex\hbox{$\sim$}}}\ht2=\grdimen\dp2=0pt

\newcommand{\lax}{$\mathrel{\copy\laxbox}$}

\newcommand{\HAR}{rHARM}
\newcommand{\dF}{{^{^*}\!\!F}}

\usepackage[usenames,dvipsnames]{color}

\usepackage{CJKutf8}
\usepackage{amsmath}
\usepackage{ctable} 
\usepackage{graphicx}
\usepackage{natbib}
\begin{document}

\title{Jet launching in resistive GR-MHD black hole - accretion disk systems}
\shorttitle{GR-MHD jet launching}
\shortauthors{Qian et al.}
\author{Qian Qian (\begin{CJK*}{UTF8}{gbsn}{钱前\!\!}\end{CJK*})\altaffilmark{1,2},
        Christian Fendt\altaffilmark{1}, 
        Christos Vourellis\altaffilmark{1,2},
}
\altaffiltext{1}{Max Planck Institute for Astronomy, Heidelberg, Germany}
\altaffiltext{2}{Fellow of the {\em International Max Planck Research School for Astronomy and Cosmic 
                 Physics at the University of Heidelberg} (IMPRS-HD).}
\email{ fendt@mpia.de, qian@mpia.de}                                   
\begin{abstract}
%
%
%
We investigate the launching mechanism of relativistic jets from black hole sources, in particular the strong winds 
from the surrounding accretion disk.
Numerical investigations of the disk wind launching - the simulation of the accretion-ejection transition - have 
so far almost only been done for non-relativistic systems.
From these simulations we know that {\em resistivity}, or magnetic diffusivity, plays an important role for the 
launching process.

Here, we extend this treatment to general relativistic magnetohydrodynamics (GR-MHD) applying the resistive GR-MHD 
code {\HAR}.
Our model setup considers a thin accretion disk threaded by a large-scale open magnetic field. 
We run a series of simulations with different Kerr parameter, field strength and diffusivity level.
Indeed we find strong disk winds with, however, mildly relativistic speed, the latter most probably due to 
our limited computational domain.

Further, we find that magnetic diffusivity lowers the efficiency of accretion and ejection, as it weakens
the efficiency of the magnetic lever arm of the disk wind. 
As major driving force of the disk wind we disentangle the toroidal magnetic field pressure gradient, however,
magneto-centrifugal driving may also contribute.
Black hole rotation in our simulations suppresses the accretion rate due to an enhanced toroidal magnetic field pressure
that seems to be induced by frame-dragging.

Comparing the energy fluxes from the Blandford-Znajek-driven central spine and the surrounding disk wind, we find that 
the total electromagnetic energy flux is dominated by the total matter energy flux of the disk wind (by a factor 20).
The kinetic energy flux of the matter outflow is comparatively small and comparable to the Blandford-Znajek
electromagnetic energy flux.
\end{abstract}

%
\section{Introduction}
What powers the energy source of active galactic nuclei (AGN) and drives extra-galactic relativistic jets is one
of the fundamental questions in astrophysics that is not yet fully answered.
The common understanding is that relativistic jets originate from the black hole accretion system 
consisting of a central (super-)massive black hole surrounded by an accretion disk that carries a 
strong magnetic field (see e.g.~\citealt{2015SSRv..191..441H}).

Seminal papers have suggested that magnetohydrodynamic jets can be driven magnetocentrifugally by the rotation of 
the inner accretion disk ( in the following denoted as Blandford-Payne mechanism, \citealt{1982MNRAS.199..883B};
see also \citealt{1986ApJ...301..571P, 2007prpl.conf..277P}),
or gain their energy from the magnetic field that is rooted in the ergosphere of a rotating black hole 
(in the following denoted as Blandford-Znajek mechanism, \citealt{1977MNRAS.179..433B}).
A third scenario was introduced by \citet{1996MNRAS.279..389L} describing jets as magnetic towers driven mainly by 
the toroidal magnetic pressure gradient, in difference to the Blandford-Payne mechanism mention above in which the 
poloidal magnetic field component play the major role.
Numerical modeling have modeled these jets as growing twisted (helical) magnetic fields together with the currents 
that they carry \citep{1995ApJ...439L..39U}.

Unfortunately, the above mentioned processes can hardly be proved by {\em direct} observational evidence as relativistic jet sources 
are mostly detected in synchrotron emission and do not deliver direct information about e.g. mass fluxes or velocities
that are the outcome of MHD simulations.
Also, the observational resolution is not sufficient to disentangle which of the possible jet driving mechanisms plays 
which role.
Due to its proximity, M87 is the single exception for which - depending on wavelength - down to 6 Schwarzschild radii
can be resolved.
Studies of the M87 black hole environment have indicated a spine-sheath structure on pc-scales with about 0.1 pc 
resolution (see e.g. \citealt{2007ApJ...668L..27K}).
Indication for a Blandford-Znajek driving of the M87 jet has been proposed comparing the evolution of observed jet radii 
along the jet with numerical simulations \citep{2012ApJ...745L..28A, 2016ApJ...833...56A}.
Recent work suggests that a turbulent mass loading of the disk jet of M87 may be triggered by coronal reconnection events  
\citep{2017A&A...601A..52B},
while \citet{2016A&A...595A..54M} recover a two-dimensional velocity field of the 100-1000 pc-scale jet propagation. 
Actual observational activities such as the event horizon telescope envisage to resolve the black hole shadow and the 
very close black hole environment on the scale of a Schwarzschild radius 
\citep{2017ApJ...838....1A, 2017arXiv170104955A}.

To treat the jet launching problem for a black hole accretion system, the governing general relativistic
magneto-hydrodynamic(GR-MHD) equations have to be solved. 
A number of GR-MHD codes have been developed over the past decade
(see e.g. \citealt{1999ApJ...522..727K, 2003ApJ...589..444G, 2003ApJ...589..458D, 2006ApJ...641..626N, 2007A&A...473...11D}) that 
integrate the GR-MHD equations in time accurately and finally allow a realistic numerical simulation of 
magnetized accretion disks around black holes.

The foremost targets of these simulations has been the black hole accretion physics
\citep{2003PASJ...55L..69N, 2008NewAR..51..733N, 2012MNRAS.423.3083M}
and the Blandford-Znajek mechanism of black hole jets 
\citep{2004ApJ...611..977M, 2005MNRAS.359..801K, 2005ApJ...630L...5M, 2010ApJ...711...50T, 2011MNRAS.418L..79T, 2012MNRAS.423L..55T}. 
Furthermore, methods that to convert the simulation data of teh GR-MHD dynamics into a radiation pattern have also been 
developed.
In \citet{2009ApJ...696.1616D} a novel technique for quick and accurate calculation of null geodesics in the Kerr metric
has been presented. 
\citet{2011ApJ...743..115N} applied GR-MHD simulation results to derive both the radiative efficiency of accretion and 
the emitted spectrum assuming that essentially all of the emitted power is thermal.
These approaches are in particular interesting as they allow to connect simulation data to observations and may 
provide information on e.g. the accretion rate, that could otherwise not be derived from a pure MHD simulation.


The time evolution of thin disks has been studied by \citet{2010ApJ...711..959N} in particular considering the evolution 
of and the electromagnetic stress at the inner disk radius in the Schwarzschild case. 
In particular, the authors suggest a quantitative measure of quality for resolving the magnetorotational instability
in the disk.
Tilted thin disks and the potential onset of precession due to the Bardeen-Petersen effect were investigated by 
\citet{2014ApJ...796..103M} in 3D simulations lasting up to time scales of 13,000\,M.
Even longer times scales (70,000\,M) were treated by \citet{2016MNRAS.462..636A} investigating deviations from the 
Novikov-Thorne thin disk evolution due to strong magnetic field in magnetically arrested disks (MAD).

In general, the launching of GR-MHD disk winds and jets have not been treated in detail.
Motivated by the success of non-relativistic jet launching modeling from accretion disks 
\citep{2002ApJ...581..988C, 2007A&A...469..811Z, 2009MNRAS.400..820T, 2012ApJ...757...65S, 
2014ApJ...793...31S, 2016ApJ...825...14S},
we have made the effort to understand the launching of MHD outflows from resistive accretion disks of black holes.
Since resistivity, or magnetic diffusivity, respectively, is a necessary ingredient for long-term 
launching simulations, in \citet{2017ApJ...834...29Q} we have implemented magnetic diffusivity in the ideal MHD 
code HARM.
First, preliminary solutions of the launching problem in GR-MHD have been shown in the same paper.
In the present paper a much more detailed study of the launching of disk winds from GR-MHD disks will be presented.

So far a number of resistive relativistic MHD codes have been developed
\citep{2006ApJ...647L.123W, 2007MNRAS.382..995K, 2009MNRAS.394.1727P, 2011ApJ...735..113T, 2013PhRvD..88d4020D, 
2013ApJS..205....7M, 2013MNRAS.428...71B, 2014MNRAS.440L..41B}, 
yet, none of these codes have been applied to the launching problem of disk winds and jets.

We have detailed the implementation of magnetic resistivity into the GR-MHD code HARM \citet{2006ApJ...641..626N} in
\citet{2017ApJ...834...29Q}. 
The new code {\HAR} was applied in the context of accretion and ejection in GR-MHD - presenting test simulations
for the resistive modules of the code, investigating the MRI in resistive tori, and presenting also preliminary studies to the 
launching of disk winds. 
In the present paper, we continue to apply our new code {\HAR} to the black hole accretion system, presenting a much more
detailed study of the launching of outflows from the black hole and the accretion disk.

Our paper is structured as follows.
In Section 2 we show the basic equations that are evolved in {\HAR}. 
In Section 3 we describe the initial and boundary condition in our simulations. 
We verify the choice of initial condition in Section 4 and show the evolution of the disk outflow 
from simulations in Section 5. 
We further discuss the driving force of these disk winds in Section 6 and investigate how resistivity governs
the accretion and launching process in Section 7. 
As one of the key goals for developing {\HAR}, we compare the power of disk winds or jets to 
the power of black hole rotational energy extraction - the Blandford-Znajek mechanism - in Section 8.

\section{Model setup}
In this section, we review the equations that are used for the time evolution in the simulation code. 
We will also describe the coordinates setup, units and its normalization for our simulations. 
Here we basically follow \citet{2017ApJ...834...29Q}. 

\subsection{Basic equations}
In the following, we employ the conventional notation of \citet{1973grav.book.....M}, in particular the sign 
convention for the metric $(-,+,+,+)$.
Applying the Einstein summation convention, Greek letters have values $0,1,2,3$, while Latin letters take values $1,2,3$.
We apply the two observer frames that are defined by the co-moving observer, $u^{\mu}$, and the normal observer, $n^{\mu}$. 
The space-time of normal observer is split into the so called {"}3+1{"} form. 
The electric and the magnetic four vectors that are measured in the two frames are denoted by 
$e^{\mu}$, $b^{\mu}$ and $\mathcal{E}^{\mu}$, $\mathcal{B}^{\mu}$, respectively. 
For the normal observer frame we follow \citet{2006ApJ...641..626N} with the normal observer four velocity
$n_{\mu}=(-\alpha,0,0,0)$  and the lapse time $\alpha = 1/\sqrt{-g^{tt}}$.
Bold letters denote vectors while the corresponding thin letters with indices represent vector components.
 
For our simulations we apply the newly developed GR-MHD code {\HAR} \citep{2017ApJ...834...29Q}, evolving the resistive GR-MHD equations. 
{\HAR} is a conserved scheme and is based on the ideal MHD code HARM \citep{2003ApJ...589..444G,2006ApJ...641..626N}. 
In the context of general relativity, mass conservation is expressed by 
\begin{equation} 
\frac{1}{\sqrt{-g}}\partial_{\mu}\left(\sqrt{-g}\rho u^{\mu}\right)=0
\label{particle_conservation}
\end{equation}
with $g \equiv det(g_{\mu \nu})$ and the mass density $\rho$. 
The conservation of energy-momentum considers
\begin{equation} 
\partial_{t}\left(\sqrt{-g}T^{t}_{\,\,\mu}\right)+\partial_{i}\left(\sqrt{-g}T^{i}_{\,\,\mu}\right)
  = \sqrt{-g}T^{\kappa}_{\,\,\lambda}\Gamma^{\lambda}_{\,\,\,\mu \kappa}
\label{eq_ene-mom-cons}
\end{equation}
where $\Gamma^{\lambda}_{\,\,\nu \kappa}$ is the connection. 
Taking both the fluid part and the electromagnetic part into account, the stress-energy tensor
$T^{\mu}_{\,\,\nu}$ can be written as
\begin{eqnarray} 
T^{\mu \nu} 
    & = & \left(\rho + u + p+ b^{2} + e^{2}\right) u^{\mu}u^{\nu} +
    \left(p + \frac{1}{2}\left(b^{2} + e^{2}\right)\right) g^{\mu \nu} 
\nonumber \\
    & - & b^{\mu} b^{\nu} - e^{\mu}e^{\nu} - u_{\lambda}e_{\beta}b_{\kappa}
\left(  u^{\mu}\epsilon^{\nu \lambda \beta \kappa} + u^{\nu}\epsilon^{\mu \lambda \beta \kappa}\right)
\label{eq_str_ene_tens}
\end{eqnarray}
(see e.g. \citealt{2017ApJ...834...29Q}). 
Here, $u$ is the internal energy, $p$ denotes the gas pressure and  $b^{2}=b^{\mu}b_{\mu}$, $e^{2}=e^{\mu}e_{\mu}$. 
The Levi-Civita tensors in Equation~\ref{eq_str_ene_tens} are defined by
\begin{equation}
\epsilon_{\alpha \beta \gamma \delta} =             \sqrt{-g}[\alpha \beta \gamma \delta], \quad
\epsilon^{\alpha \beta \gamma \delta} = - \frac{1}{\sqrt{-g}}[\alpha \beta \gamma \delta],
\end{equation}
with the conventional permutation symbol $[\alpha \beta \gamma \delta]$.
With the help of Faraday tensor
\begin{eqnarray} 
F^{\mu \nu} & = & u^{\mu}e^{\nu} - e^{\mu}u^{\nu} + \epsilon^{\mu \nu \lambda \kappa} u_{\lambda} b_{\kappa}
\label{eq_fara_tens}
\end{eqnarray}
and the dual Faraday tensor
\begin{eqnarray} 
\dF^{\mu \nu} & = & -u^{\mu}b^{\nu} + b^{\mu}u^{\nu} + \epsilon^{\mu \nu \lambda \kappa} u_{\lambda} e_{\kappa},
\label{eq_dual_fara_tens}
\end{eqnarray}
we define the magnetic field four vector as 
$\mathcal{B}^{\mu} \equiv n_{\nu}\,\dF^{\nu \mu} = \alpha \dF^{\mu t}$ ($\mathcal{B}^{0}=0$) 
and the electric field four vector as 
$\mathcal{E}^{\mu} \equiv n_{\nu}F^{\mu \nu}=-\alpha F^{\mu t}$ ($\mathcal{E}^{0}=0$). 
The time evolution of magnetic field follows
\begin{equation} 
\gamma^{-1/2} \partial_{t} \left(\gamma^{1/2}\pmb{\mathcal{B}}\right) + \nabla \times \left(\alpha \pmb{\mathcal{E}} + \pmb{\beta} \times \pmb{\mathcal{B}} \right) =0
\label{B_evolution}
\end{equation}
while the evolution of the electric field follows
\begin{eqnarray} 
\gamma^{-1/2} \partial_{t} \left(\gamma^{1/2}\pmb{\mathcal{E}}\right) 
    & - & \nabla \times \left(\alpha \pmb{\mathcal{B}} - \pmb{\beta} \times \pmb{\mathcal{E}} \right)
          + \left(\alpha \pmb{v} - \pmb{\beta} \right)q \nonumber \\
    & = &  -\alpha \Gamma \left[ \pmb{\mathcal{E}}  +\pmb{v} \times \pmb{\mathcal{B}} 
          - \left(\pmb{\mathcal{E}} \cdot \pmb{v}\right) \pmb{v} \right] / \eta
\label{eq_E_evol}
\end{eqnarray}
(see \citealt{2013MNRAS.428...71B} and \citealt{2017ApJ...834...29Q} for the derivation and the numerical 
implementation details), where $\pmb{\beta}=\{\beta^{i}\}$ is the spatial shift vector
in 3+1 formalism, $q$ is the charge density, $\Gamma$ denotes the Lorentz factor\footnote{Not to be confused with the 
connection $\Gamma^{\lambda}_{\,\,\nu \kappa}$ in Equation~2, that will not be used anymore 
below.} and $\gamma = \sqrt{-g}/\alpha$ is the determinant of its spatial 3-metric. 
The $\pmb{v}$ denotes the three velocity in the normal observer frame and the variable $\eta =\eta (r, \theta)$
is the scalar magnetic diffusivity (see \citealt{2017ApJ...834...29Q}).

Our simulations are performed in 3D-axisymmetry applying modified Kerr-Schild coordinates (see below). 
A Lax–Friedrichs Riemann solver is used together with a simple first-second scheme for time evolution. 
As was mentioned above, {\HAR} is a conserved scheme. 
The inversion scheme in {\HAR} that transforms conserved quantities to primitive quantities follows the 
description in \citet{2017ApJ...834...29Q}.

\subsection{Numerical grid}
\label{num_grid_setup_subsec}
The computational domain in our simulations is an axisymmetric half sphere. 
In the radial direction, the computational domain ranges from $r_{in}=r_{\rm H}$ to $r_{out}=80$, where $r_{\rm H}$ 
is the radius of the event horizon. 
The angle $\theta$ ranges from $0$ to $\pi$. 

The numerical integrations are carried out on a uniform grid in {\em modified Kerr-Schild coordinates},
$x_{0}$, $x_{1}$, $x_{2}$, $x_{3}$, where $x_{0}=t$, $x_{3}=\phi$ stay the same as in Kerr-Schild 
coordinates, while the radial and $\theta$ coordinates are interrelated as
\begin{eqnarray} 
r &=& R_{0}+e^{x_1},
\nonumber \\
\theta &=& \pi x_2 + \frac{1}{2} (1-h) \sin(2\pi x_2).
\label{MKS_r_theta}
\end{eqnarray}
Different $R_{0}$ and $h \in [0,1]$ will return different concentration of grid resolution in radial and $\theta$ direction. 
The simulations in this paper take $R_{0}=0$ and $h=0.3$.

\subsection{Units and normalization}
The typical units are used throughout the simulations applying $GM=c=\mu_{0}=1$, which sets for the length unit 
the gravitational radius $r_{\rm g} \equiv GM/c^{2}$ and for the time unit the light travel time 
$t_{\rm g} \equiv GM/c^{3}$ over the length unit.
$\mu_{0}$ is the permeability of the vacuum.
The black hole spin is characterized by the Kerr parameter $0<a<1$ (0 for non-rotating black hole). 
The event horizon is located at $r_{\rm H} = 1 + \sqrt{1-a^{2}}$.


Further simulation parameters are the plasma-beta $\beta$ and the gas entropy (parameter $\kappa$) of the
initial condition (see Section \ref{init_cond_sec} for further details).
We note that the plasma moves as a test mass in the fixed space time.
Thus, in code units the mass density is normalized to unity and can be scaled in principle to any 
astrophysical density.
The gas pressure, the internal energy and the field strength then follow from the choice of $\kappa$, $\beta$
and the polytropic index $\gamma_{\rm G}$.
In this paper we present all densities, field strengths, mass and energy fluxes in code units. 
Snapshots of these variables are shown at certain time steps in Kerr-Schild coordinates.

\begin{figure}
\centering
\includegraphics[width=6.5cm]{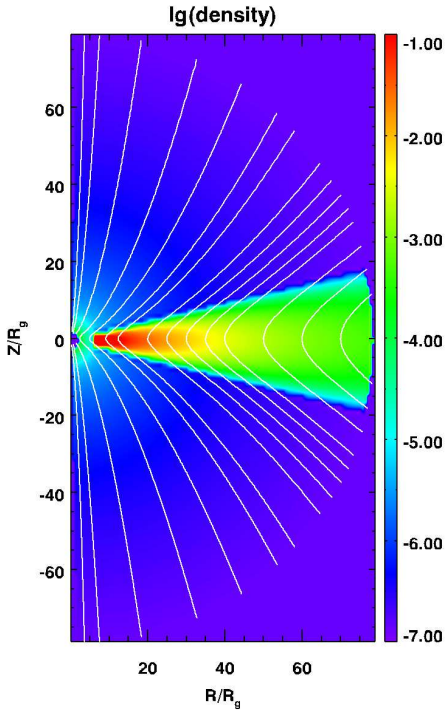}
\caption{The typical initial condition, here shown for simulation {\em D1}.
Shown is a snapshot for density (color coding) and poloidal magnetic field lines (while lines) at 
time $t=0$. 
The white solid lines in the plot show the structure of the large scale magnetic field lines. 
Here and in the following, snapshots are time slices in Kerr-Schild coordinates.
}
\label{astro_dis_init_example_img}
\end{figure}

\section{Boundary and initial conditions}
\label{init_cond_sec}
The simulations presented in this paper consider the time evolution of a thin accretion disk around a 
black hole threaded by an inclined open poloidal field. 
In this section, we will describe the boundary and initial conditions that are employed in our simulations 
and give the definition of accretion and ejection rates which are used to identify the accretion and outflow 
process in the system.

We apply outflow conditions at the inner and outer radial boundary, along which all primitive 
variables are projected into the ghost zones while forbidding inflow at inner and outer boundary.
Along the axial boundary we apply the original HARM reflection conditions, where the primitive variables 
are mirrored into the ghost zones.
The initial electric field is chosen to be equal to the ideal MHD value, 
$\pmb{\mathcal{E}}=-\pmb{v} \times \pmb{\mathcal{B}}$.

\begin{table}
\caption{Parameter choice in the thin disk simulations. The table shows the 
maximum magnetic diffusivity (as defined in Equation~\ref{astro_dis_eta_init_eq}), 
the plasma beta $\beta$, 
the scale height of the diffusivity profile $\chi$ compared to the pressure scale height 
  (see Section \ref{astro_disk_initial_condition_mag_subsec}), 
the black hole spin parameter $a$, and 
the grid size that is used in the simulations.
The radius of the {\em physical} grid size is always $r=80$.
The plasma beta is $\beta = 10$ for all simulations except {\em D0} with $\beta = 10^8$, thus approaching 
the hydrodynamic limit.
}
\begin{center}
  \begin{tabular}{ c | c | c | c | c | c | c }
               & $\eta_{0}$ & $\beta$ & $\chi$  & a  & grid size     \\
     \noalign{\smallskip}   \hline
    \hline  \noalign{\smallskip}  
    {\em D0}   & $10^{-12}$  &$10^{8}$ & $0.8$  &  $0$  & 128x128   \\
    {\em D1}   & $10^{-12}$ &10  & $0.8$  &  $0$  & 128x128   \\
    {\em D2}   & $10^{-6}$  &10  & $0.8$  &  $0$  & 128x128   \\
    {\em D3}   & $10^{-5}$  &10  & $0.8$  &  $0$  & 128x128   \\
    {\em D4}   & $10^{-4}$  &10  & $0.8$  &  $0$  & 128x128   \\
    {\em D5}   & $2 \times 10^{-4}$  &10  & $0.8$ &  $0$  & 128x128   \\
    {\em D6}   & $5 \times 10^{-4}$  &10  & $0.8$ &  $0$  & 128x128   \\
    {\em D7}   & $10^{-3}$  &10  & $0.8$  &  $0$  & 128x128   \\
    {\em D8}   & $10^{-3}$  &10  & $0.8$  &  $0$  & 256x256   \\
    {\em D9}   & $5 \times 10^{-3}$  &10  & $0.8$ &  $0$  & 128x128   \\
    {\em D10}  & $10^{-2}$  &10  & $0.8$  &  $0$  & 128x128   \\
    {\em D11}  & $10^{-3}$  &10  & $1.0$ &  $0$  & 128x128   \\
    {\em D12}  & $10^{-3}$  &10  & $1.0$ & $0.1$ & 128x128   \\
    {\em D13}  & $10^{-3}$  &10  & $1.0$ & $0.2$ & 128x128   \\
    {\em D14}  & $10^{-3}$  &10  & $1.0$ & $0.5$ & 128x128   \\
    {\em D15}  & $10^{-3}$  &10  & $1.0$ & $0.9375$ & 128x128 \\
  \end{tabular}
  \end{center}
\label{astro_dis_sim_table}
\end{table}

\subsection{The thin accretion disk initial condition}
\label{astro_disk_initial_condition_subsec}
We apply a Keplerian rotation with the Paczy\'{n}ski-Wiita approximation for the disk velocity profile
\citep{1980A&A....88...23P}, 
with the angular velocity
\begin{equation}
\Omega = r^{-3/2} \left(\frac{r}{r - R_{\rm pw}}\right),
\label{astro_dis_PW_omega_eq}
\end{equation}
and  a smoothing length scale $R_{\rm pw} = 1.0$.
We choose the Paczy\'{n}ski-Witta profile mainly for simplicity. 
The disk-outflow system will evolve into a new dynamical equilibrium anyway, thus with a new distribution 
of the physical quantities. 
This rotation profile is reasonably stable, and, thus, allows to disentangle the effects of the magnetic 
field on the disk structure and the wind launching (see the discussion for simulation {\em D0} below). 

For the disk density and gas pressure distribution, we apply the solution known from non-relativistic simulations
of jet launching \citep{2002ApJ...581..988C}, where
\begin{equation}
\rho (r,\theta) = 
\frac{R_{\rm ck}^3}{\left(R_{\rm ck}^2 + r^2 \right)^{3/2} } 
\left(1 - \left(\gamma_{\rm g} - 1 \right)\frac{\cos^2\theta}{2 \epsilon_{\rm D}^2} \right)^{1 / (\gamma_{\rm g} -1)},
\label{astro_dis_thindisk_rho_eq}
\end{equation}
where $R_{\rm ck}$ is used as a smoothing length for the gravitational potential.
A natural choice is $R_{\rm ck}=R_{\rm pw}$.
The polytropic index for the gas law is $\gamma_{\rm g}=4/3$,  Equation~(\ref{astro_dis_thindisk_rho_eq}),
in difference to the non-relativistic simulations.
The parameter $\epsilon_{\rm D} = H/r$ is the disk aspect ratio defined by the local disk height $H(r)$
and radius $r$. 
Initially, we apply $\epsilon_{\rm D}=0.1$
and set the coronal initial condition above the disk surface where
\begin{equation}
\theta < \arccos \left( \sqrt{\frac{2 \epsilon_{\rm D}^2}{\gamma_{\rm g} - 1}} \right),
\end{equation}
and similarly in the lower hemisphere.

The initial inner disk radius is located at $r=6$, which corresponds to the innermost stable circular orbit for a 
non-rotating black hole. 
At this radius one orbital period $T$ corresponds to 77 time units $t_{\rm g}$, thus $T=77$.
Inside the disk the density is defined as in Equation~(\ref{astro_dis_thindisk_rho_eq}) and the gas pressure follows 
the polytropic equation of state 
$ p (r,\theta) = \kappa \rho^{\gamma_{\rm g}}$,
where $\kappa$ parametrizes the gas entropy.
We apply $\kappa=10^{-3}$ for the disk in all simulations presented in this paper.

As mentioned before, the grid setup follows the description in Section \ref{num_grid_setup_subsec}. 
With the grid size $128 \times 128$ (see Table \ref{astro_dis_sim_table}) the region (initially) inside the disk inner 
boundary - the plunging region - is resolved by 38 grid cells. 
In the $\theta$-direction the disk region is resolved by 48 grid cells. 
The choice of grid resolution is sufficient to resolve the dynamics within the disk inner radius, inside the disk, and
also the launching region of the outflow.
On the other hand, the outflow area is not very well resolved.
Due to the long run time for these diffusive simulations we are so far restricted to a lower resolution for the coronal
region.
However, the physically interesting accretion region is well resolved.

\subsection{Coronal initial condition and floor model}
Outside the disk, we prescribe an initial corona that is in reasonable ''hydrostatic equilibrium".
The radial profile of the initial corona follows from the gas law and the same polytropic index
as for the disk $\gamma_{\rm G} = 4/3$.
The density ratio between disk and corona is $10^{-4}$ at the inner disk radius.
The corresponding $\kappa$ is larger for the corona, $\kappa = 1$, in order to be able to provide a pressure
equilibrium along the disk surface.
Thus the coronal gas has a higher entropy.
We note that the disk corona (the initial gas distribution above the disk) will be swept
out quickly by the disk wind.
This setup, but applying $\gamma_{\rm g} = 5/3$, is widely used in 
non-relativistic simulations of jet launching from accretion disks 
\citep{2007A&A...469..811Z, 2012ApJ...757...65S,2014ApJ...793...31S}.

However, in difference to non-relativistic simulations, a hydrostatic corona that touches the black
hole horizon cannot remain in equilibrium and will collapse to the black hole quickly.
General relativistic modelling therefore requires a floor model for those regions of the
simulations that are not affected by mass loading and outflow activity.
This is a general issue of all GR-MHD simulations published so far.
Consequently, the treatment of the axial outflows for rotating black holes must be taken with care.
The axial mass flux that are considered may be dominated by the floor model, potentially affecting the induction
also of the toroidal magnetic field.
Still, the axial jet - a Poynting flux dominated structure - is indeed a realistic feature, a region close to the
force-free limit generated by the magnetic field anchored in the ergosphere
\citep{1977MNRAS.179..433B, 2009MNRAS.394L.126M, 2010ApJ...711...50T}.

In principle, one may consider a floor model that follows the same polytropic index $\gamma_{\rm g}$ as 
for the disk.
However, this must not necessarily be the case.
We find it beneficial for most simulations to apply a radial profile for the floor model that is flatter.
In this paper, we assume the original profile applied in HARM with $\rho(r) \sim r^{-3/2}$ and correspondingly 
for the internal energy $u \propto \rho /r$.
The maximum density for the floor density is $10^{-5}$ times the (initial) disk density at the (initial) 
inner disk radius.
The radial profile would correspond to a polytropic index $\gamma_{\rm g} = 5/3$.
The numerical advantage is that the density decreases less with radius compared with a gas law with
index $4/3$ and thus the quantity $\rho/B$ that is essential for convergence of the relativistic MHD 
code remains at a larger value in particular along the rotational axis, where there is no physical mass 
injection from the disk.
The floor density and internal energy is only activated if the physical values fall below
a certain threshold at this point in space.

\begin{figure}
\centering
\includegraphics[width=3.5cm]{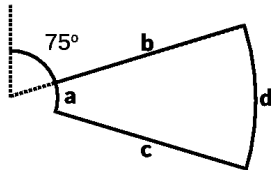}
\caption{Illustration of the control volume used to integrate the accretion and ejection rates.
Surfaces {\em a} and {\em d} consider accretion, while Surfaces {\em b} and {\em c} consider 
ejection.
}
\label{astro_dis_accre_eject_area_example_img}
\end{figure}

\subsection{Initial magnetic field and diffusivity}
\label{astro_disk_initial_condition_mag_subsec}
The initial magnetic field is purely poloidal and follows a distribution commonly applied in 
non-relativistic jet launching simulations \citep{2007A&A...469..811Z, 2012ApJ...757...65S}.
The initial field is defined by the vector potential 
\begin{equation}
\label{astro_dis_B_init_eq}
A_{\phi}(r,\theta ) = B_{\rm p,0} \left( r \sin\theta \right)^{3/4} \frac{m^{5/4}}{\left( m^2 + \tan^{-2}\theta \right)^{5/8}}.
\end{equation}
The parameter $B_{\rm p,0}$ determines the strength of the initial magnetic field and is determined by the choice of the plasma
$\beta \equiv p_{\rm gas}/p_{\rm mag}=8\pi p/\pmb{B}^{2}$. 
The parameter $m$ defines the inclination angle of the magnetic field lines along the initial disk surface. 
We apply $m=0.4$.
Naturally, the field inclination and thus the launching angle for the outflow changes during the simulation.

In our simulations, the magnetic diffusivity is not constant in the computational domain. The profile follows
\begin{equation}
    \eta(r,\theta) = \eta_{0}  \exp\left[- 2 \left(\frac{\alpha}{\alpha_{\eta}}\right)^2\right],
\label{astro_dis_eta_init_eq} 
\end{equation}
which depends on $\theta$ and symmetric to the disk mid-plane while decreasing exponentially with distance 
from the disk mid-plane.
Here, $\alpha \equiv \pi/2 -\theta$ is the angle towards the disk mid-plane,  and 
$\alpha_{\eta} \equiv \arctan(\chi \cdot \epsilon_{\rm D})$ 
is the angle defining the scale height of the diffusivity profile. 
$\chi$ is a scale parameter (see below). 
Equation~(\ref{astro_dis_eta_init_eq}) is a Gaussian profile over $\theta$, whose maximum value is determined by 
parameter $\eta_{0}$ (see also the model setup in \citealt{2012ApJ...757...65S}). 
Thus, the choice of $\eta_0$ decides the peak of the Gaussian and can be treated as the indication of the diffusive 
level, while the choice of $\chi$ controls the width of the Gaussian. 
In this paper, we apply $\chi=0.8$ and $\chi=1.0$ for our simulations (see Section \ref{astro_dis_survey_instruct_subsec}).  

The strength of the magnetic field and the magnetic diffusivity are essential for mass accretion and jet launching.
In the simulation this is controlled by the parameters $\eta_{0}$, the maximum diffusivity, the plasma-beta $\beta$.
The field inclination parameter $m$ plays a role for the initial jet formation until advection of magnetic flux
changes the field distribition.
These parameters are shown in  Table \ref{astro_dis_sim_table}.
Figure \ref{astro_dis_init_example_img} shows the initial density and magnetic field structure profile of simulation
{\em D1}. 

\subsection{Simulation parameters}
\label{astro_dis_survey_instruct_subsec}
Table \ref{astro_dis_sim_table} shows our choice of simulation parameters.
They mainly cover two surveys strategies. 
Simulations {\em D1} - {\em D7} together with simulations {\em D9} and {\em D10} are 
supposed to investigate the disk outflow evolution under different levels of magnetic diffusivity 
(see Section \ref{astro_dis_acc_sec}). 
Simulations {\em D11} - {\em D15} investigate the influence of different 
black hole rotation periods (see Section \ref{astro_dis_wind_vs_BZ_sec}). 
Additionally, the comparison between simulations {\em D7} and {\em D8} demonstrate the influence of the 
grid resolution, namely the impact of numerical diffusivity (see Appendix \ref{astro_dis_high_resolution_sec}).  
Simulation {\em D0} is a control simulation to examine the choice of initial density and angular 
velocity, where the behavior of a very weakly magnetized thin disk is investigated.
As discussed later, simulation {\em D0} is the only case where no disk outflow is observed.

\subsection{Measuring the accretion rate and ejection rate}
\label{astro_dis_acc_eject_illu_sec}
The mass accretion rate and ejection rate are essential parameters that quantify the evolution of the disk-jet system.
We calculate them applying a control volume as shown in Figure \ref{astro_dis_accre_eject_area_example_img}. 
The accretion rate is numerically integrated following
\begin{eqnarray} 
\dot{M}_{\rm acc}(r) = \int_{\theta_{1}}^{\theta_{2}} 2 \pi \rho(r,\theta) u^{r}(r,\theta) \sqrt{-g} d\theta, 
\label{astro_dis_acc_def_eq}
\end{eqnarray}
while for the ejection rate we apply
\begin{eqnarray} 
\dot{M}_{\rm eje}(\theta) = \mp \int_{r_{1}}^{r_{2}} 2 \pi \rho(r,\theta) u^{\theta}(r,\theta) \sqrt{-g} dr.
\label{astro_dis_eject_def_eq}
\end{eqnarray}
Here we consider poloidal mass fluxes that flow across the specific surfaces vertically. 
The minus sign is applied for surface {\em b},  while the plus sign denotes surface {\em c}.

\begin{figure}
\centering
\includegraphics[width=4.2cm]{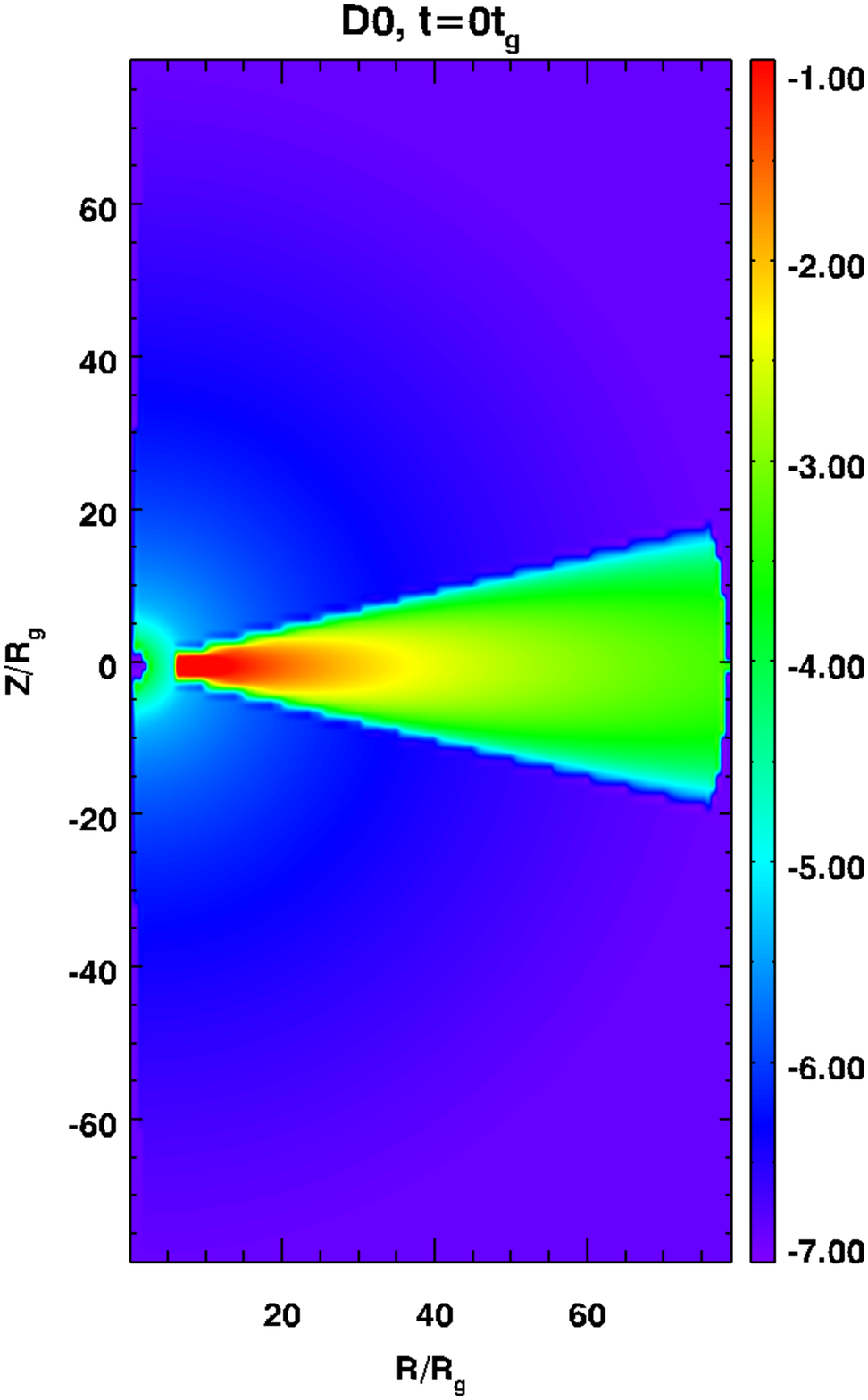}
\includegraphics[width=4.2cm]{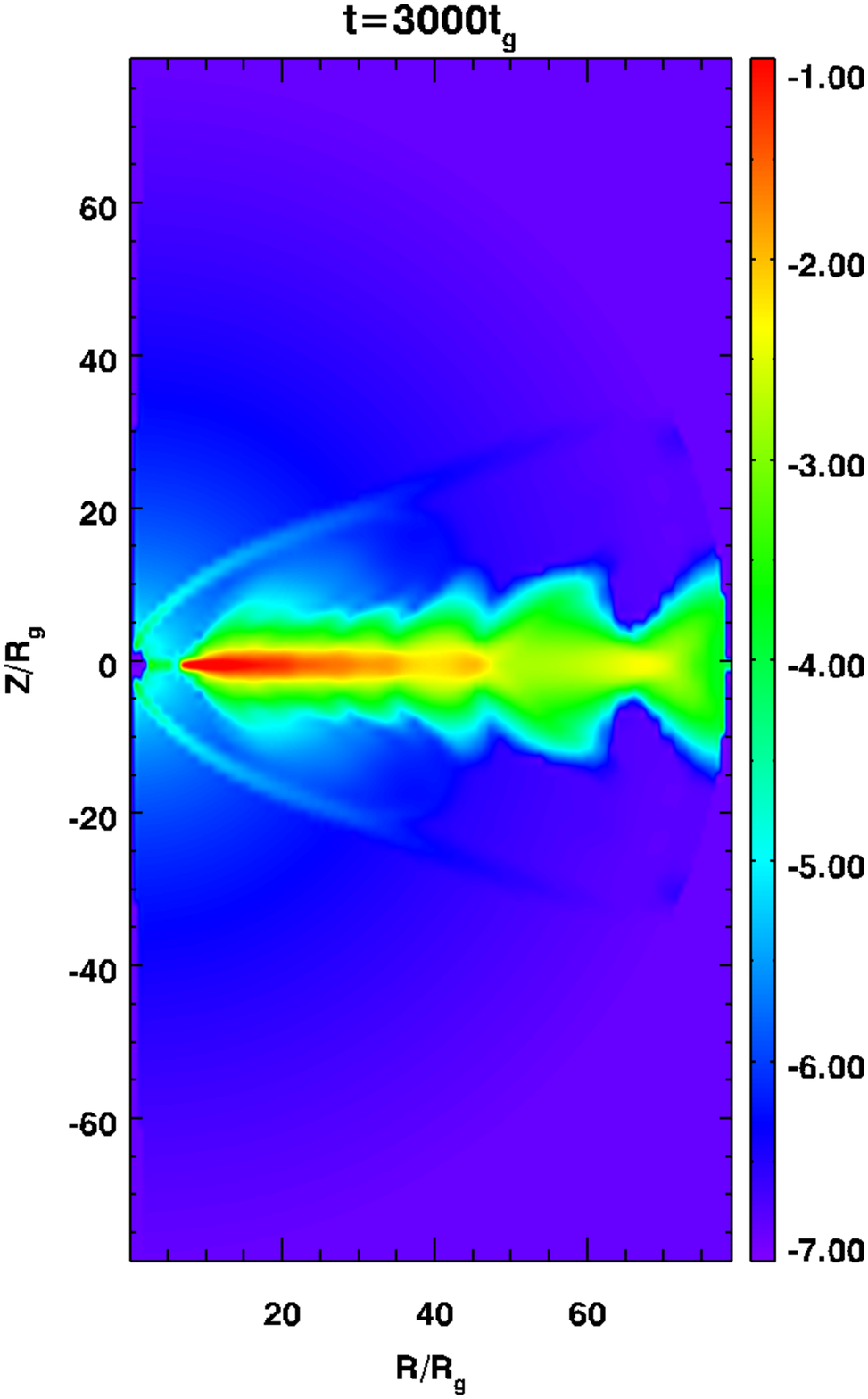}
\caption{Snapshots of the density distribution for simulation {\em D0} at $t=0$ (left), $t=1000$ (middle) and $t=3000$ 
(lower right). No field lines are shown as $\beta=10^{8}$.
Apart from some turbulent structure along the disk surfaces, the accretion disk basically kept its disk-like shape 
during the time evolution and no outflow stream originates from the disk. 
}
\label{astro_dis_weak_B_rho_img}
\end{figure}

We integrate the radial accretion rate along surfaces {\em a} and {\em d} using 
Equation~(\ref{astro_dis_acc_def_eq}), while at 
surface {\em b} and {\em c}, the poloidal ejection will be measured uisng Equation~(\ref{astro_dis_eject_def_eq}).  
We set surface {\em a} at $r=6$ (disk inner boundary), while surface {\em d} is further outside and is not 
of concern in this paper. 
Surface {\em b} is set along $\theta_{1}=75^{\circ}$ (the disk initial upper boundary) and surface {\em c} along
$\theta_{2}=105^{\circ}$ (the disk initial lower boundary). 
The ejection process in the simulations is basically symmetric to the disk mid-plane, thus the ejection rates 
presented in the data analysis are the sum of $\dot{M}_{\rm eje}$ at surfaces {\em b} and {\em c}.

A negative mass flux rate at radius $a$ and $d$ means radial accretion, while a positive value indicates radial motion 
outwards. 
For the ejection, a positive ejection rate at surfaces $b$ and $c$ indicates disk wind outflow, while a negative value 
refers to slow mass concentration towards disk mid-plane. 
Both terms ``accretion rate" and ``inner accretion rate" are used to indicate the accretion rate at the disk inner 
boundary $r=6$.

\begin{figure}
\centering
\includegraphics[width=6.2cm]{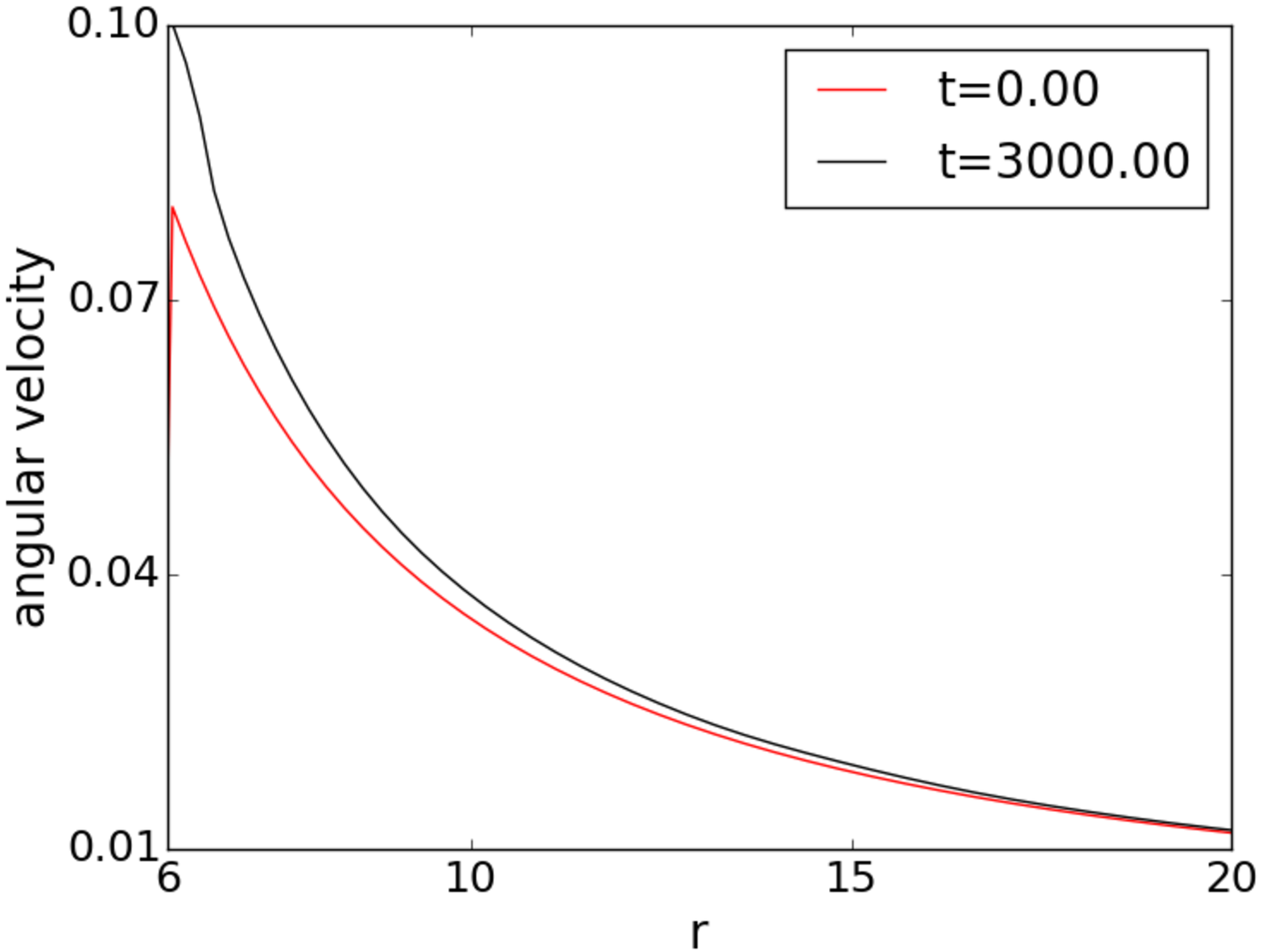}
\includegraphics[width=6.2cm]{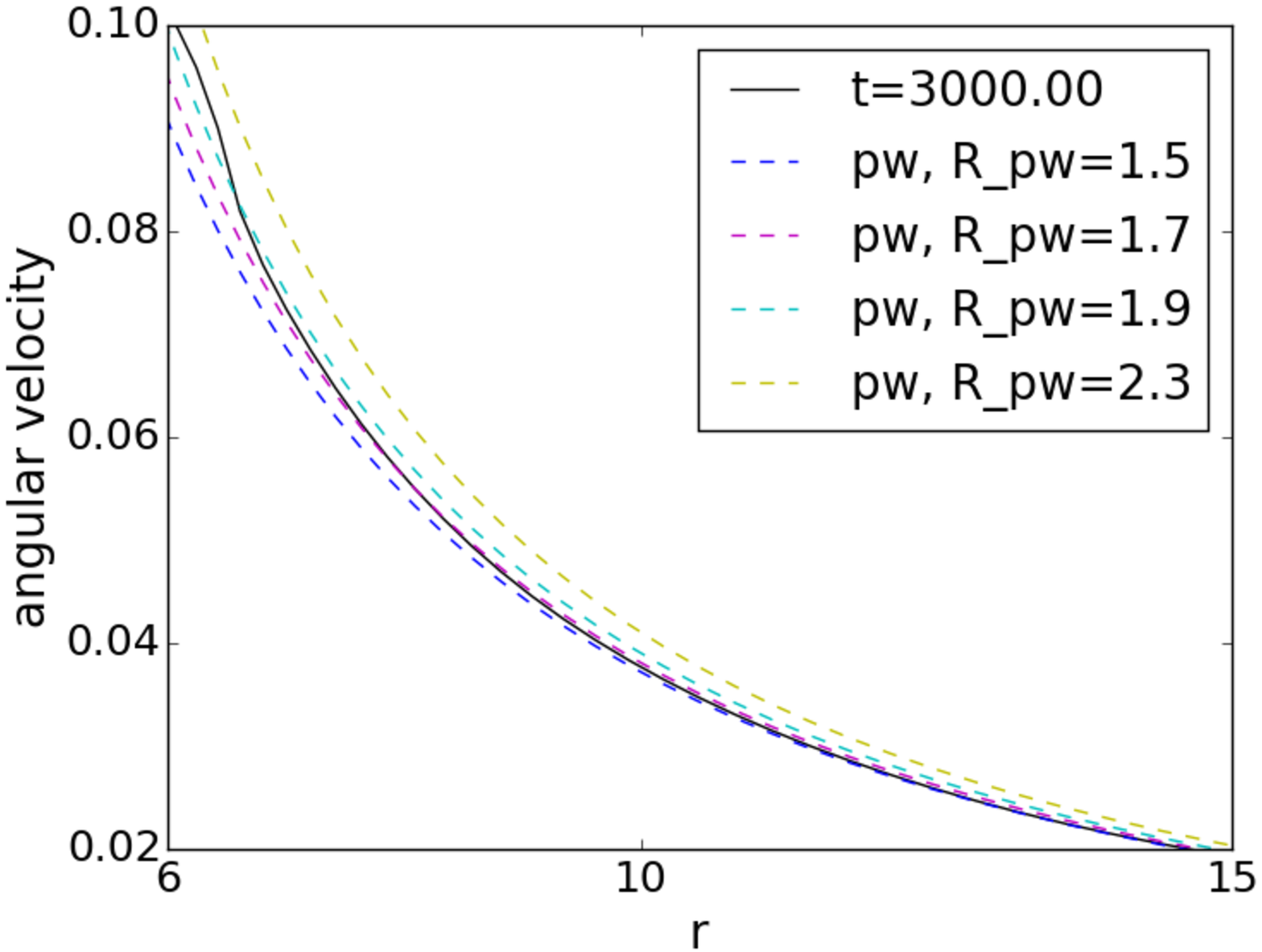}
\caption{The disk angular velocity profile for simulation {\em D0} at time $t=0 $ and $t = 3000$ (top).
The red curve at $t = 0$ also stands for the Paczy\'{n}ski-Witta profile with $R_{pw}=1.0$. 
Throughout this plot, we know that the initial condition given in Equation~(\ref{astro_dis_PW_omega_eq}) is very close 
to the angular velocity at $t=3000$, where the disk has evolved into a steady state.
Simple fits of $R_{pw}$ in the initial angular velocity (bottom).
We note that the plot starts at $r=6$ since this is the disk inner radius.}
\label{astro_dis_weak_B_R_fit_img}
\end{figure}

\begin{figure}
\centering
\includegraphics[width=3.5in]{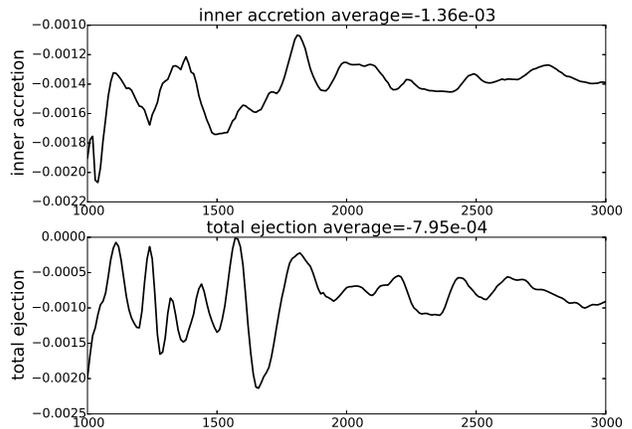}
\caption{Accretion (upper panel) and ejection (lower panel) rates for simulation {\em D0} from $t=1000$ to $t=3000$. 
The average values in the plots are taken from $t=2000$ to $t=3000$. 
The negative ejection rate implies a slow concentration of mass towards the disk mid-plane. 
}
\label{astro_dis_weak_B_acc_eject_img}
\end{figure}

\section{Weakly magnetized disk}
\label{astro_dis_weak_B_sec}
In this section, we examine the evolution of a weakly magnetized disk. 
This is interesting to investigate by two reasons.
Firstly, it allows us also to investigate the stability of our thin disk initial condition.
Secondly, a weakly magnetized disk is subject to the magneto-rotational instability (MRI, see \citealt{1991ApJ...376..214B})
that influences angular momentum transport and, therefore, accretion.
We will see further in Section \ref{astro_dis_morph_sec} that the presence of a strong magnetic field is crucial for the jet 
ejection process. 

\subsection{Disk structure evolution}
We first discuss simulation {\em D0} considering an accretion disk that is only weakly magnetized. 
With an initial $\beta = 10^{8}$ at the inner disk radius, the magnetic field is dynamically negligible.  

In Figure \ref{astro_dis_weak_B_rho_img} we show the density evolution for simulation {\em D0}.
While the disk evolution is overall smooth and stable, we observe a number of discontinuous structures developing
along the disk surface and also inside the disk.
We attribute this to the choice of our initial density distribution.

We argue that the density profile Equation~\ref{astro_dis_thindisk_rho_eq} taken from non-relativistic simulations
does not fit the relativistic Paczy\'{n}ski-Witta rotation profile.
This holds in particular in the vicinity close to the black hole.
We note that the density profile implies a gas pressure profile that triggers the overall force-balance in the 
disk.
With the Paczy{\'n}ski-Witta rotation the centrifugal forces that determine the non-relativistic pressure profile
are not in balance anymore.

In Figure \ref{astro_dis_weak_B_R_fit_img} we show the time evolution of the angular velocity profile along the disk mid-plane. 
We compare the initial angular velocity profile with that for $t=3000$. 
Since the angular velocity profile rarely evolves after $t \sim 2000 $, we may thus consider the profile at 
$t=3000$ as in a quasi steady state. 
We find that in general the initial profile matches quite well with the profile at later times when the disk has evolved to
a new dynamical equilibrium.
Nevertheless, there is discrepancy in the inner disk, in particular for $r<10$.
This discrepancy will then lead to a radial motion (lack of angular momentum support) which further triggers the density 
perturbation. 

We also compare the angular velocity profile of simulation {\em D0} at $t=3000$ to the Paczy{\'n}sky-Witta rotation 
profile  Equation~\ref{astro_dis_PW_omega_eq} for different smoothing parameters $R_{pw}$.
It turns out that no matter which $R_{pw}$ we choose for the initial rotation profile, the disk rotation always evolves into
a profile that is slightly different from a pure PW profile close to the inner boundary.
An initial PW parameter $R_{pw} \sim 1.7$ would fit best to the long-term evolution, and thus may reduce the impact of 
the initial discrepancies. 
However, apart from the initial perturbation patterns, the accretion disk basically keeps its disk-like shape 
as expected.

\subsection{How is accretion supported?}
\label{astro_accretion_discussion_subsec}
%

%

\begin{figure}
\centering
\includegraphics[width=3.5in]{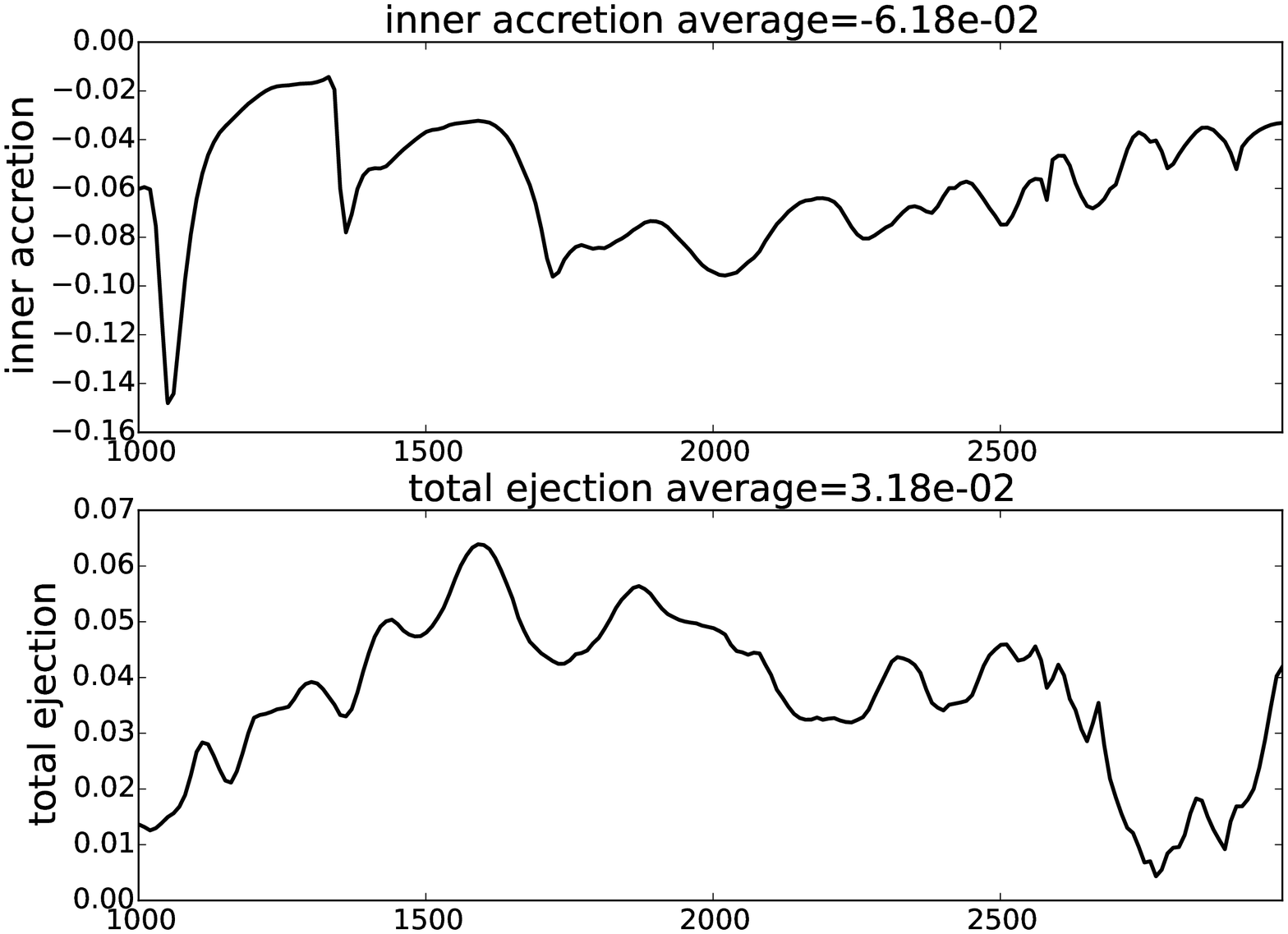}
\caption{Accretion (upper panel) and ejection (lower panel) rates for simulation {\em D1} from $t=1000$ to $t=3000$. 
The average values are taken from $t=2000$ to $t=3000$. 
The positive ejection rate implies the generation of disk wind.
}
\label{astro_dis_D1_acc_eject_img}
\end{figure}

%

\begin{figure*}
\centering
\includegraphics[width=3.2in]{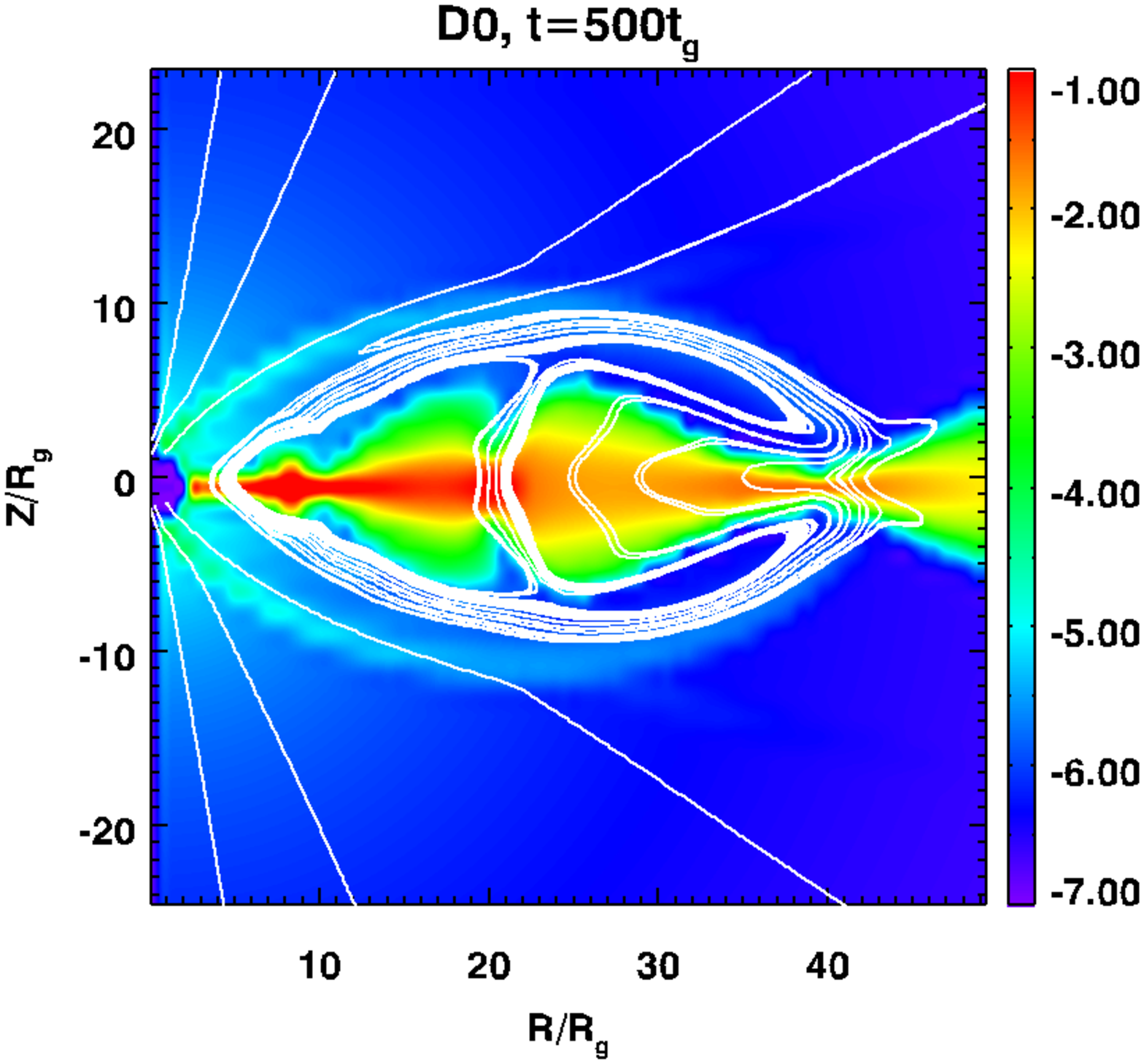}
\includegraphics[width=3.2in]{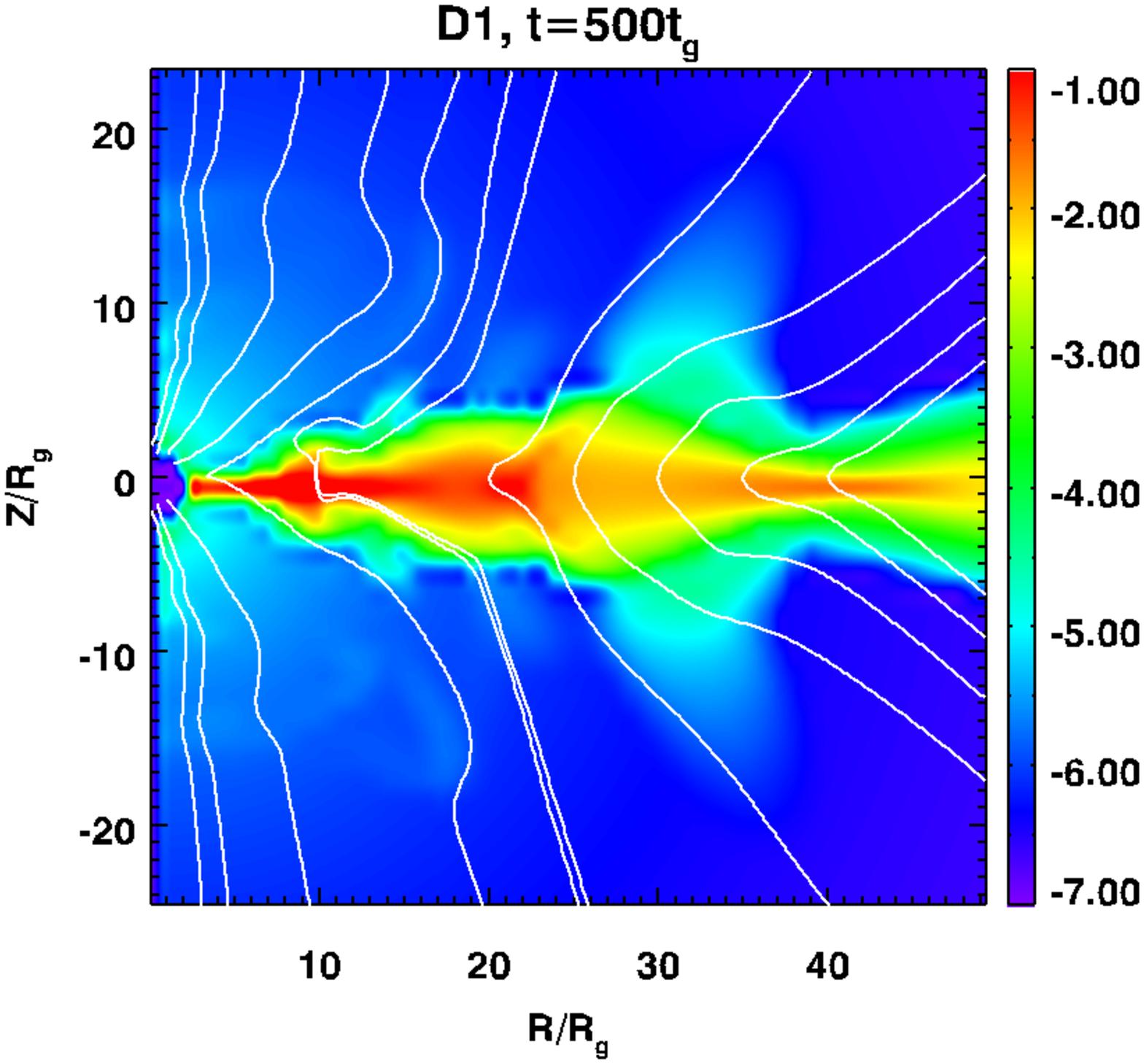}
\caption{Density and magnetic field structure for simulations {\em D0} (left) and {\em D1} (right) at t=500.
}
\label{astro_dis_D0D1_MRI_leverarm_discussion_img}
\end{figure*}

Here we discuss the question how accretion is initiated and maintained in our thin disk simulations
that start with a Keplerian Paczynsky-Wiita disk rotation law.

In the following we will consider the angular momentum transport mechanism in simulations {\em D0} and 
{\em D1}.
In Figure~\ref{astro_dis_weak_B_acc_eject_img} we show the accretion and ejection rates for simulation {\em D0}
(integrated according to the definitions in Section~\ref{astro_dis_acc_eject_illu_sec}). 
We concentrate on the time interval $t=1000-3000$, during which the overall disk evolution can be considered as in
quasi-steady state.
The time $t=3000$ corresponds to $\sim 39$ orbital periods at the initial disk inner radius $r=6$.
We thus consider the mass accretion rate at $r=6$ as an indicator of the dynamical evolution inside the 
disk.
As shown in Figure \ref{astro_dis_weak_B_acc_eject_img} there is continuous accretion of matter with an 
accretion rate $\dot{M}_{\rm acc} = 1.36 \times 10^{-3}$ from $t=2000 $ to $t=3000$ at this 
radius. 

Accretion, however, implies that the angular momentum must have been removed from 
the disk material.
We may consider two potential angular momentum transport mechanisms here.
One is magnetic braking by the lever arm of the large-scale field, where the lever arm is usually
defined by the foot-point of the magnetic field line and the Alfv\'en radius.
The other is the magneto-rotational instability (MRI), where the angular momentum is transported locally but
also by a magnetic torque\citep{1991ApJ...376..214B}.

Which of the two mechanisms is operating, depends on the structure and strength of the poloidal magnetic field.
In Figure~\ref{astro_dis_D0D1_MRI_leverarm_discussion_img} (upper left) we show the field line 
distribution for simulation {\em D0} at $t=500$ (at this time we could in principle still catch the MRI growth
in linear regime inside the disk). 
Due to the weak initial field strength (plasma-$\beta$ = $10^{8}$), the initial field structure has collapsed
and is now mainly confined to the disk.
We can see that on larger scales the magnetic field lines form {\em loops} within the disk. 
As these loops connect differentially rotating parts of the disk, a magnetic torque arises
that removes angular momentum from the inner disk, and the disk material will subsequently accrete.
However, due to the weak field the magnetic torque provides only a weak angular momentum transport.

On the other hand, weakly magnetized disks can be subject to the magneto-rotational instability 
\citep{1991ApJ...376..214B}. 
It is therefore interesting to check whether in our simulations we see any indication for the MRI.
A typical signature might be wave-like wiggles that grow along the originally vertical disk
magnetic field.
For our chosen field strength and numerical resolution, there is no chance that simulation
{\em D0} can capture the growth of the fastest growing MRI mode with wavelength $\lambda_{\rm MRI} \propto B/\rho$.
Since in principle other MRI modes with wavelengths above a critical wavelength may grow as well (however slower), 
we may hypothesize that simulation {\em D0} should be able to capture such wave modes if their wavelengths 
are larger than the resolution limit.
We note that the disk region within $75^{\circ} < \theta < 105^{\circ}$ is resolved by 48 grid cells in $\theta$ 
direction after all.
Naturally, only those modes may be found that (i) are resolved by the grid resolution, and (ii) that have a
wavelength that fits into the disk height.

As a matter of fact, so far, we do not find strong indication for the MRI in our simulations.
We observed a tangled magnetic field component in the disk, but this more likely arises from the
unsteady disk evolution.

Another way to transport angular momentum is - similarl to the MRI - by turbulence that is, however,
induced by the unsteady evolution of the accretion disk.
For example, our initial disk structure is only marginally stable and initially develops a hydrodynamic 
substructure that also leads to a tangled magnetic field within the disk.
This field - connecting differentially rotating foot points will be able to remove angular momentum locally
and thus, allow for accretion.
The efficiency of this process depends on the magnetic field strength.

Due to the low magnetic field strength the lever arm of a disk wind cannot play a substantial role
for angular momentum exchange in simulation {\em D0}.
Here, the only option is the tangled disk field arising from the unsteady disk evolution.
Still, due to the weakness of the field, the process is inefficient and we do not expect large accretion
rates in simulation {\em D0}.

Simulation {\em D1} has initial conditions very similar to simulation {\em D0}. 
The only difference is the initial magnetic field strength, measured by the initial plasma-beta, 
$\beta = 10$ for {\em D1} and $\beta=10^{8}$ for {\em D0} (see Table \ref{astro_dis_sim_table}). 
In Figure~\ref{astro_dis_D1_acc_eject_img}, we show the accretion and ejection rates of simulation {\em D1}. 
Comparing it to Figure~\ref{astro_dis_weak_B_acc_eject_img}, we find that both the average accretion rate 
for simulation {\em D1} are almost 40 times larger than those for simulation {\em D0}. 
Furthermore, we see a positive averaged ejection rate in simulation {\em D1} which indicates clear outflows.
This demonstrates that a strong magnetic field is essential for launching for accretion and ejection. 

In order to figure out what drives the accretion of matter in the stronger field case, we also plot 
the magnetic field structure for simulation {\em D1} in Figure \ref{astro_dis_D0D1_MRI_leverarm_discussion_img}. 
In the right panel, we see that the magnetic field, unlike the simulation {\em D0}, keeps its open
field structure during the disk evolution on large scale. 
However, the field lines close to the disk mid-plane are dragged towards the black hole by advection.
As a result, the initial field becomes bent even more with time.
Thus, the magnetic field connects differentially rotating parcels of disk material and therefore
exchanges angular momentum between them.

We finally conclude that in our weak-field simulation {\em D0}, the matter accretion results most likely from the 
angular momentum exchange
by a tangled magnetic field arising from the non-steady evolution of the accretion disk.
We expect this process to decrease on very long time scales when a steady hydrodynamic disk structure
has evolved.
In the strong-field simulations, the magnetic field that is tangled by advection over a disk height may
provide a lever arm incentive for the angular momentum transport inside disk and hence may dominate the accretion 
process.

\begin{figure*}
\centering
\includegraphics[width=2.in]{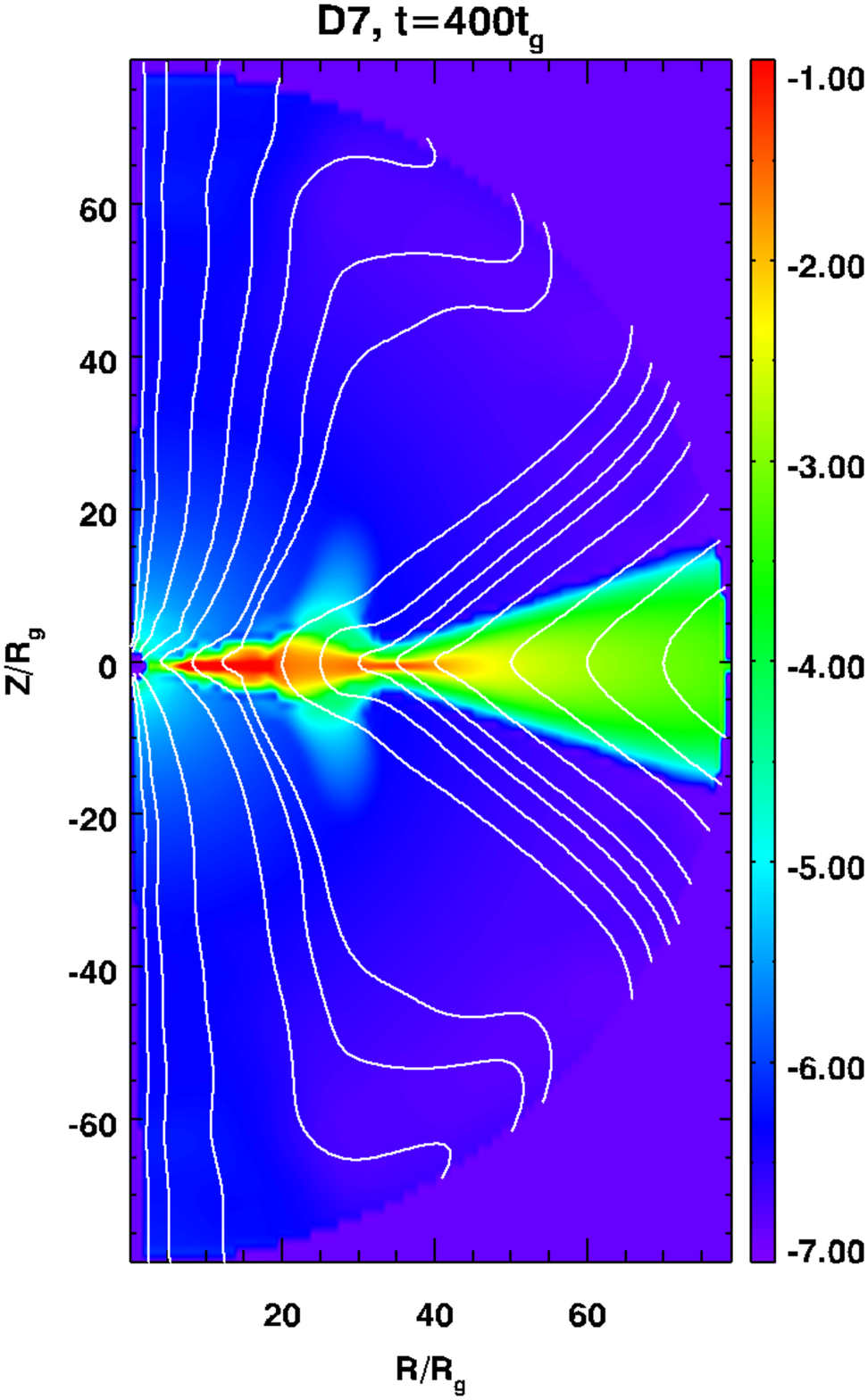}
\includegraphics[width=2.in]{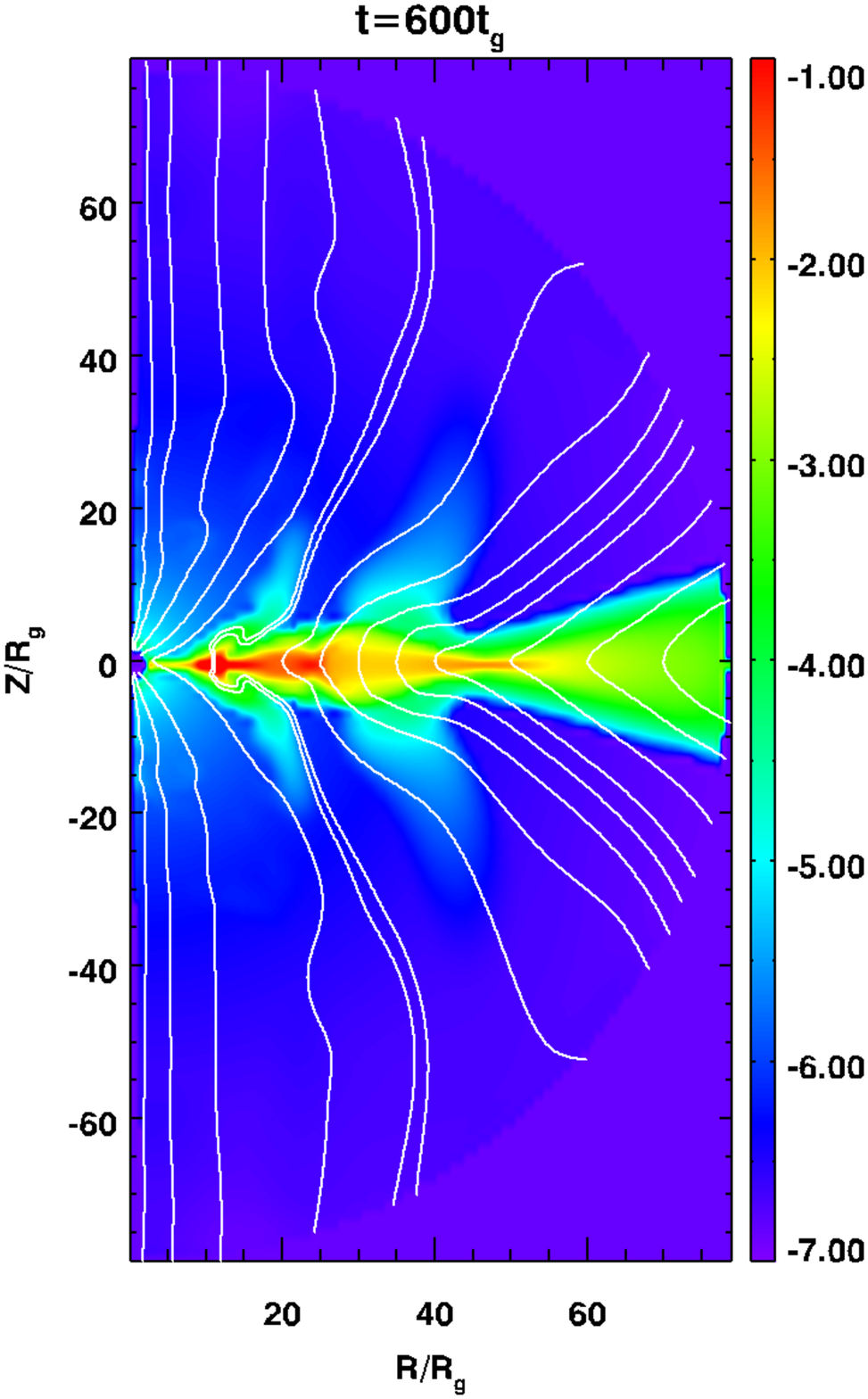}
\includegraphics[width=2.in]{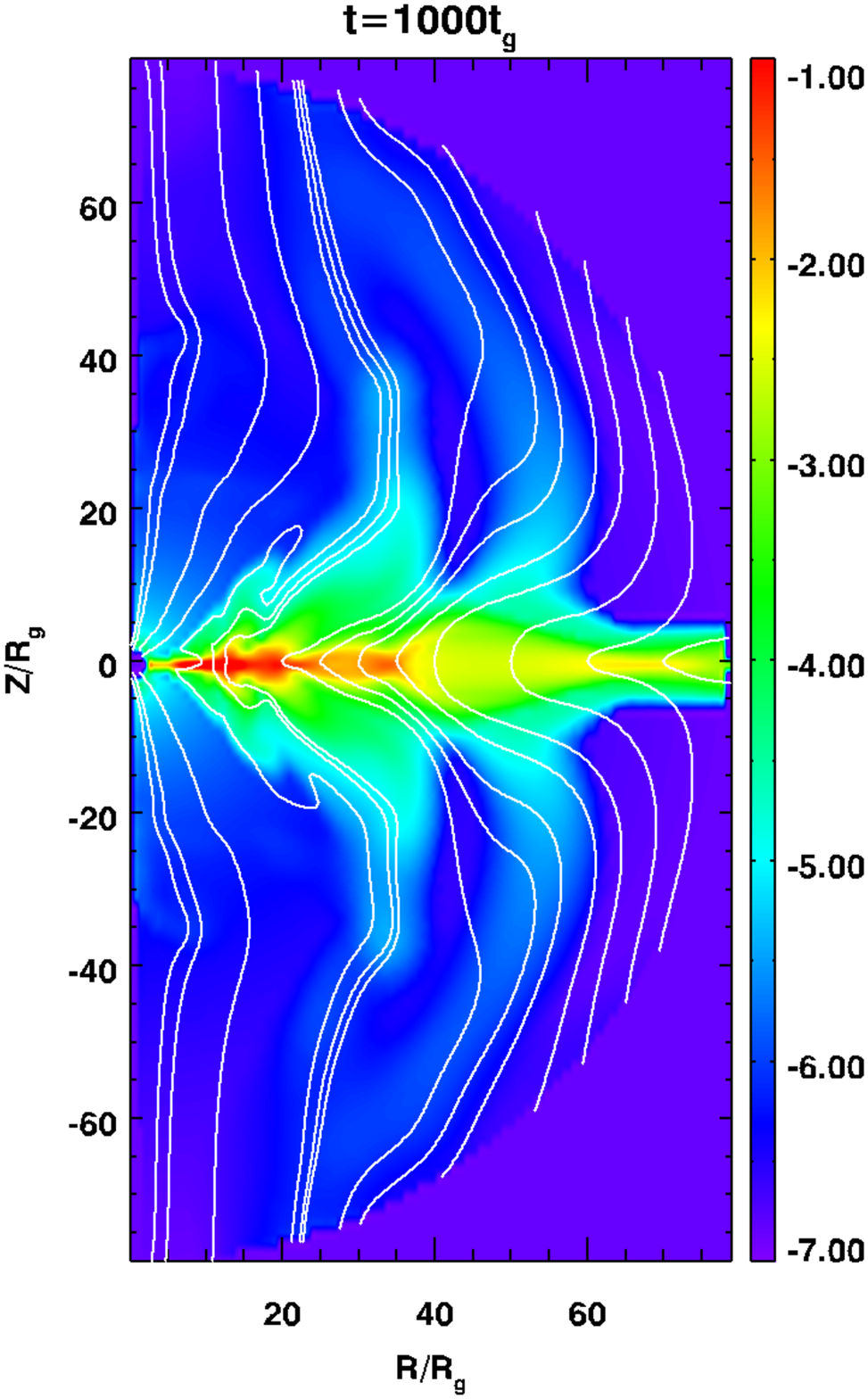}

\includegraphics[width=2.in]{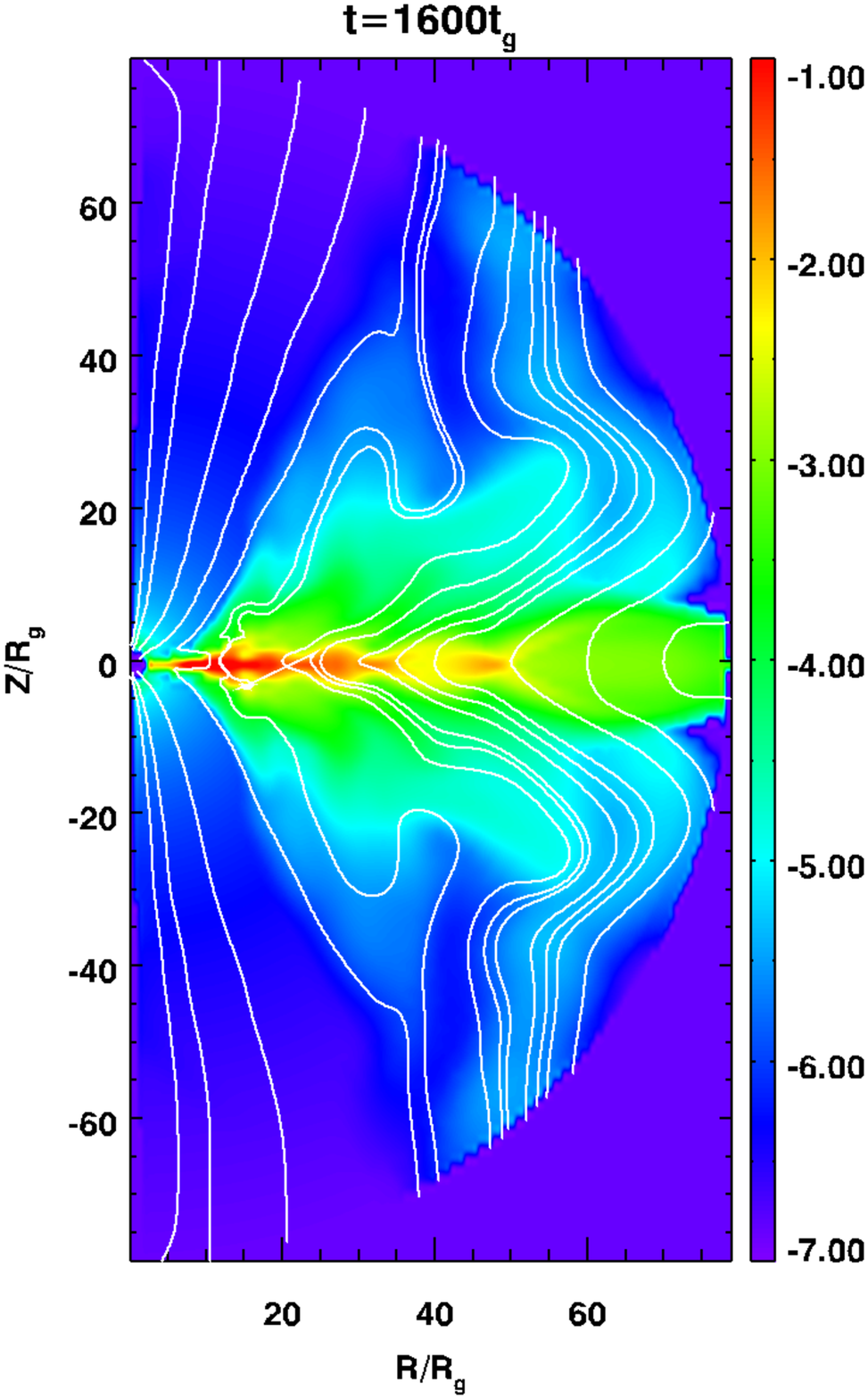}
\includegraphics[width=2.in]{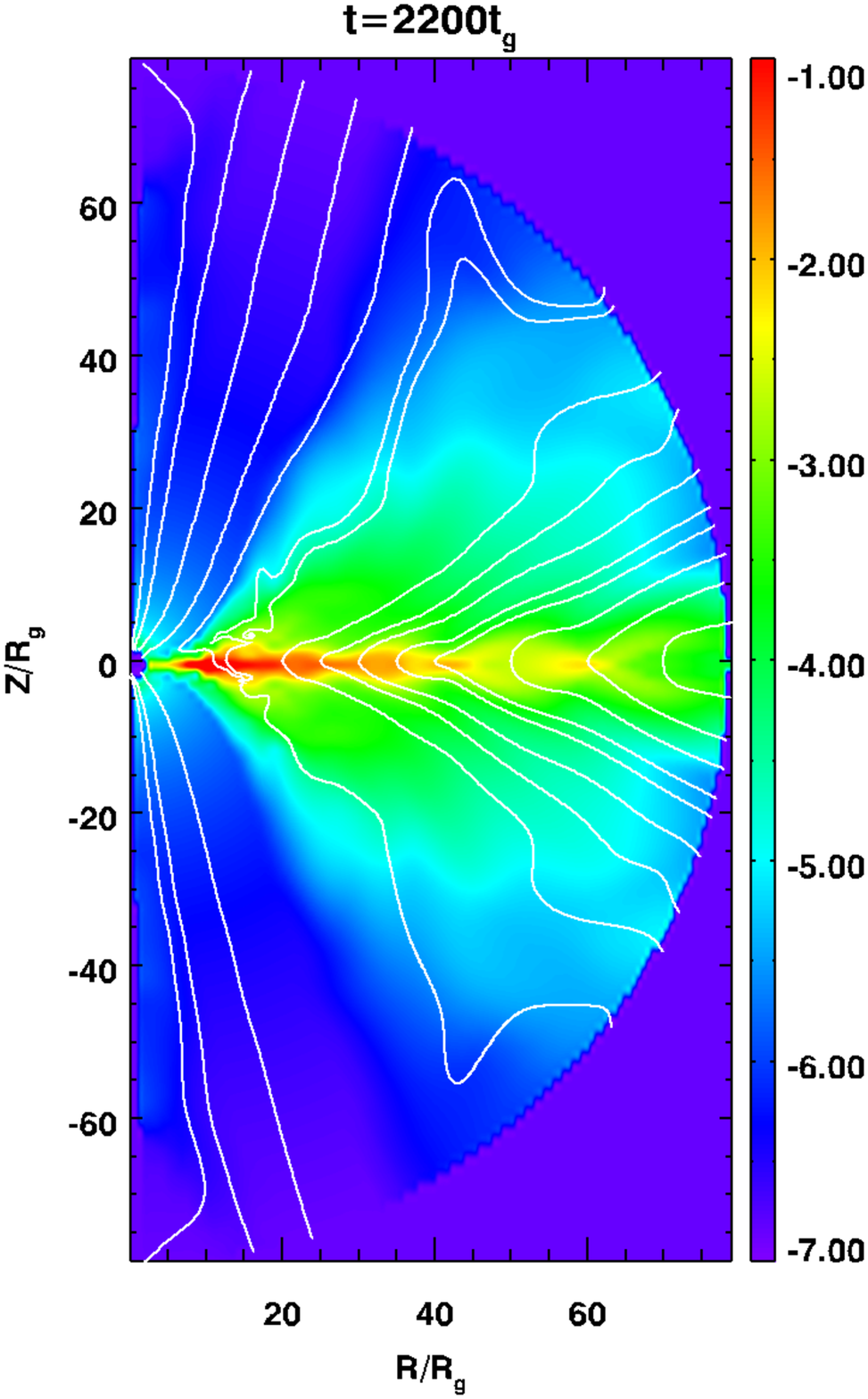}
\includegraphics[width=2.in]{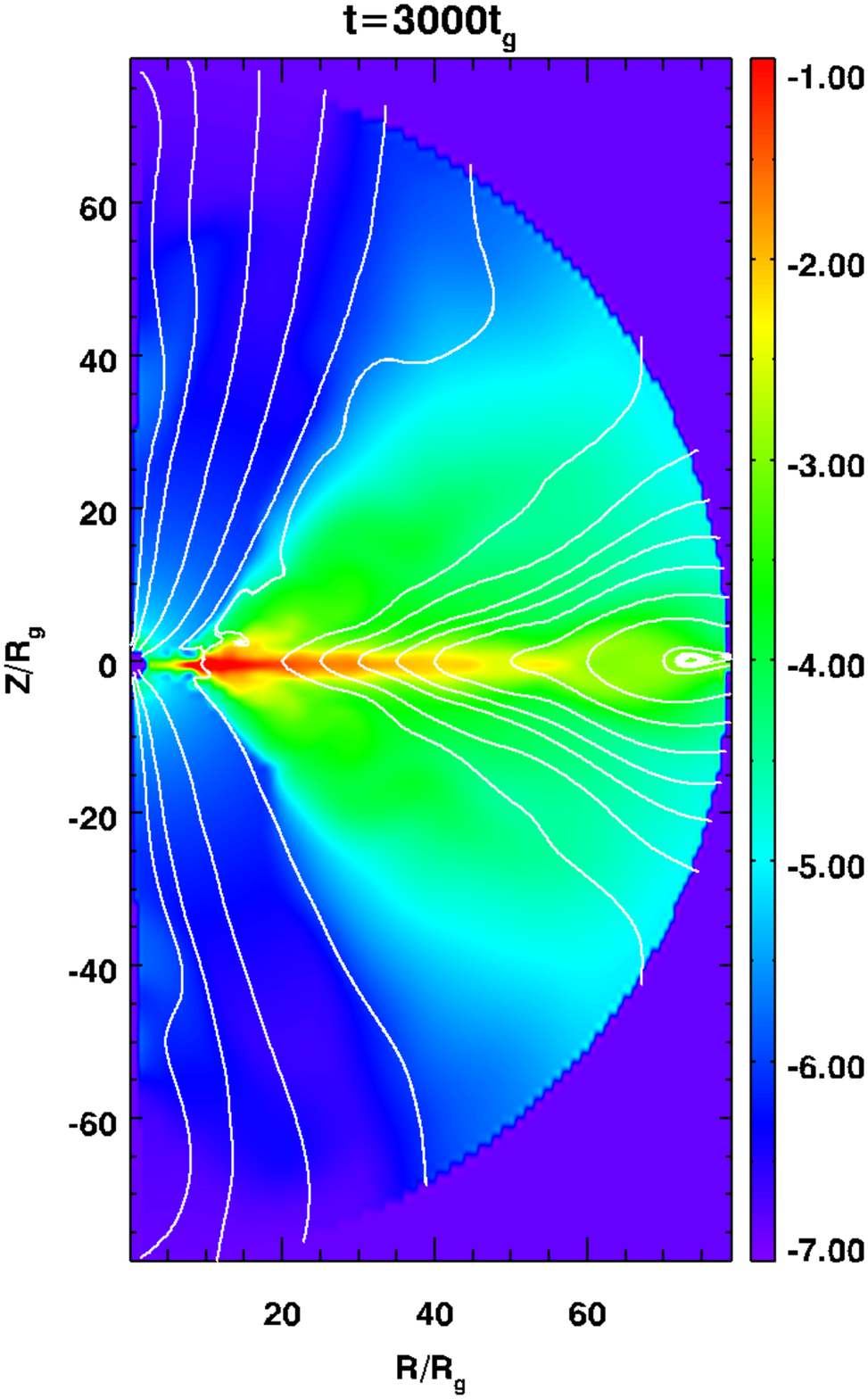}
\caption{Snapshots of density and poloidal magnetic field of simulation {\em D7} at $t=400$ (top left), $t=600$ (top right),
$t=1000$ (middle left), $t=1600$ (middle right), $t=2200$ (bottom left) and $t=3000$ (bottom right). 
The snapshots are presented in Kerr-Schild coordinates. }
\label{astro_dis_morph_D6_rho_img}
\end{figure*}

\begin{figure*}
\centering
\includegraphics[width=2.in]{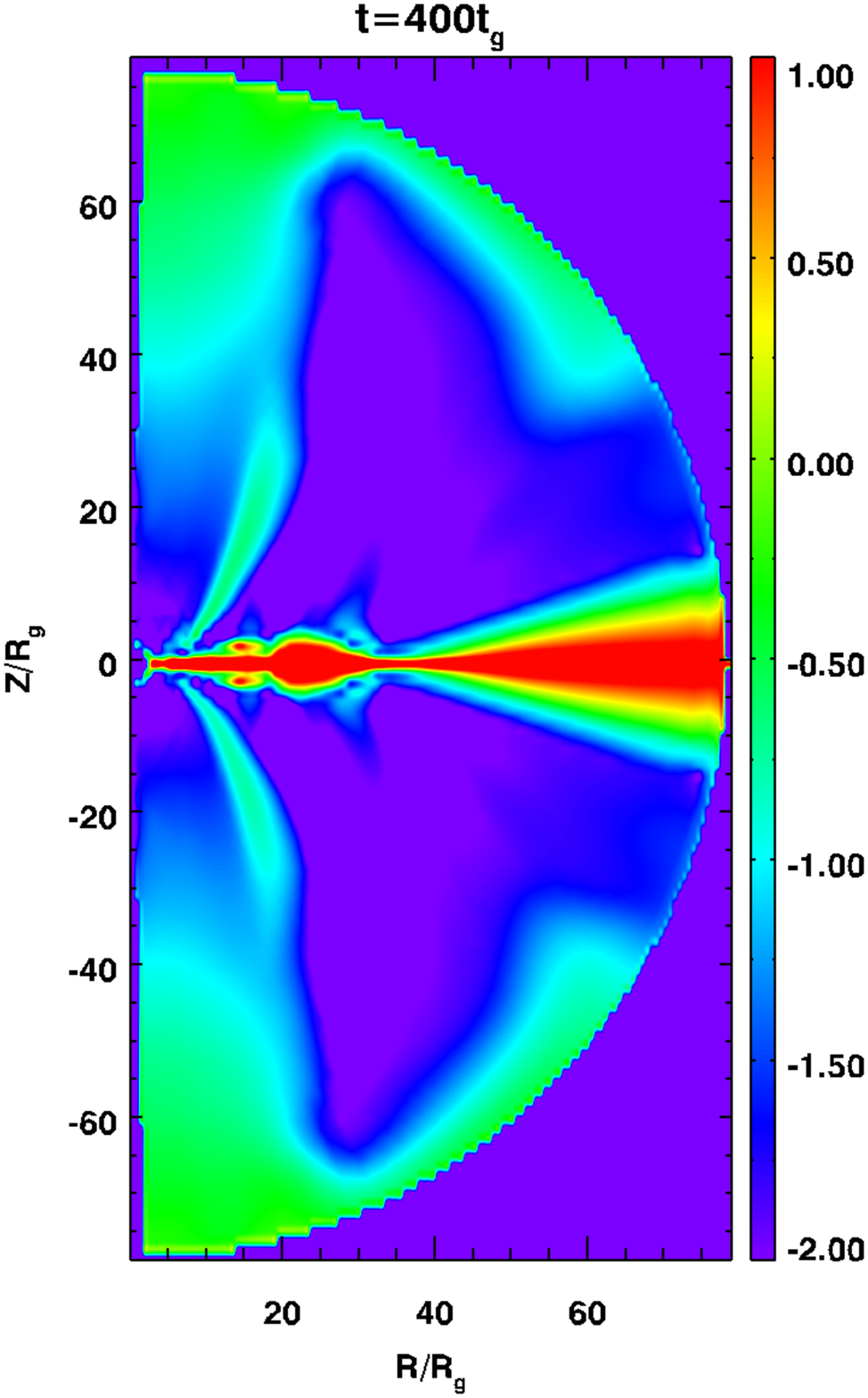}
\includegraphics[width=2.in]{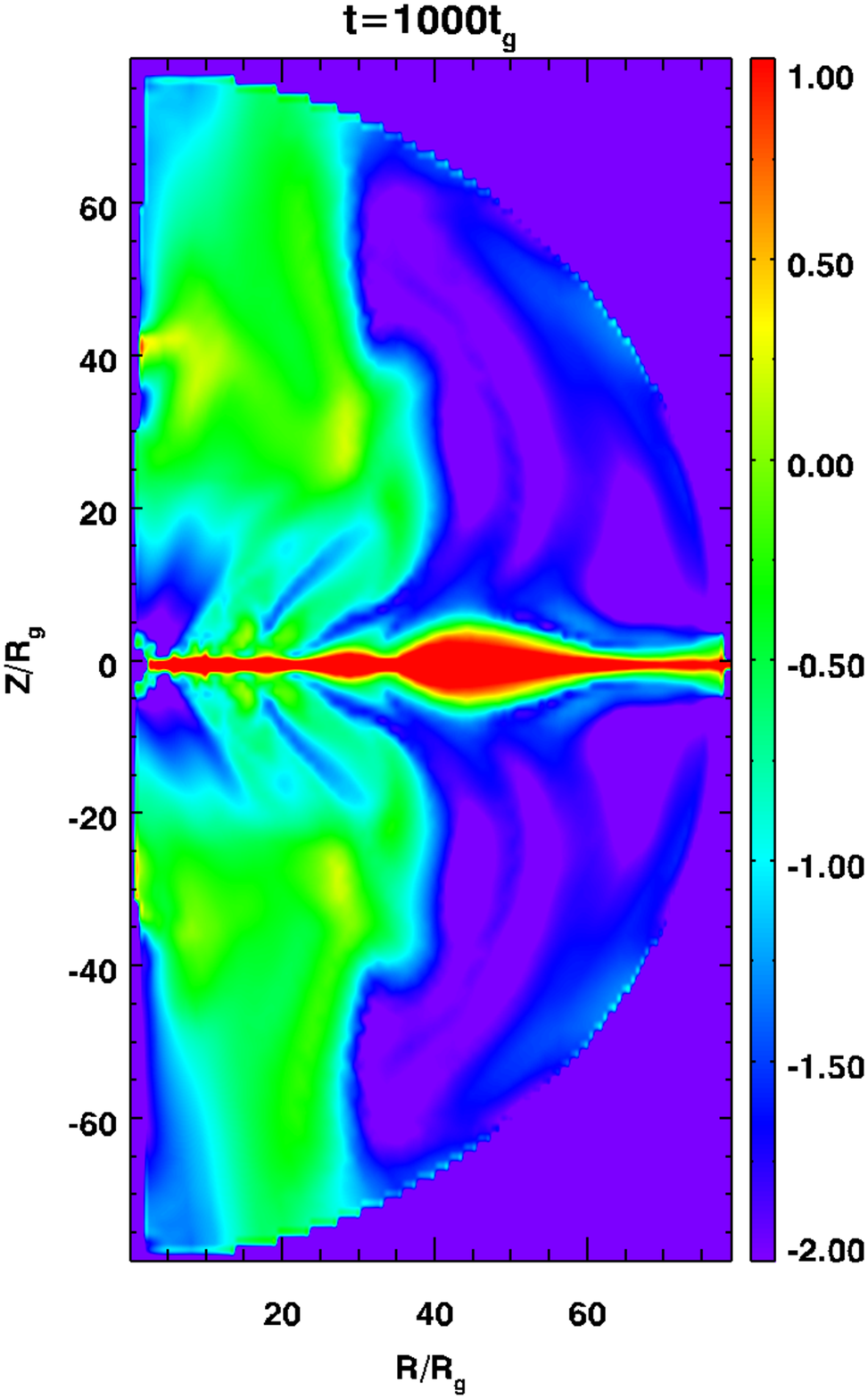}
\includegraphics[width=2.in]{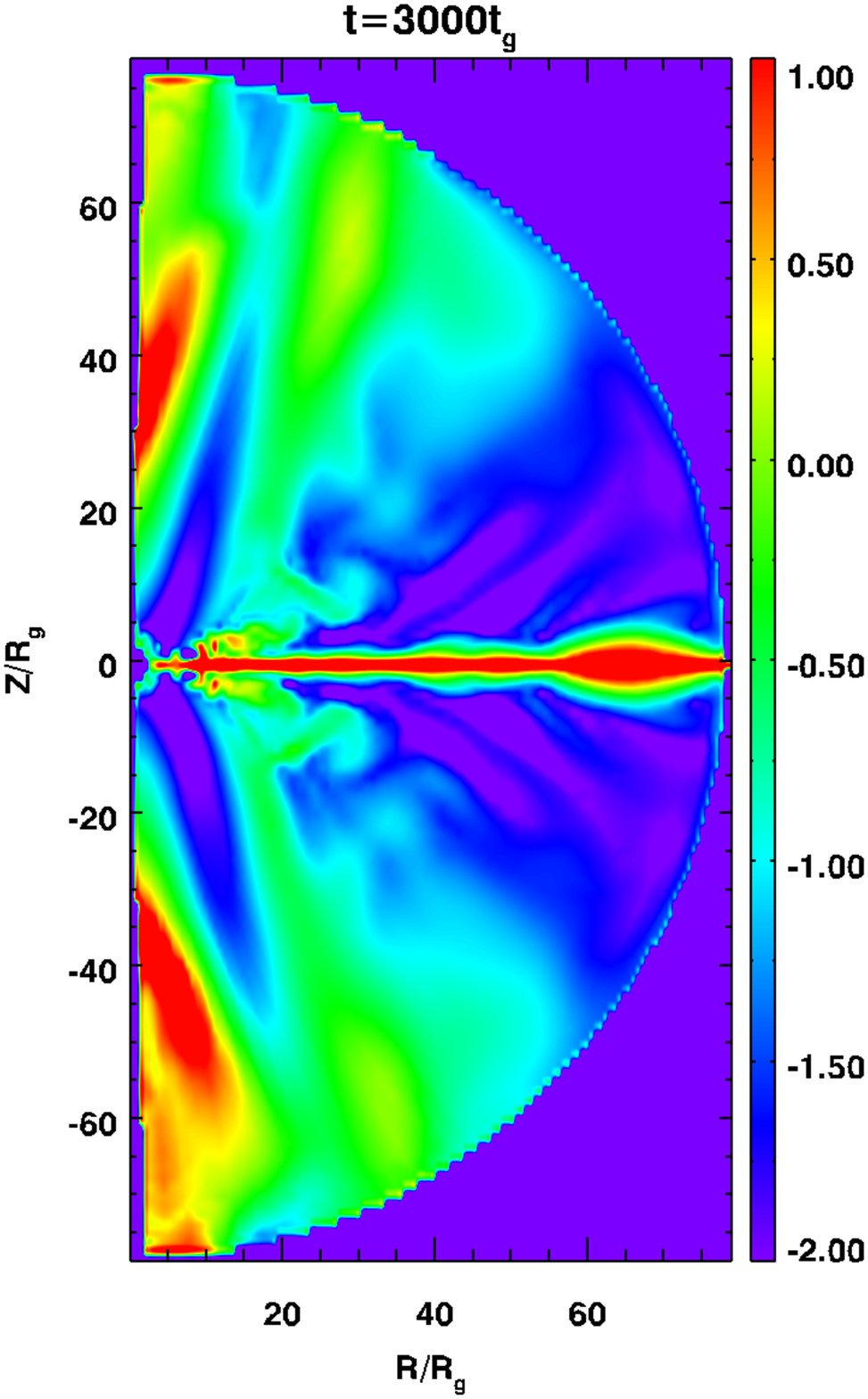}

\includegraphics[width=2.in]{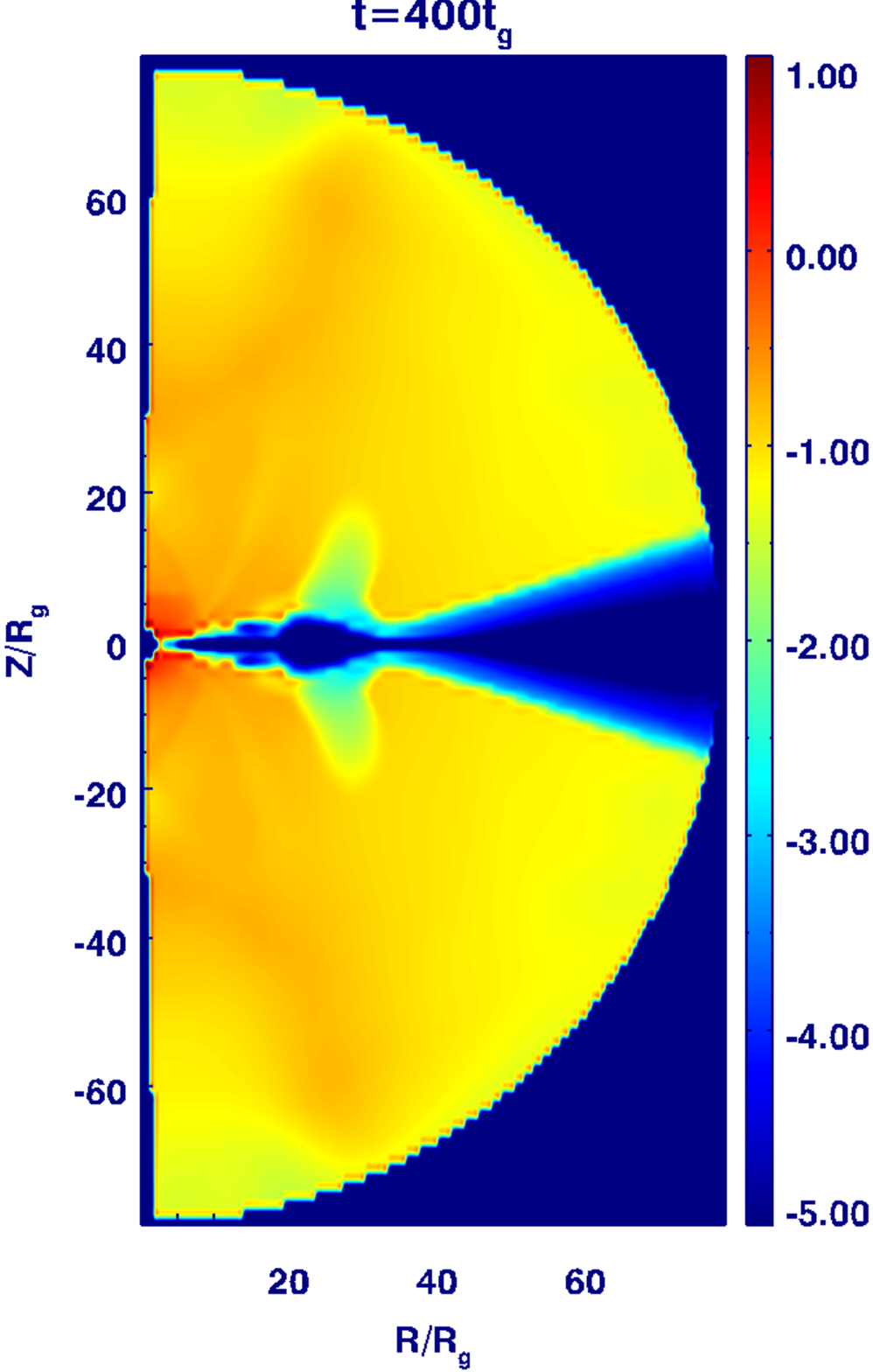}
\includegraphics[width=2.in]{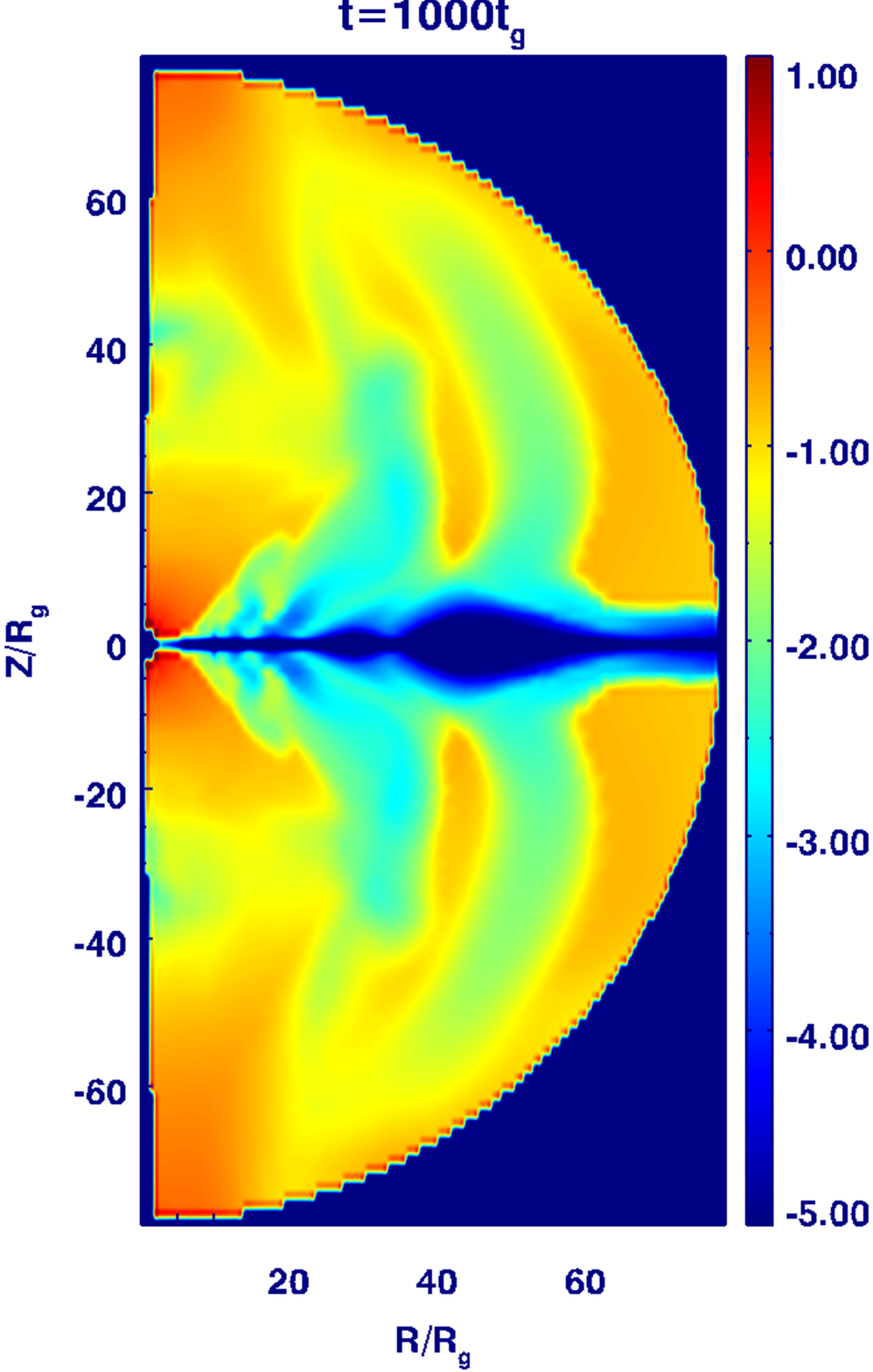}
\includegraphics[width=2.in]{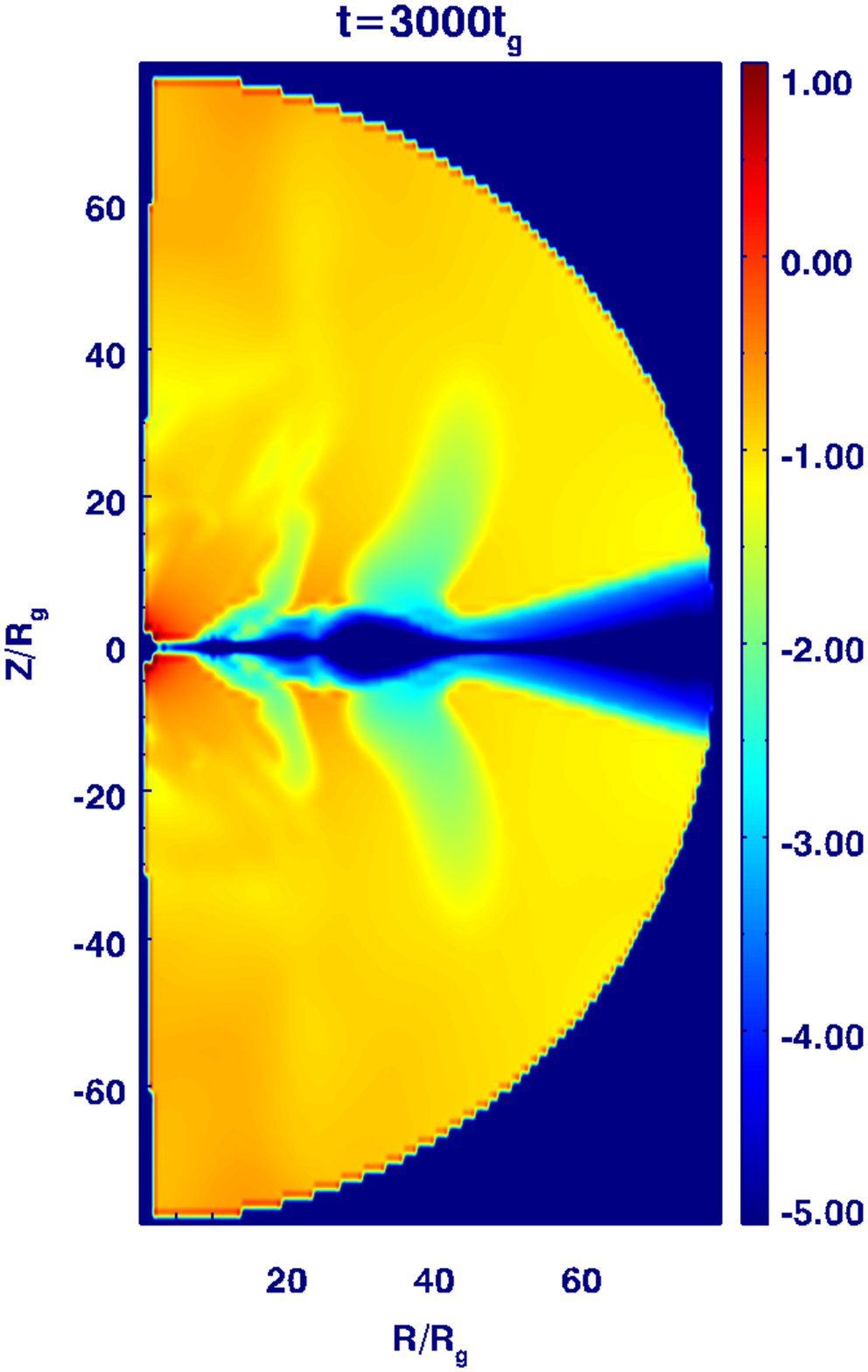}
\caption{Snapshots of the plasma-beta $p/B^2$ (top) and magnetization $B^2/{\rho}$ (bottom) for simulation {\em D7} 
at $t=400$, $t=1000$ and $t=3000$ (from left to right). 
The snapshots are presented in Kerr-Schild coordinates.}
\label{fig_D7_beta_sigma}
\end{figure*}

%
%
%


\begin{figure}
\centering
\includegraphics[width=6.cm]{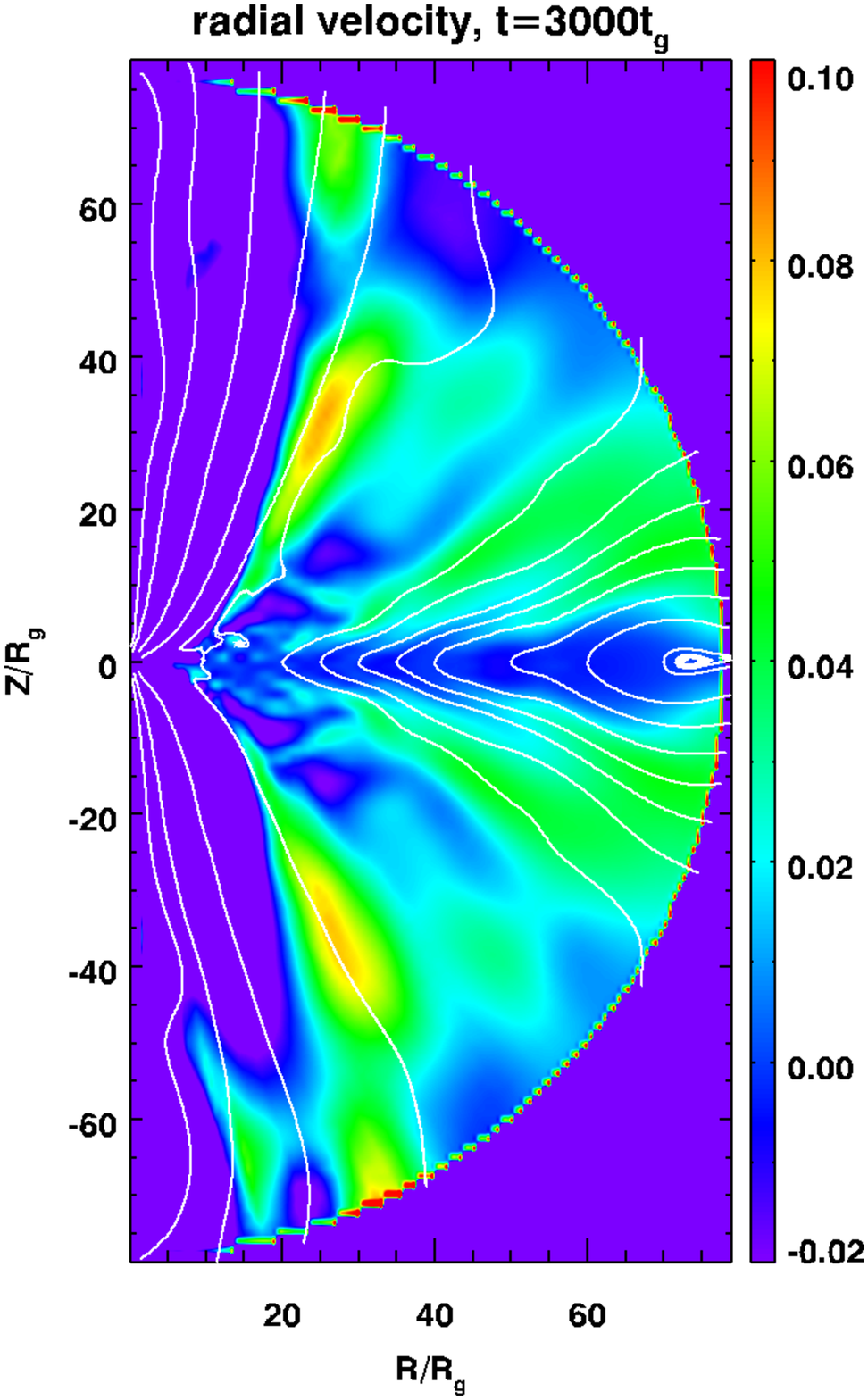}
\caption{
Distribution of the radial velocity component $u^{r}$ for simulation {\em D7} at $t=3000$.
}
\label{astro_dis_D6_ur_img}
\end{figure}

\section{Disk outflows}
\label{astro_dis_morph_sec}
From non-relativistic launching simulations
\citep{2002ApJ...581..988C,2007A&A...469..811Z,2012ApJ...757...65S,2014ApJ...793...31S} we know that
magnetically diffusive accretion disks may drive strong outflows that collimate into high-speed jets.
Particular examples are simulations by \citet{2016ApJ...825...14S} that were run up to several 100,000 inner disk 
rotations, some of them even invoking a mean-field disk dynamo that generates the jet-launching magnetic field
\citep{2014ApJ...796...29S}.
The question we like to discuss in this section is whether we can find the same phenomenon - magneto-centrifugally 
driven disk outflows and jets - also from relativistic disks around black holes.
In fact we find outflows in all of our simulations, except for simulation {\em D0} that is only very weakly magnetized.
The simulations with strong magnetic plasma $\beta=10$ all show outflows, the evolution of which share many common features. 

In Appendix~\ref{astro_dis_high_resolution_sec} we will present a resolution study for our simulation approach.
As we will see, the overall disk-outflow evolution is well treated by our choice of resolution, in particular
we may safely compare our simulations of similar resolution.
Certain physical parameters may vary, however, and their absolute values may have to be taken with care.

In the following we will give a description of the general outflow morphology. 
We consider simulation {\em D7} as our ``reference simulation".

\begin{figure*}
\centering
\includegraphics[width=2.3in]{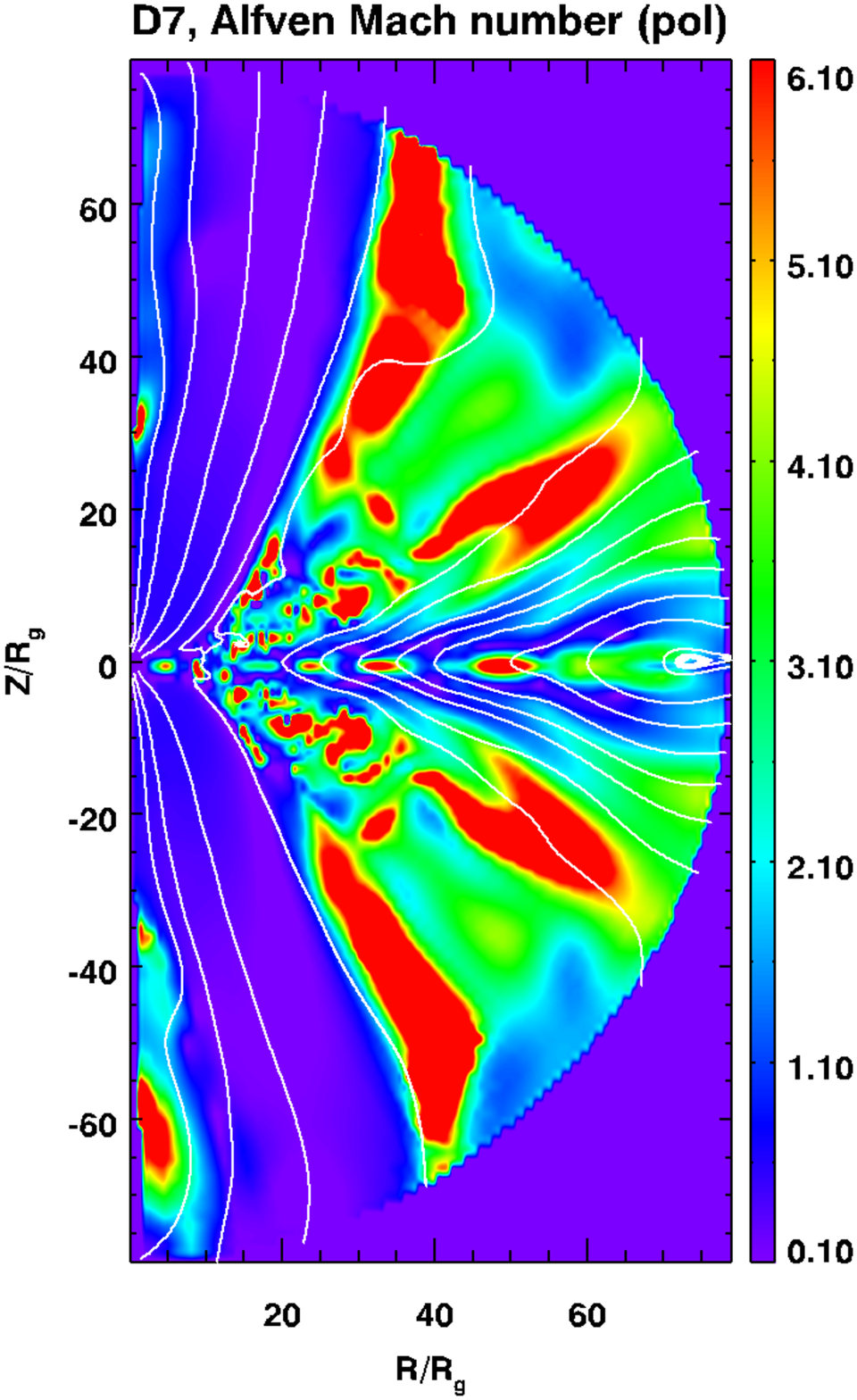}
\includegraphics[width=2.3in]{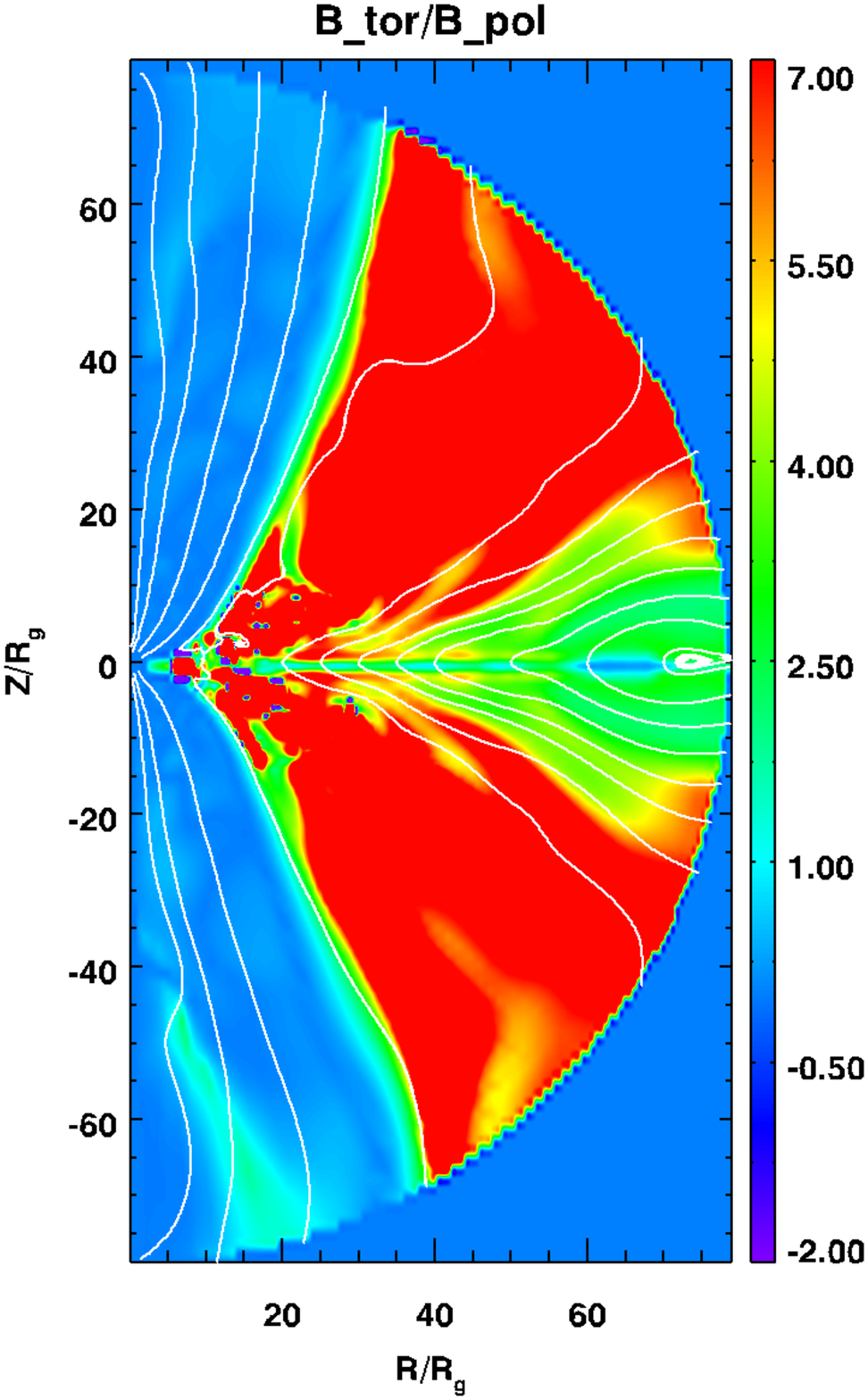}
\includegraphics[width=2.3in]{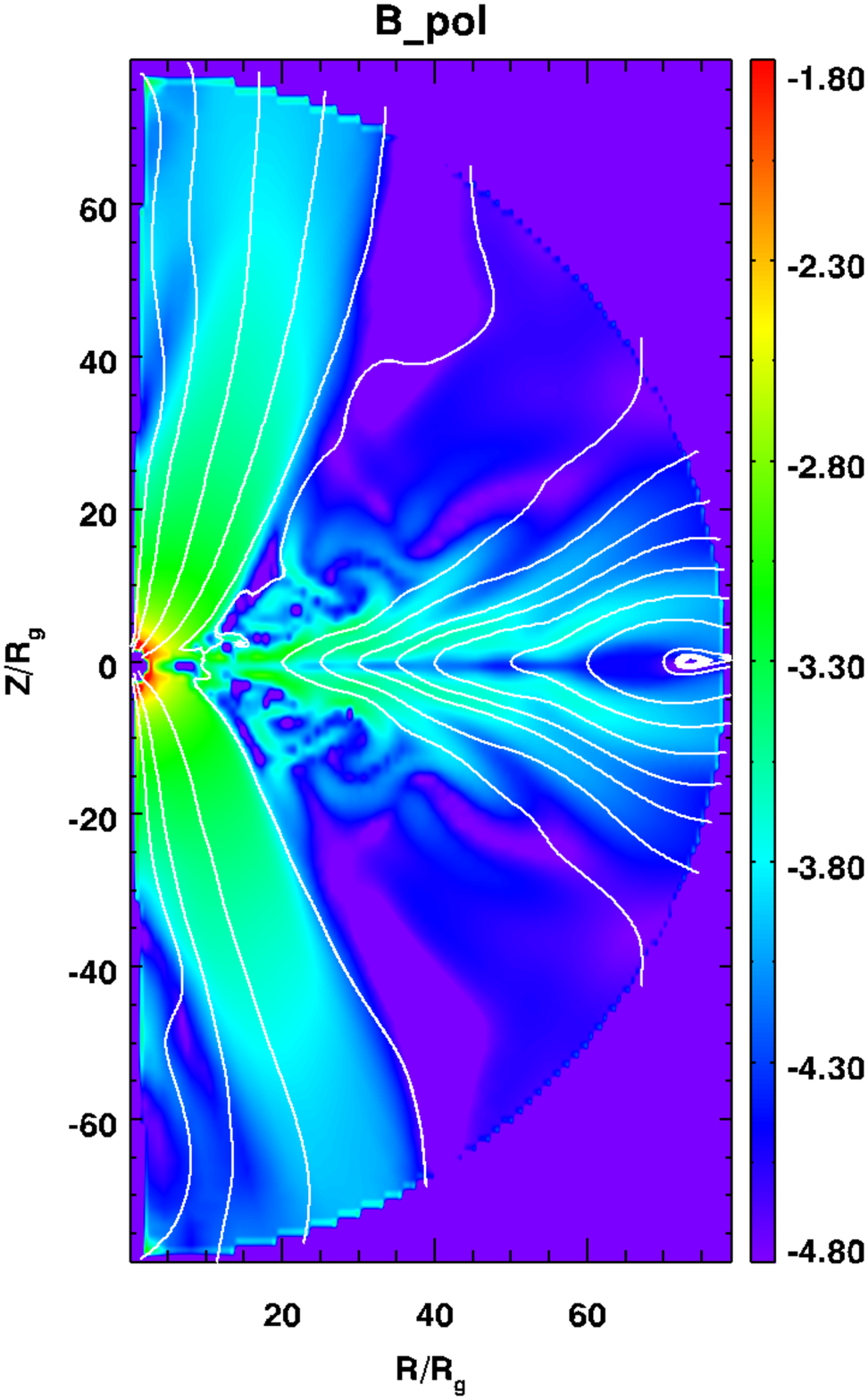}
\caption{
The magnetic field characteristics for simulation {\em D7} at $t=3000$.
Shown is the poloidal Alfv\'{e}n Mach number (left), ratio of toroidal field strength divided by poloidal field 
strength (middle) and poloidal field strength (log scale, right). 
The disk outflow becomes almost immediately super-Alfv\'{e}nic after leaving the disk surface
with a toroidal magnetic field larger than the poloidal field. 
}
\label{astro_dis_D6_B_img}
\end{figure*}

\begin{figure*}
\centering
\includegraphics[width=2.3in]{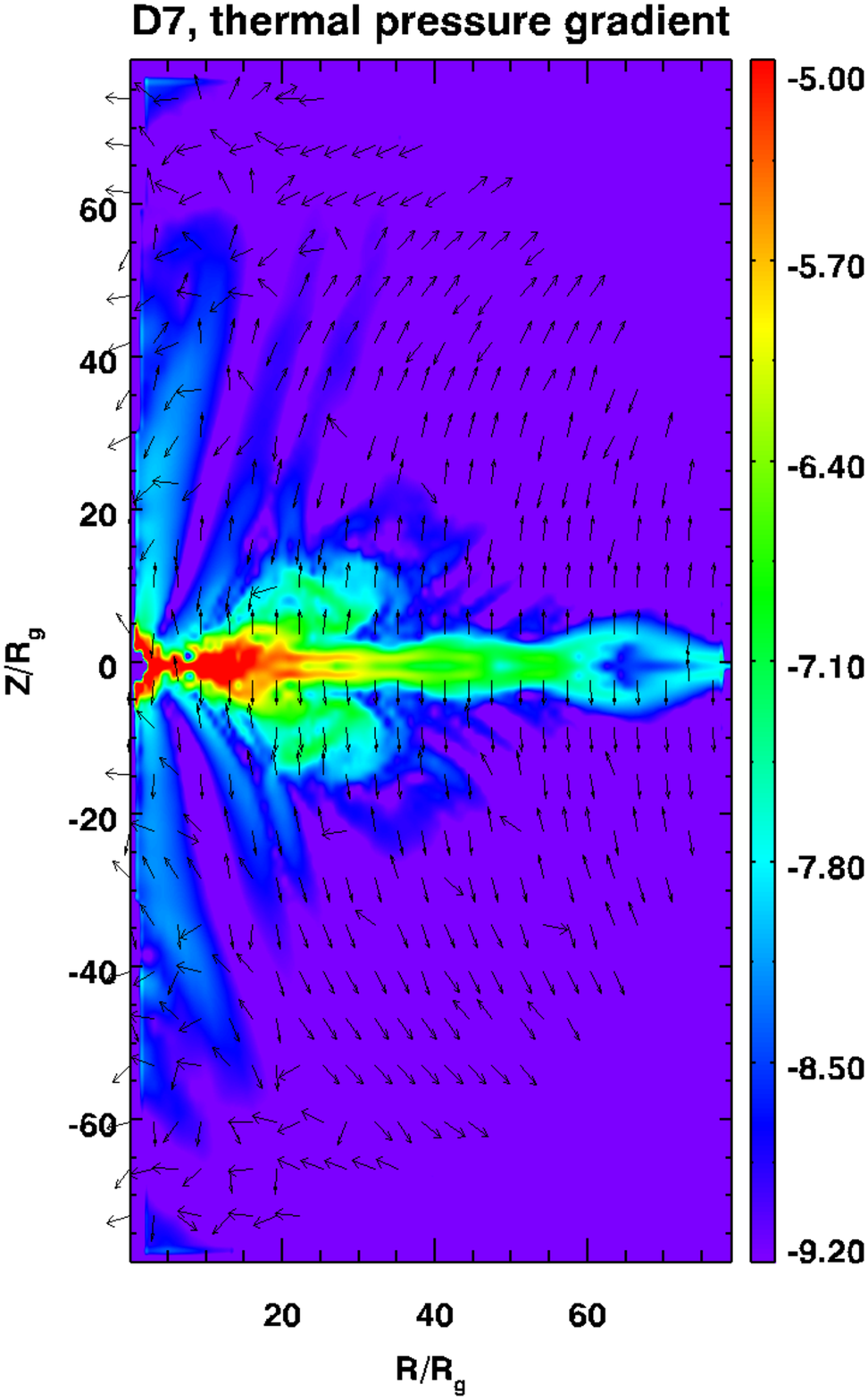}
\includegraphics[width=2.3in]{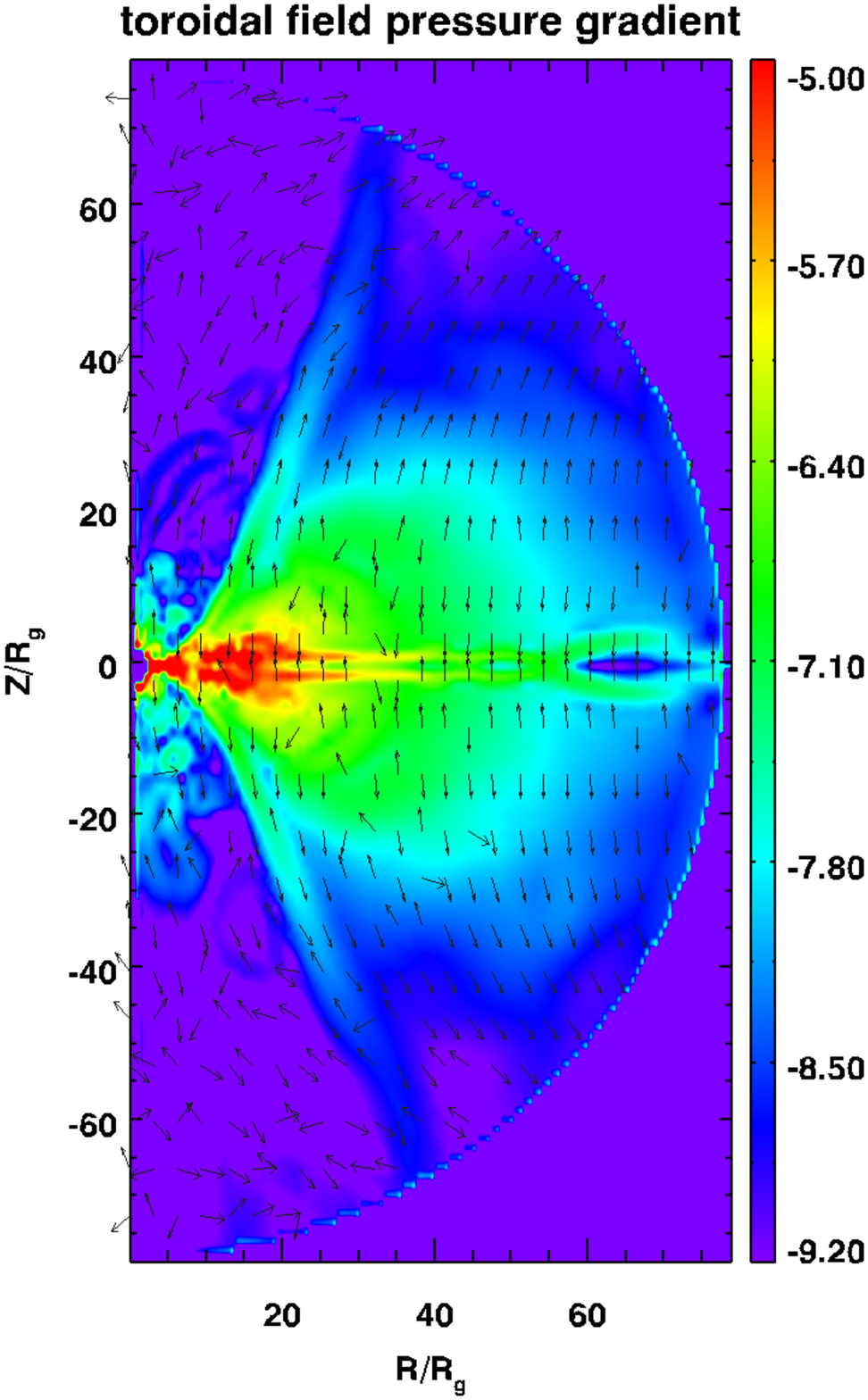}
\caption{
Pressure gradients in simulation {\em D7} at $t=3000$.
Thermal pressure gradient ({\em left}) and
toroidal magnetic field pressure gradient ({\em right}).
Arrows show the direction of the forces. 
}
\label{astro_dis_D6_pre_col_img}
\end{figure*}

\subsection{Disk winds}
As an exemplary simulation we now investigate the disk winds launched in simulation run {\em D7}.
In Figure \ref{astro_dis_morph_D6_rho_img}
we show density snapshots for six time steps.
The time evolution of the accretion disk shows some wave-like structures or ``density condensations" that seem
to move from the inner disk to the outer disk. 
As discussed in Section \ref{astro_dis_weak_B_sec}, we attribute this feature as a consequence of the chosen initial 
condition with the disk evolving into a new dynamical equilibrium, now also under the influence of a strong magnetic field. 

We may clearly identify outflow structures leaving the disk surface. 
The outflows first originate from the density condensations close to the disk inner boundary.
When this density pattern moves outwards, new outflows are generated from them at larger radii.
New density concentrations appear in the inner disk and again launch outflows in the inner disk.
Figure \ref{astro_dis_morph_D6_rho_img} demonstrates how outflows are repeatedly generated and triggered
by the density condensations in the disk.

Besides the density and magnetic field structure shown in Figure~\ref{astro_dis_morph_D6_rho_img}, 
also the evolution of the plasma-beta $\beta = p/B^2$ and the 
magnetization $\sigma \equiv/ B^2  \rho$ provides insight in the physics of jet formation.
This is shown in Figure~\ref{fig_D7_beta_sigma}.
We clearly see the accretion disk as gas pressure dominated.
In the plasma-beta plots, the purple areas in the disk corona are dominated by the magnetic field pressure.
Similarly for the magnetization distribution, that shows a magnetically dominated coronal region.
This is indeed essential as only for a (relatively) strong magnetic field we may expect relativistic 
disk winds.

Disk and corona are initially in pressure equilibirum.
Initially, the coronal density is low and the magnetization is strong ($t=400$).
When the outflow is launched, higher density material is ejected from the disk and the magnetization 
lowers (see $t=1000, 3000$) for the regions interacting with the disk.
The central area is still highly magnetized (red colored axial region). 
We note, however, that simulation {\em D7} considers a non-rotating black hole with no physical outflow
being launched from the black hole.
The low density in this area is maintained by the floor model.

Comparison of the snapshots in Figure~\ref{fig_D7_beta_sigma} directly shows the growth of the disk outflow
as the area of lower magnetization increases in time.
Also the area of higher plasma-beta increases when the outflow emerges from the disk.
Interestingly, the outflow from the outer disk maintains a lower plasma beta.
The reason is that the disk gas pressure and gas density decrease faster with radius
compared to the magnetic field.
Thus, this part of the outflow may benefit from magnetic pressure driving, however the time scale
of the simulation is not sufficient to reach a new dynamical equilibrium in these outer parts of
the disk\footnote{We note that at $r=40$ we have performed only about one disk revolution at $t=5000$}.

Below we will discuss the launching and acceleration mechanism of our disk winds. 
The discussion just above fully supports the view of {\em magnetically driven} disk winds as
we observe a plasma-beta $\beta$ \lax 1 (the green areas of the wind).

The panels of Figure~\ref{fig_D7_beta_sigma} showing the magnetization again demonstrate the 
difficulty of performing relativistic MHD simulations.
In general, relativistic MHD codes have problems with their inversion schemes for high magnetization areas $\rho/B < 1$.
%
These are in particular the axial regions close to the axis (plotted in red for $\rho/B^2$).
On one hand this is probably the most relativistic area of interest in the domain where we expect
the Blandford-Znajek driven Poynting jet to operate.
On the other hand this area can often only be numerically treated for long-term by involving a floor model.
So far, there seems no way out of this concerning a numerical treatment.

The outflows that originate at larger radii show a certain degree of collimation towards the 
axis (see e.g. upper right panel of Figure \ref{astro_dis_morph_D6_rho_img} for $t=1000$). 
This collimation does not seem to be the typical MHD collimation process by toroidal magnetic field tension.
We have therefore ran a simulation (not listed in Table \ref{astro_dis_sim_table}) for which the outer boundary was 
located at $R_{out}=160$ (double size compared to {\em D7}). 
In fact, the jets in that test simulation just keep growing outwards at $t = 1000$ with no sign of collimation at 
radius $r = 80$. 
We thus can attribute the collimation we see in Figure \ref{astro_dis_morph_D6_rho_img} to the outflow boundary condition 
applied.
Before outflows reach the boundary, the region outside outflows is ``vacuum" which provides a large pressure gradient. 
However, we have continuous density condition at outer boundary. 
When outflows arrive at the boundary, their densities are projected into the ghost zone, which diminish the pressure 
gradient between ``inside" and ``outside", thus influence the propagation of outflows. 
For future work on disk jet collimation we suggest to modify the standard outflow conditions in order tho avoid artificial 
collimation (see e.g. \citealt{2010ApJ...709.1100P, 2011ApJ...737...42P}).

On the long term the individual streams keep growing in magnitude and also in launching area
and finally merge to a single large-scale bipolar disk wind originating from all over the disk surface.
Between  $t = 2000$ and $t = 3000$ we measure a time averaged ejection rate for simulation {\em D7} of $7\times 10^{-3}$.

\begin{figure}
\centering
\includegraphics[width=2.in]{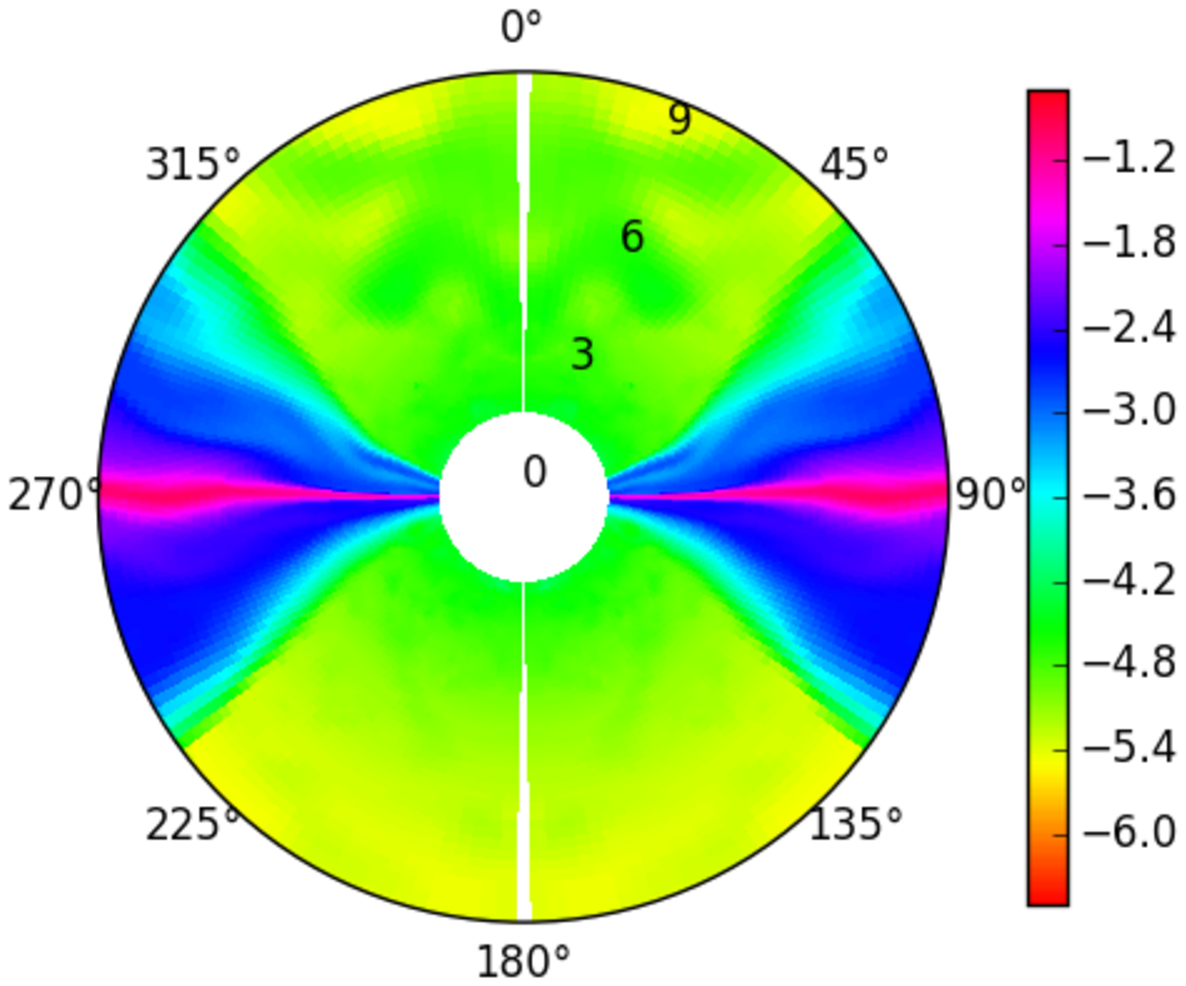}
\includegraphics[width=2.in]{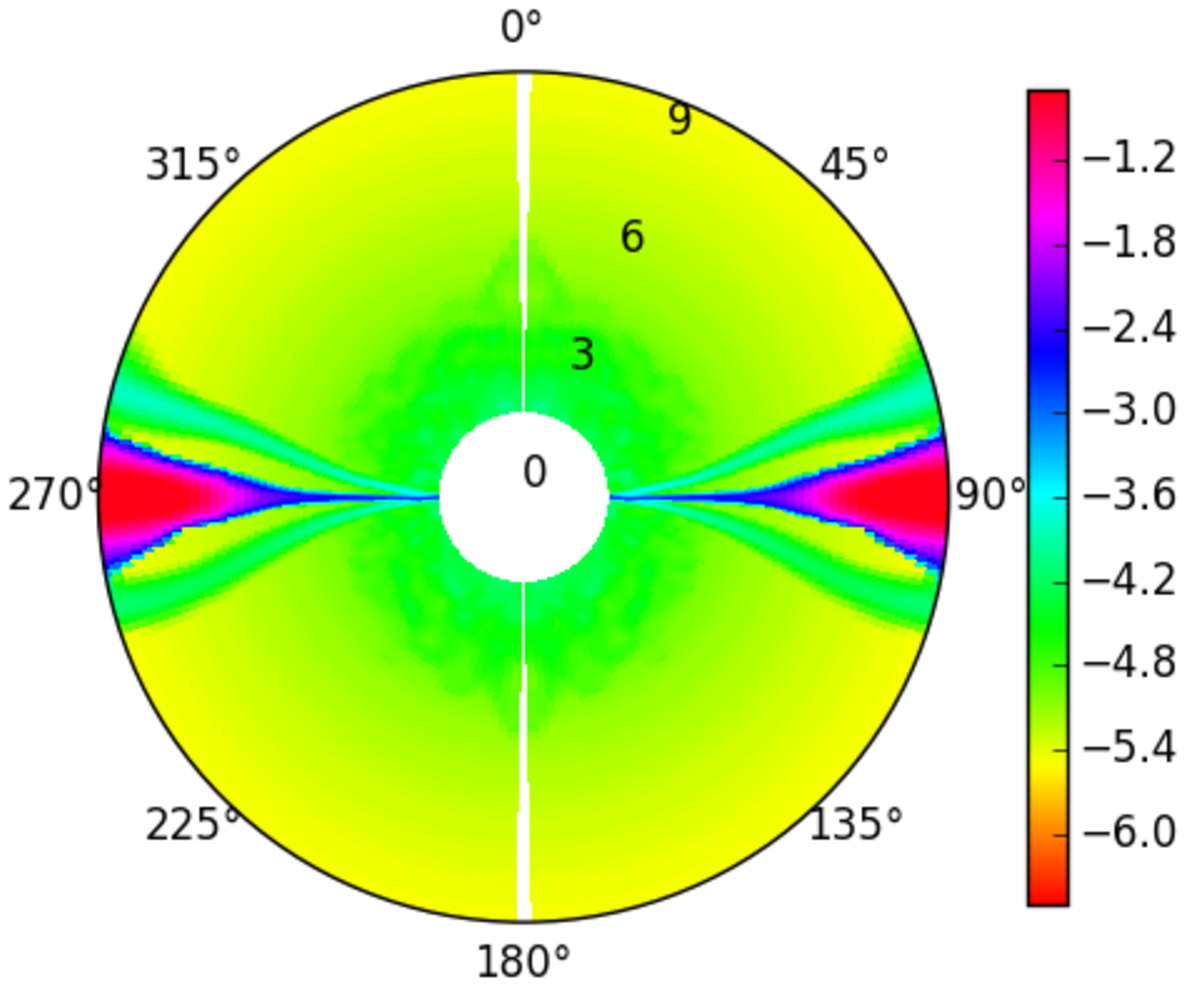}
\caption{
Density snapshots for simulations {\em D1} (upper panel) and {\em D7} (lower panel) at $t=3000$ (with the left 
hemisphere just mirrored from the right hemisphere).
Shown is a zoomed-in sub-panel of the whole simulation grid, he radial positions of $r = 3.0, 6.0, 9.0 $ are superimposed.
Efficient accretion in simulation {\em D1} connects the disk and the horizon (plunging region), 
while in {\em D7} an disk inner edge is pronounced and remains at $r=6$ with only a thin accretion stream towards 
the horizon.
}
\label{astro_dis_morph_D1D6_rho_zoom_img}
\end{figure}

\subsection{Wind radial velocity}
In Figure~\ref{astro_dis_D6_ur_img} the radial velocity distribution for simulation {\em D7} is shown at $t=3000$. 
The accretion velocity (directed inwards) is generally much smaller than the wind velocity (directed outwards).
We find that at this time the disk wind has only reached a moderate outward radial velocity $u^{r} \lesssim 0.1$. 
With the exception of minor turbulent patterns in the wind, negative radial velocities exist only in the accretion. 
The typical accretion velocity is $v_{r} \sim 10^{-3}$.

The outflow velocity we measure in other simulation runs  show a similar pattern.
However, we find a higher outflow speed for a lower magnetic diffusivity than in simulation {\em D7}. 
For simulation {\em D1} the maximum wind velocity at $t=3000$ exceeds $0.3$. 
The physical reason for this is the stronger coupling between magnetic field and matter as we will discuss in 
Section~\ref{astro_dis_drive_sec} where we investigate the driving mechanism of the disk outflows.
%

\subsection{Disk evolution at the inner edge}
When changing the magnetic diffusive level (parameter $\eta_{0}$) in the disk, an interesting behavior of disk inner
boundary is observed.
As was mentioned in 
the introduction, the presence of magnetic diffusivity allows accreting matter to pass through magnetic field lines 
freely and accrete onto the black hole. 
However, if the disk is not diffusive (like in simulation {\em D1}), accreting matter will drag the field lines together
towards the black hole, destroying the structure of the initially well ordered field lines. 
The influence of magnetic diffusivity on the disk morphology can be clearly seen in
Figure~\ref{astro_dis_morph_D1D6_rho_zoom_img}. 
The upper panel shows the density snapshot of simulation {\em D1} at $t=3000$. Without diffusivity,
the massive flow accretes with the field lines and pushes the disk inner edge towards the black hole. 
In this case, the accretion disk connects directly to the black hole horizon. 
As is shown in Figure~\ref{astro_dis_morph_D1D6_rho_zoom_img} lower panel (simulation {\em D7}), the magnetic diffusivity, whose value 
peaks at the disk mid-plane, allows an accreting flow at the disk mid-plane without disturbing the disk inner boundary. 
The disk inner edge stays then exactly at the initial radius $r_{in}=6$. 
Nevertheless, the massive accretion in simulation {\em D1} implies a stronger angular momentum transport inside the 
disk. 
We will discuss this point in the following. 

\begin{figure*}
\centering
\includegraphics[width=2.in]{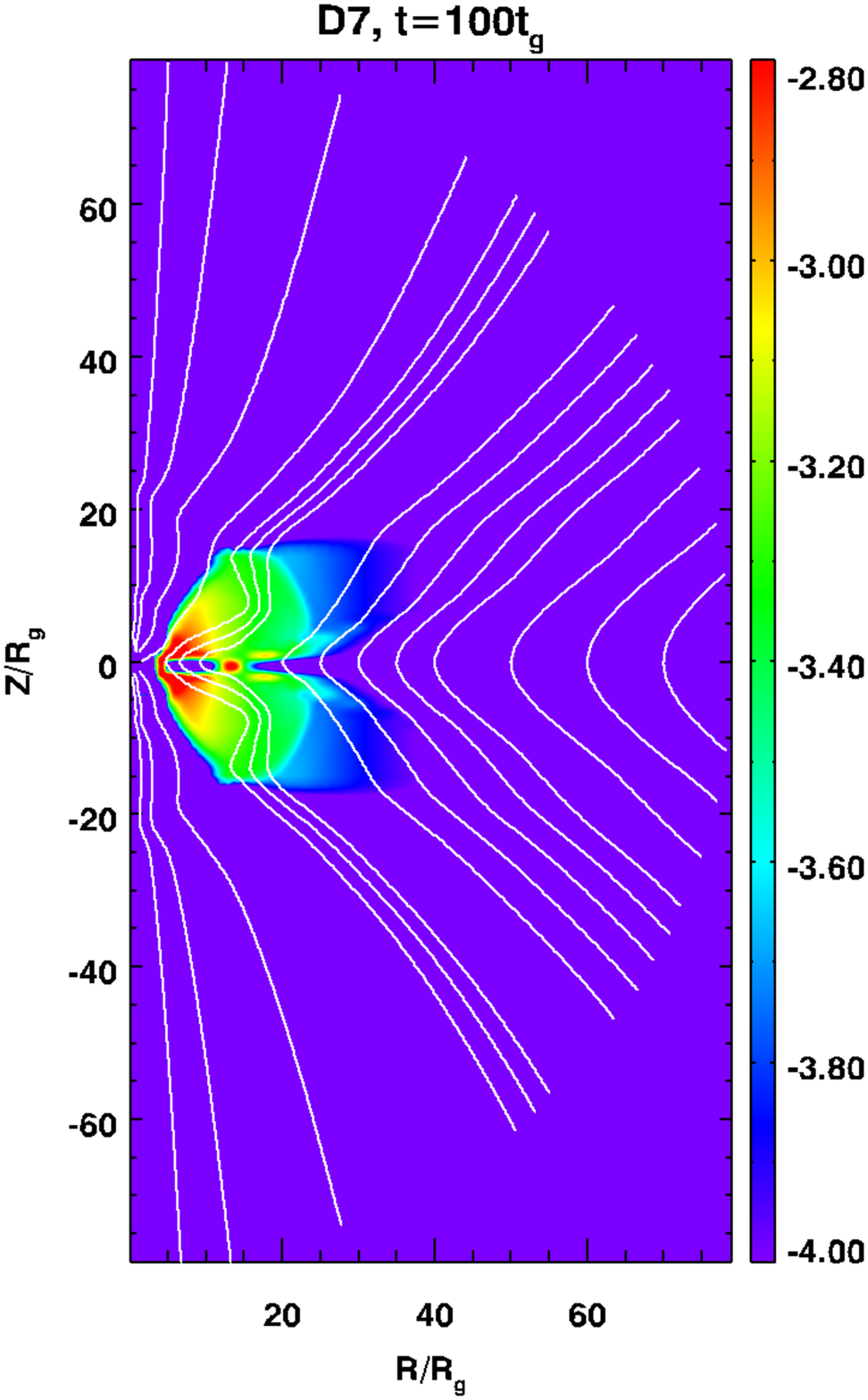}
\includegraphics[width=2.in]{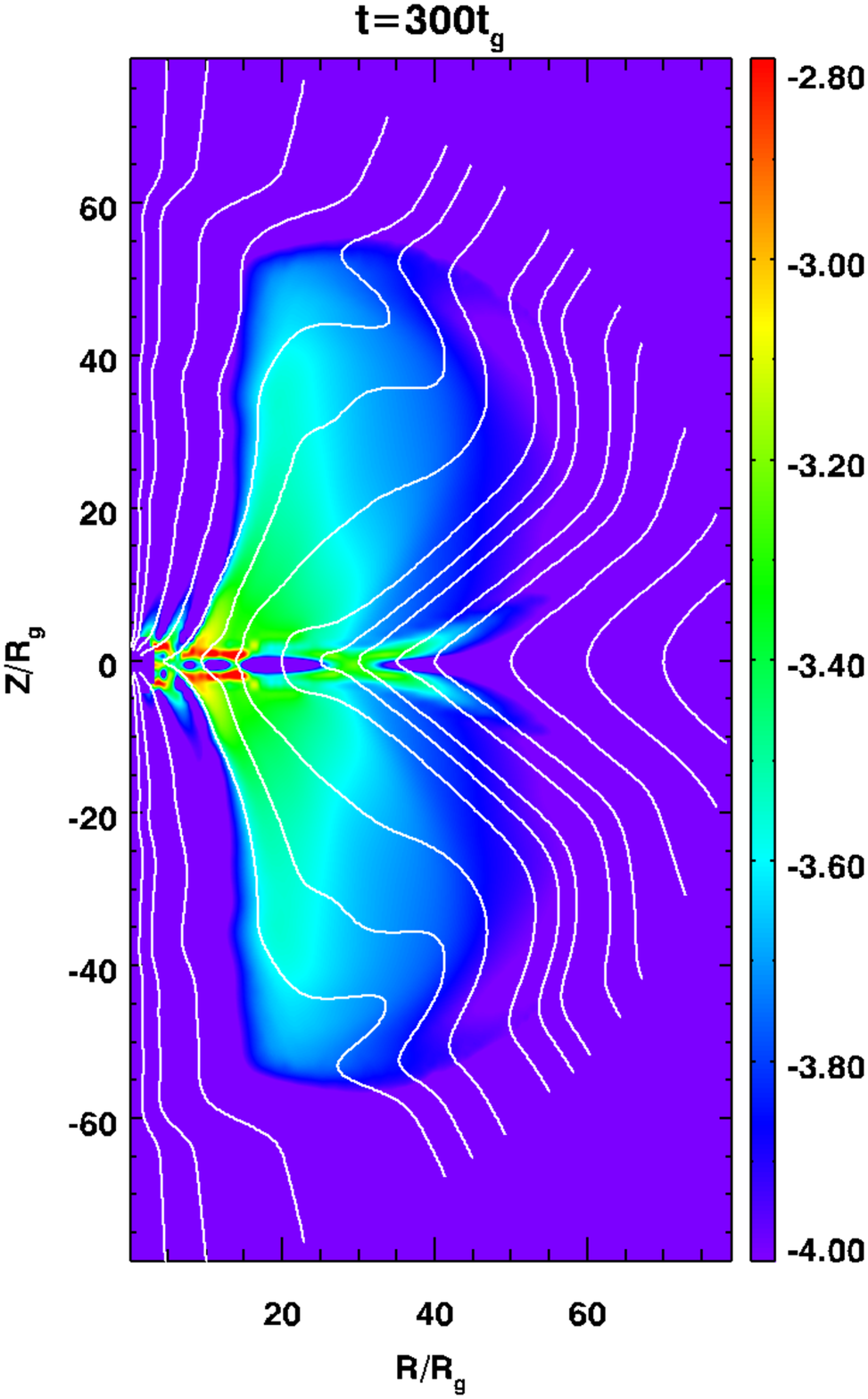}
\includegraphics[width=2.in]{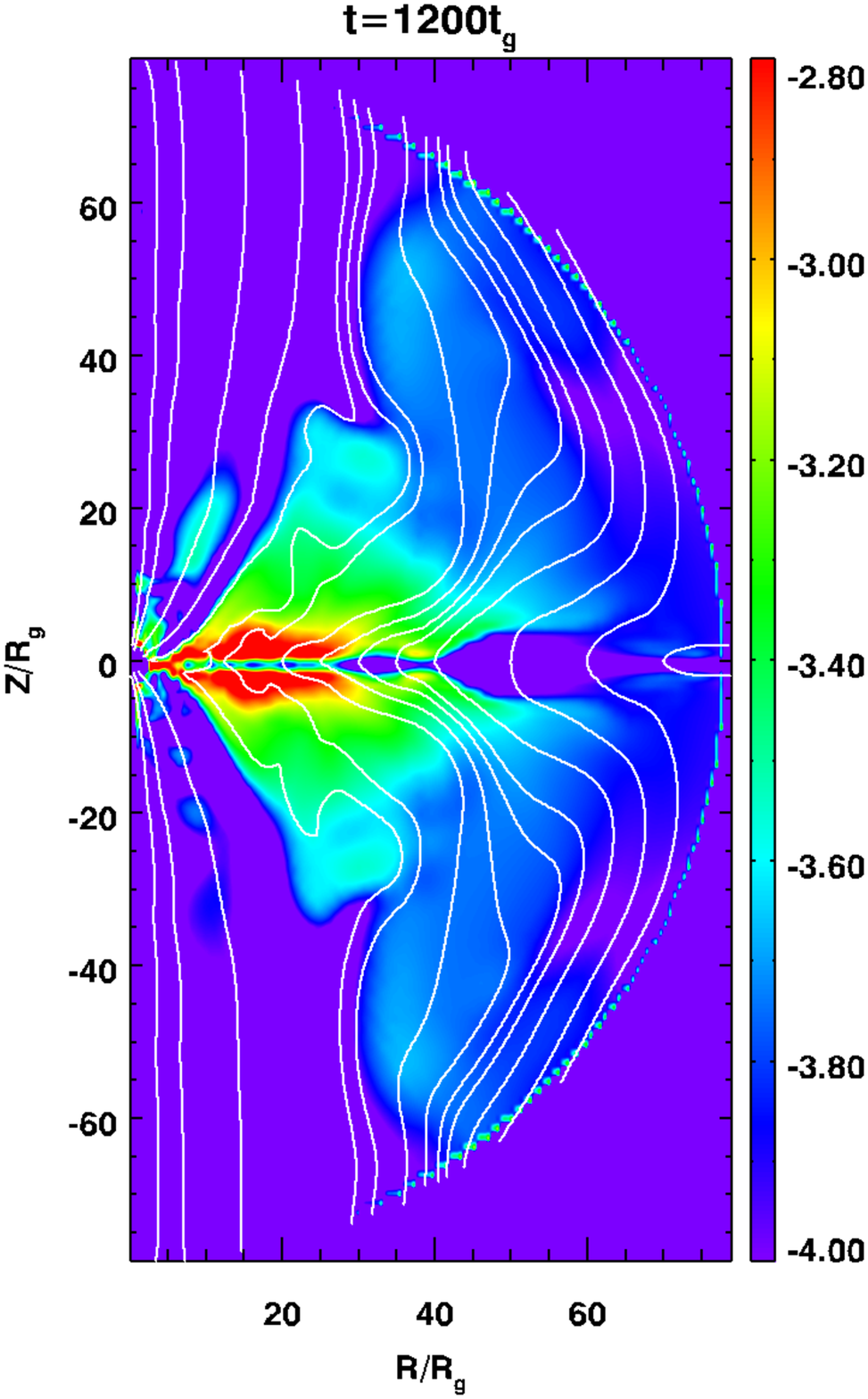}
\caption{Time evolution of the toroidal field strength (log scale) for simulation {\em D7} at $t=100$ (left), $t=300$ 
(middle); $t=1200$ (right), see also the lower left plot in Figure \ref{astro_dis_D6_B_img} for $t=3000$. 
}
\label{astro_dis_D6_Bphi_tower_img}
\end{figure*}

\section{The driving force of disk winds}
\label{astro_dis_drive_sec}
In Section \ref{astro_dis_morph_sec}, we described the morphology of the disk-driven outflow in simulation {\em D7}. 
In this section, we investigate the possible mechanisms that may drives the wind.
A clear way to disentangle the driving mechanism would be to calculate and compare the different force components 
along and across the magnetic field lines. 
Such an analysis have been done in \citep{1997ApJ...482..712O,2002A&A...395.1045F,2010ApJ...709.1100P}, but
requires a steady state where the physical quantities in the simulation region are only position dependent. 
This condition is well satisfied in \citet{2010ApJ...709.1100P}, where the simulation regions do not include the 
accretion disk. 
This approach has also been applied in non-relativistic disk-jet simulations for which the disk-outflow
structure has evolved into a quasi-steady state (see e.g. \citealt{2012ApJ...757...65S}).

However, in our case, none of the simulations listed in Table \ref{astro_dis_sim_table} have reached such a 
(quasi-) steady state.
Thus, we will mainly analyze the pressure distribution of different origin and thus discuss conceptually the role 
of the (magneto-) centrifugal force, the magnetic force and the thermal pressure force. 
In the following, we concentrate on simulation {\em D7}.

\subsection{Poloidal Alfv\'en Mach number}
If the disk wind is driven {"}magneto-centrifugally{"} (Blandford-Payne mechanism), we expect a 
poloidal\footnote{The letter 'p' in the subscript denotes the poloidal component of a vector, e.g. 
$B_{\rm p} =\sqrt{B^{r}B_{r}+B^{\theta}B_{\theta}}$ or $u_{\rm p}=\sqrt{u^{r}u_{r}+u^{\theta}u_{\theta}}$, 
while the toroidal component is denoted by a subscript $\phi$.} magnetic field to dominate the region close 
above the disk surface where the wind is centrifugally accelerated along the poloidal magnetic field. 
This can be quantified by the poloidal Alfv\'{e}n Mach number\footnote{We note that since $\mu_{0}\equiv1$ in HARM, 
the factor $4\pi$ will be omitted in the following.} of the flow,
$ M_{\rm A,p} = \sqrt{h \rho u_{\rm p}^2 / B_{\rm p}^2}$, 
with the specific enthalpy
\begin{eqnarray} 
h = \frac{\gamma_{\rm g}}{\gamma_{\rm g} -1}\frac{p}{\rho} + 1. 
\label{astro_dis_drive_enthal_eq}
\end{eqnarray}
with the gas pressure $p$.
The poloidal Alfv\'{e}n Mach number measures the kinetic energy in terms of magnetic energy. 
A super-Alfv\'{e}nic value, $M_{\rm A,p}>1$ would implies a weak poloidal magnetic field that is dominated by 
a large kinetic flux. 
Under such conditions, the driving of the outflow by the Blandford-Payne mechanism would be inefficient.

In Figure~\ref{astro_dis_D6_B_img} (left) we show the poloidal Alfv\'{e}n Mach number for simulation 
{\em D7} at $t=3000$. 
We see that except some small regions in the outflow that show some turbulent motion, the disk wind becomes almost 
immediately super-Alfv\'{e}nic after leaving the disk surface (defined in Section \ref{astro_dis_acc_eject_illu_sec}).
In contrast, Blandford-Payne outflows start with sub-Alfv\'enic velocity and subsequently supersede the 
(local) Alfv\'en speed and the (local) fast magneto-sonic speed.
Inside the outflow region at the height of $z \simeq 30$ from the disk mid-plane we find typical values of $M_{\rm A,p} > 5$. 

In comparison, in the relativistic jet formation simulations of \citet{2010ApJ...709.1100P} the accelerating
disk wind stays sub-Alfv\'{e}nic out to altitudes of $\sim 20$ times the disk inner radius.
This difference may intrinsically arise from the difference between the choice of the accretion disk and 
injection boundary condition. 
In particular, in \citet{2010ApJ...709.1100P} a relatively strong poloidal magnetic field was assumed that
led to rather strong outflow velocities.
In the present simulation, the disk field cannot really be fixed, but evolved in interrelation with the disk
hydrodynamics.
In {\HAR} simulations, the fluid element inside the disk has only toroidal rotation initially. 
By the injection boundary condition in \citet{2010ApJ...709.1100P}, the inflow possesses an initial poloidal velocity 
that can keep magnetic field lines from bending to toroidal directions due to the ideal MHD assumption.
As the flow is already super-Alfv\'{e}nic very close to the disk surface we conclude that the outflow is not 
accelerated magneto-centrifugally.
This holds at least for the time scales considered.

Magneto-centrifugal driving of disk winds and jets has been confirmed in non-relativistic launching
simulations with an otherwise similar setup as applied in this paper
\citep{2002ApJ...581..988C, 2007A&A...469..811Z, 2009MNRAS.400..820T, 2012ApJ...757...65S, 2016ApJ...825...14S}.
In particular, a similar plasma-beta is applied in these simulations for the initial magnetic field, $\beta = 10-100$.
However, note that the run time of the non-relativistic simulations are much longer than typical simulations
in GR-MHD.
For example, \citet{2014ApJ...793...31S, 2016ApJ...825...14S} have run jet launching simulations for more than
200,000 revolutions of the inner disk.
In these simulations, steady-state jet launching is typically established after 500 revolutions of the launching 
point of the inner jet.
This, however, is corresponding to a time period of $t \sim 40,000$ in our GR-MHD setup and is
clearly beyond the times scales we are currently reaching with {\HAR}.
The early evolution is still affected by the expansion of torsional Alfv\'en waves, a situation that we may still
be experiencing in our GR-MHD setup.

Besides a longer run time, one may also think about applying a stronger initial magnetic field and thereby extend the 
sub-Alfv\'{e}nic and hence produce magneto-centrifugally driven winds.
It is well known that small plasma-beta is a challenge for all MHD codes, in particular for the case of GR-MHD. 

\subsection{Magnetic field pressure}
\label{astro_dis_drive_magnetic_pressure_subsec}
We now estimate at the strength of the Lorentz forces acting in the wind system. 
As the magnetic diffusivity has a Gaussian profile that peaks at mid-plane and decays quickly towards the disk 
corona, we may apply the ideal MHD condition, $\pmb{E} = - \pmb{v} \times \pmb{B}$ in the regions outside the disk. 
For the current density $\pmb{j}$, we have\footnote{The factors $4\pi$ and $c$ do not show up because of the 
magnetic field normalization in HARM and {\HAR}.}
$
\pmb{j} + \partial_{t}\pmb{E} = \nabla \times \pmb{B},
$
where we have included the displacement current.
However, since the outflow velocity of the disk wind is not very high, $v\sim 0.1\,c$, we may neglect this term for 
calculating the Lorentz force, and end up with the usual term
\begin{eqnarray} 
\pmb{F}_{L} &=& \pmb{j} \times \pmb{B} = -\nabla \pmb{B}^{2} + (\pmb{B} \cdot \nabla)\pmb{B}.
\end{eqnarray}

In Figure~\ref{astro_dis_D6_B_img} (middle) we show the ratio of toroidal to poloidal field strength (amplitude), namely 
the ratio $\sqrt{B^{\phi}B_{\phi}}/\sqrt{B^{p}B_{p}}$. 
Except the area close to the disk mid-plane where toroidal field cannot be induced by the disk rotation, we clearly 
see that the toroidal field dominates the poloidal field strength in almost the entire outflow region. 

The poloidal field strength is shown in Figure \ref{astro_dis_D6_B_img}(right). 
The field strength decreases quickly in radial direction for the outflow region,
while along the axial region the poloidal field is stronger and more widely distributed.
Part of the magnetic flux in the axial region has been advected from the disk.
This area looks ideal for a Blandford-Znajek jet driving, however, the simulation considered applies $a=0$.

Considering all information from Figure~\ref{astro_dis_D6_B_img}, we find that magneto-centrifugal driving of
the disk outflow is unlikely.
Instead, the toroidal magnetic field seems to play a major role in driving the disk wind.
This is a situation that is more comparable to a tower jet \citep{1996MNRAS.279..389L}, see below.

\subsection{A magnetic pressure driven tower jet?}
\label{astro_dis_drive_thermal_pressure_subsec}
\label{astro_dis_tower_jet_subsec}

Also thermal pressure can drive the outflow.
Here we compare the thermal pressure gradient with the toroidal magnetic field pressure gradient
(as we have seen above, the poloidal field is rather weak).
In Figure~\ref{astro_dis_D6_pre_col_img} we show these force components. 

Inside the accretion disk and close to the disk surface the amplitude of the thermal pressure gradient is 
comparable to that of the toroidal field pressure gradient inside the disk. 
We thus think that jet launching - the lifting of material from the disk into the outflow - is maintained by
both force components.

We see that the thermal pressure force is always directed away from the disk. 
However, the thermal pressure gradient decays faster above the disk and quickly becomes less than the magnetic pressure 
gradient. 
In the disk wind region, the toroidal magnetic pressure force is clearly dominating and is directed outwards. 
Figure~\ref{fig_D7_beta_sigma} indicates a plasma-$\beta$ below unity for the disk wind,
suggesting again magnetic pressure being dominant (but note that forces are defined by the gradients).

Therefore, we argue that the thermal pressure contributes to the outflow launching near the disk surface, 
but the wind acceleration is due to the toroidal magnetic field pressure in simulation {\em D7}.

For $B_{\phi} > B_{\rm p}$ we may ignore the poloidal field terms in the Lorentz force and thus consider for
the tension force
$(\pmb{B} \cdot \nabla)\pmb{B}$
only the components
\begin{equation} 
(\pmb{B} \cdot \nabla \pmb{B})_{r} = -\frac{{B_{\phi}}^{2}}{r}, \,\,\,
(\pmb{B} \cdot \nabla \pmb{B})_{\theta} = -\frac{\cot \theta {B_{\phi}}^{2}}{r}
\end{equation}
The radial tension force is always pointing to the center black hole (minus sign). 
The magnitude of the radial tension force is about one order of magnitude smaller than the pressure gradient 
force, thus not contributing to the launching or the acceleration process significantly. 
The $\theta$ component of the tension force points counterclockwise in the upper hemisphere and clockwise in the lower 
hemisphere, respectively.
Since $\cot \theta$ is tiny close to the disk surface, the tension force $\theta$-component does not 
contribute to the launching process. 
In the disk wind region, its amplitude is comparable to the radial tension force, hence much smaller then the 
magnetic pressure force as well. 
We note that these two components may be not negligible far from the disk and may ultimately govern the collimation 
of the disk wind into a jet.

In Figure~\ref{astro_dis_D6_Bphi_tower_img} we show the time evolution of toroidal field strength. 
We clearly see the expansion of the toroidal field along the outflow as it is induced by the disk
rotation and inertial forces of the outflow.
We have discussed above that the toroidal magnetic field is larger than the poloidal component and also that
its pressure force dominates the thermal pressure force.
We thus conclude that the toroidal magnetic field plays the leading role in the driving of the disk wind generation. 
For these reasons, we interpret the outflow we see in simulation {\em D7} to be very likely the base of a tower jet.

Such jets - driven by the pressure gradient of the toroidal magnetic field - were predicted by \citet{1996MNRAS.279..389L}. 
Early numerical studies in ideal MHD simulation (no disk evolution included) similarly report that the disk rotation
twists the initial poloidal magnetic field, building up jet outflows as 
``growing towers of twisted magnetic field together with the currents that they carry" \citep{1995ApJ...439L..39U}. 

For all our simulations with a non-rotating black hole (simulations {\em D1}-{\em D15}), we find that the 
disk wind driving mechanism is similar to simulation {\em D7}, thus a driving force that is governed by the
toroidal magnetic field pressure gradient with only little contribution from a magneto-centrifugal acceleration.
Nevertheless, different level of magnetic diffusivity will impact on the efficiency of the growth of these magnetic
tower winds and will thus influence the accretion and ejection rates (see Section \ref{astro_dis_acc_sec}).

\section{Impact of diffusivity on accretion and ejection rates}
\label{astro_dis_acc_sec}
In this section, we discuss how the magnitude of the magnetic diffusivity affects the accretion and ejection processes 
of the disk. 
We will consider the accretion and ejection rates as well as the accretion/outflow amplitudes to quantify the outflow 
efficiency.
 
\begin{table*}
\caption{ 
Time averaged accretion and ejection rates for simulations {\em D2} - {\em D7}, {\em D9}, and {\em D10}. 
Time averaging is done from $t = 2000$ to $t = 3000$. 
The ejection efficiency $\xi$ is defined in Equation~\ref{astro_dis_outflow_efficiency_eq}.
}
\begin{center}
  \begin{tabular}{ c | c | c | c | c | c | c | c | c| c}
    simulation & {\em D1}  & {\em D2}  & {\em D3}  & {\em D4} & {\em D5}  & {\em D6} & {\em D7}  & {\em D9}  & {\em D10}   \\
     \noalign{\smallskip}   \hline
    \hline  \noalign{\smallskip}  
    $\eta_0$  & $10^{-12}$ & $10^{-6}$ & $10^{-5}$ & $10^{-4}$ & $2\times10^{-4}$ & $5\times10^{-4}$ & $10^{-3}$ & $5\times10^{-3}$ & 
                                                                                                          $5\times10^{-2}$ \\  
    $-\dot{M}_{\rm acc}$ & $6.2\times10^{-2}$ & $5.8\times10^{-2}$ & $5.6\times10^{-2}$ & $7.7\times10^{-2}$ & $5.9\times10^{-2}$ &  
                                     $3.9\times10^{-3}$  & $2.0\times10^{-3}$ & $1.7\times10^{-3}$ & $1.0\times10^{-3}$\\   
    $\dot{M}_{\rm eje}$ & $3.2\times10^{-2}$ & $3.4\times10^{-2}$ &  $3.5\times10^{-2}$ &  $3.0\times10^{-2}$ & $3.7\times10^{-2}$ &             
                                    $1.7\times10^{-2}$  &  $7.0\times10^{-3}$ &  $3.8\times10^{-3}$ &  $2.9\times10^{-3}$ \\
    $\xi$  & 0.5 &   0.6  &   0.6  &   0.4  &  0.6  &   4.3  &  3.6 &  2.3 & 2.9  \\

  \end{tabular}
  \end{center}
\label{astro_dis_survey_1_acc_eje_table}
\end{table*}


\begin{figure}
\centering
\includegraphics[width=6.cm]{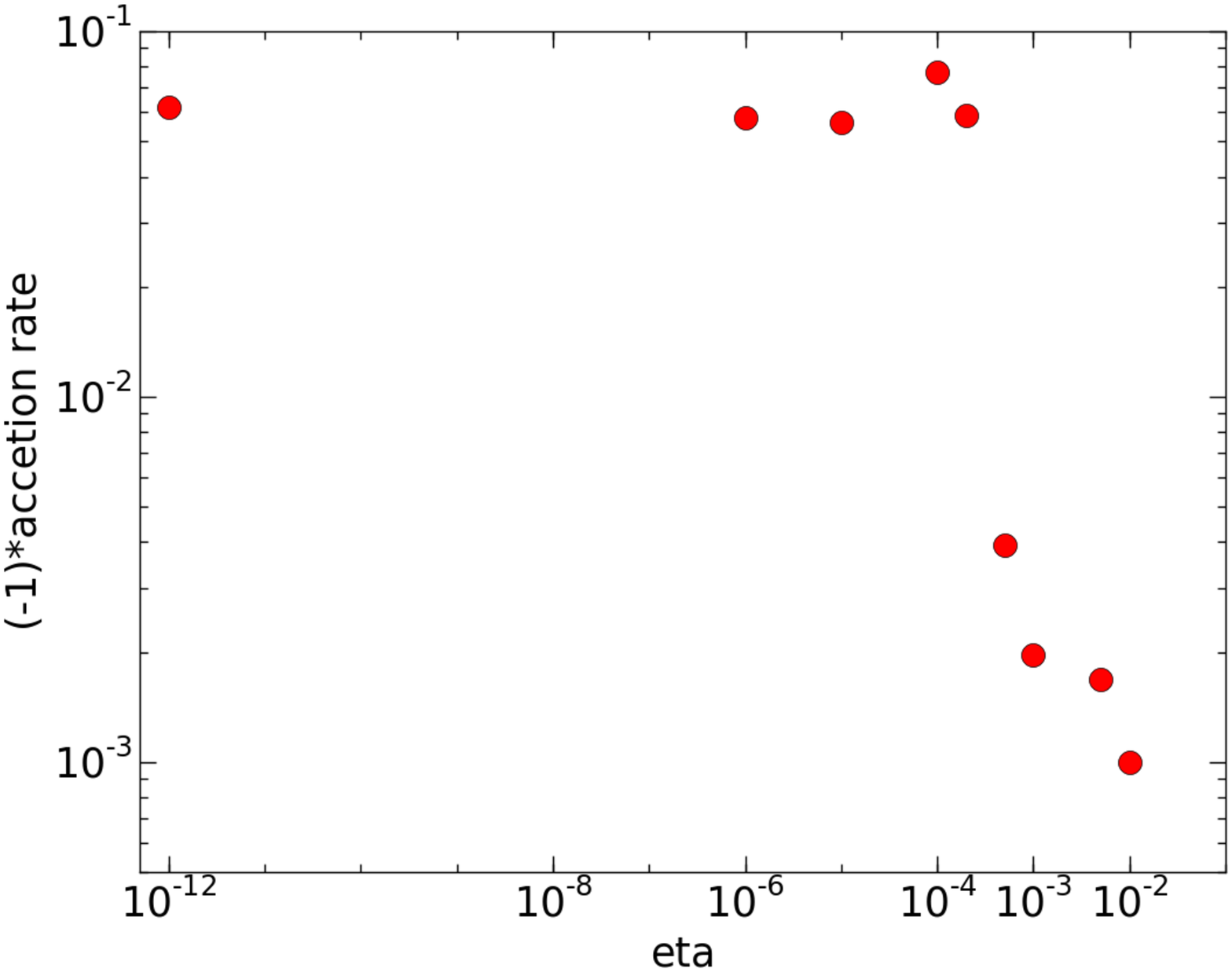}
\includegraphics[width=6.cm]{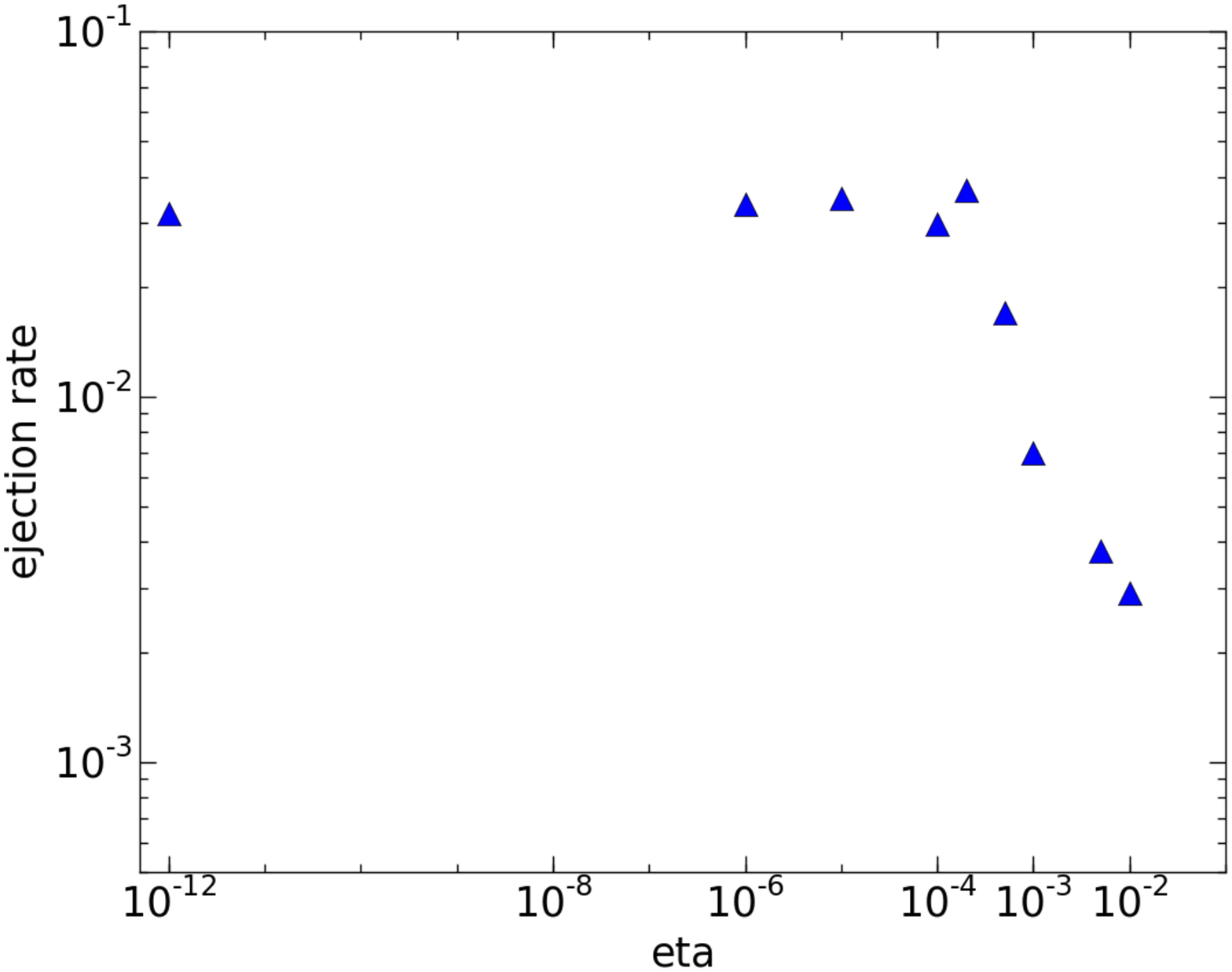}
\caption{Time averaged accretion ({\em top}) and ejection rate ({\em bottom}) for simulation 
runs {\em D1} - {\em D7} and {\em D9}, {\em D10}, considering a (normalized maximum diffusivity) of
$\eta_{0} = 10^{-12}, 10^{-6}, 10^{-5}, 10^{-4}, 2\times10^{-4}, 5\times10^{-4}, 10^{-3}, 5\times10^{-3},$ 
and $10^{-2}$, respectively.
The rates are averaged within the time interval $t=2000$ to $t=3000$.
We note that {\em negative} accretion rates are used in the upper panel in order to plot them with log-scale.
}
\label{astro_dis_acc-eje-eta_relation_img}
\end{figure}


\begin{figure*}
\centering
\includegraphics[width=7.cm]{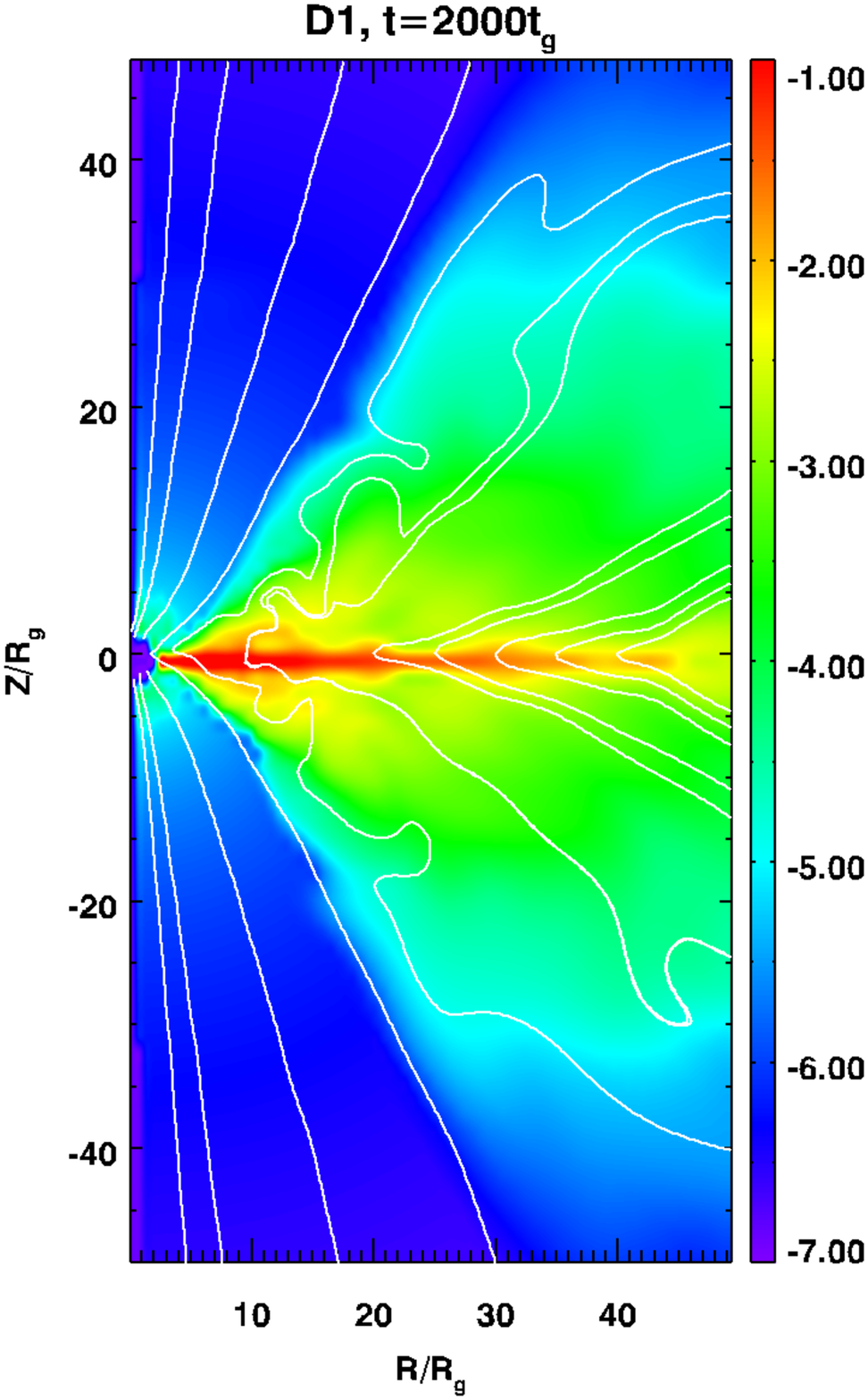}
\includegraphics[width=7.cm]{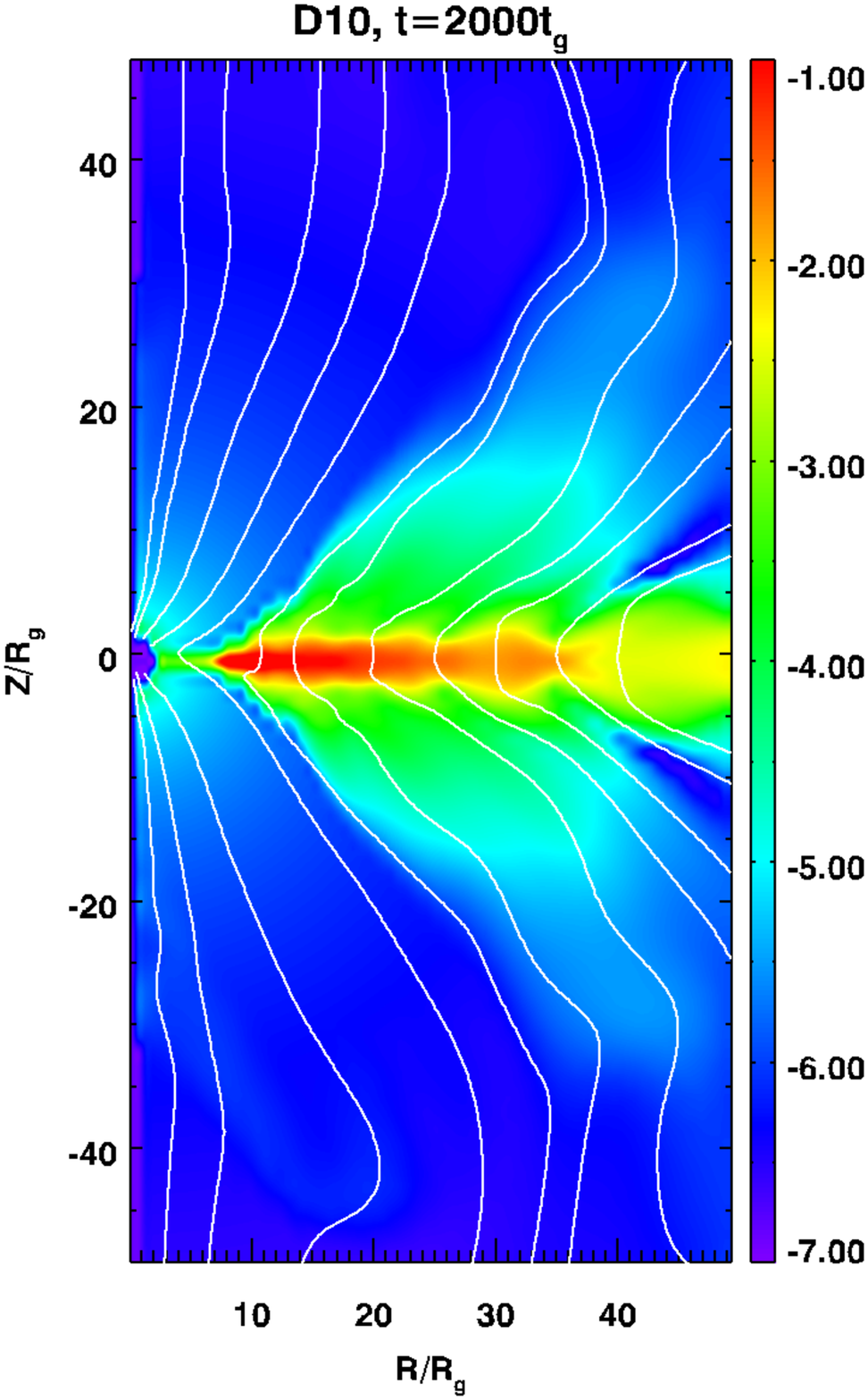}
\caption{Density and magnetic field line distribution (zoomed-in) for simulation {\em D1} (left) and {\em D10}
(right) at $t=2000$.  
In simulation {\em D1} the field lines inside the disk have a more turbulent distribution (especially within $r<20$),
most probably due to the MRI. 
}
\label{astro_dis_D1D17_field_lines_img}
\end{figure*}

\subsection{Accretion and ejection versus resistivity}
Simulations {\em D1} - {\em D7} and {\em D9}, {\em D10} start from the same initial conditions except for the 
level of magnetic diffusivity $\eta_{0}$ (see Table \ref{astro_dis_sim_table}). 
As will be discussed below, the numerical diffusivity is not negligible in the regime $\eta_{0} < 10^{-4}$ 
according to our simulation results.
This depends of course on resolution and on the model setup.
In our comparison we will therefore focus on simulations with $\eta_{0} > 10^{-4}$.
In the following we will compare time averaged accretion and ejection rates for these simulations
(see Table~\ref{astro_dis_survey_1_acc_eje_table}).
Time averages were taken from $t=2000$ to $t=3000$.

Figure \ref{astro_dis_acc-eje-eta_relation_img} shows the mass fluxes in relation to the diffusivity level.
Essentially, we find that the mass fluxes do not really differ for $\eta_{0} < 10^{-4}$. 
We interpret this result as indication of a numerical diffusivity of $\eta_{\rm num} \sim 10^{-4}$ for
choice of our grid size $128\times128$ (see also \citealt{2017ApJ...834...29Q}).
We note that our {\em diffusive} code {\HAR} actually allows us to measure the numerical diffusivity level in our
simulations that naturally affects also the original, ideal GR-MHD version of the code\footnote{Both,
$\eta_{0}$ and  $\eta_{\rm num}$ refer to the initial inner disk radius of the simulation.}.

In the physically interesting regime of $\eta_{0} > 10^{-4}$ when physical magnetic diffusivity is dominating, 
we clearly see that the accretion
rates are suppressed with increasing diffusivity level (see Figure~\ref{astro_dis_acc-eje-eta_relation_img} upper panel). 
We understand this as follows.
A higher diffusivity leads to lower coupling between matter and magnetic field lines, thus suppresses the efficiency of 
angular momentum transport through the magnetic torque - either of the MRI or of the magnetic lever arm 
(see Section \ref{astro_accretion_discussion_subsec}). 

The time averaged inner ejection rates show a strong correlation with the accretion rates in the regime of $\eta_{0} > 10^{-4}$
(see Figure~\ref{astro_dis_acc-eje-eta_relation_img}, lower panel).
Similar as for the accretion rates (see upper panel), the ejection of disk winds weakens for higher magnetic diffusivity.
We believe that the reason lies again in weaker coupling between matter and magnetic field due to diffusivity.
We have argued above that the disk winds in our simulations are mainly driven by the toroidal magnetic field.
The toroidal field is vanishing initially and is induced by the disk rotation and also by the inertia of the outflow
material.
Thus, similar to our arguments for the accretion rate,  
we understand that with higher diffusivity the field lines that penetrate the disk are less coupled to the material
and a weaker toroidal field component is induced.
As a result, a weaker toroidal field pressure is available for the disk wind ejection and acceleration.
In this sense, the increasing magnetic diffusivity also suppresses the ejection process in the system.

\subsection{Efficiency of outflow launching}
\label{astro_dis_outflow_eff_subsec}
We have discussed above the influence of magnetic diffusivity on the ejection process.
In general magnetic diffusivity lowers the ejection rate. 
However, we have noticed that the magnetic diffusivity may lead to a higher outflow {\em efficiency}. 
Here, we define the outflow efficiency as 
\begin{eqnarray} 
\xi=\left|\frac{\dot{M}_{\rm eje}}{\dot{M}_{\rm acc}}\right|
\label{astro_dis_outflow_efficiency_eq}
\end{eqnarray}
where $\dot{M}_{eje}$ and $\dot{M}_{acc}$ are the total ejection rate and the inner accretion rate, respectively,
defined in Section~\ref{astro_dis_acc_eject_illu_sec}. 
We note that outflow efficiencies $\xi > 1$ are possible.

In Table~\ref{astro_dis_survey_1_acc_eje_table} we show the outflow efficiency $\xi$ for the different 
simulations averaged from $t=2000$ to $t=3000$. 
In general, the outflow efficiency of disks with $\eta_{0} > 10^{-4}$ is much larger than that 
of weakly diffusive disks with $\eta_{0} < 10^{-4}$.

At this stage we may only speculate about the reason for this trend.

One possibility may be the field structure just above the disk.
The region above the accretion disk evolves as in ideal MHD. 
Material is frozen to the field lines and either flows along the field or needs to ''push" the magnetic field component that is
perpendicular to the motion.
We find that the simulations with lower disk magnetic diffusivity
result in a more tangled field structure (as discussed above most probably
due to the not fully stable initial disk hydrodynamics).
One may argue that the tangled field is advected upwards along with the outflow and then lowers the efficiency of 
acceleration.

Also, mass loading is known to be governed by the physical conditions at the disk surface or the sonic
surface respectively.
In our case of a super Alfv\'enic injection, Lorentz forces will play a major role.
We conjecture that for a tangled magnetic field also the  Lorentz forces are tangled
and thus less efficient in launching.

In Figure~\ref{astro_dis_D1D17_field_lines_img} we show the field line structures for simulation {\em D1} (left panel) 
and {\em D10} (right panel) at $t=2000$. 
In the non-diffusive simulation {\em D1} (outflow efficiency $\xi=0.5$), the matter flow inside the disk becomes 
turbulent, also disturbing the smooth structure of field lines.
The disturbance propagates outwards, leading to a field structure that is suppressing efficient acceleration.
On the other hand, the field lines in simulation {\em D10} (outflow efficiency $\xi=2.9$) still keep their 
original smooth structure at $t=2000$.

Another possibility for the observed interrelation between diffusivity and mass loading can be ohmic 
heating. 
This will result in a somewhat higher sound speed at the disk surface for the more diffusive disk.
From steady state MHD wind theory and non-relativistic simulations it is well known that the 
sound speed at the disk surface influences the mass loading of the wind, what would lead to a 
relation similar to what we observe.

In summary, the disk evolution under low magnetic diffusivity leads to a tangled magnetic field structure
that we expect to lower the outflow acceleration efficiency.
On the other hand, a high magnetic diffusivity will decouple matter and magnetic field, leading to a lower acceleration 
efficiency.
Both processes compete with each other, we find that for our setup the smoothing of field structure 
for higher diffusivity is eventually winning.
This holds both for the magnetic field pressure gradient that is then less efficient in pushing the matter away 
from the disk surface, and also for a magneto-centrifugal driving.

Overall we find in our simulations that the efficiency of the launching and acceleration process does not simply 
increase linearly with the diffusivity level (see Table~\ref{astro_dis_survey_1_acc_eje_table}). 
For our model setup and the parameters chosen, the combined action of outflow driving forces have an efficiency 
that seems to peak at $\eta_{0} \sim 5\times10^{-4}$.  
Further studies are needed in order to understand the accretion-ejection interrelation in detail.

\section{Rotating black holes}
\label{astro_dis_wind_vs_BZ_sec}
As one of our key goals for developing {\HAR}, we now study the behavior of the accretion-ejection system and 
compare the ejection properties with those of a jet launched  by a rotating black hole.
We will focus on the results from simulations {\em D11} - {\em D15} (see Table \ref{astro_dis_sim_table}). 

In \citet{2017ApJ...834...29Q}, we reported convergence issues for long simulation time scales in {\HAR}. 
This problem does not affect simulations with Schwarzschild black holes,
however, we cannot reach long simulation times beyond say $t=1500$ for high Kerr parameters, such as for {\em D15}. 
For this reason, in the following we will limit our discussion to simulation results before $t=800$.
Longer simulations with high black hole spin will be published elsewhere.

\subsection{Influence of black hole spin on accretion and ejection}
Simulations {\em D11} - {\em D15} apply the same initial conditions except for their black hole spin parameter $a$, 
that is $0$, $0.1$, $0.2$, $0.5$ and $0.9375$, respectively.

\begin{figure*}
\centering
\includegraphics[width=2.in]{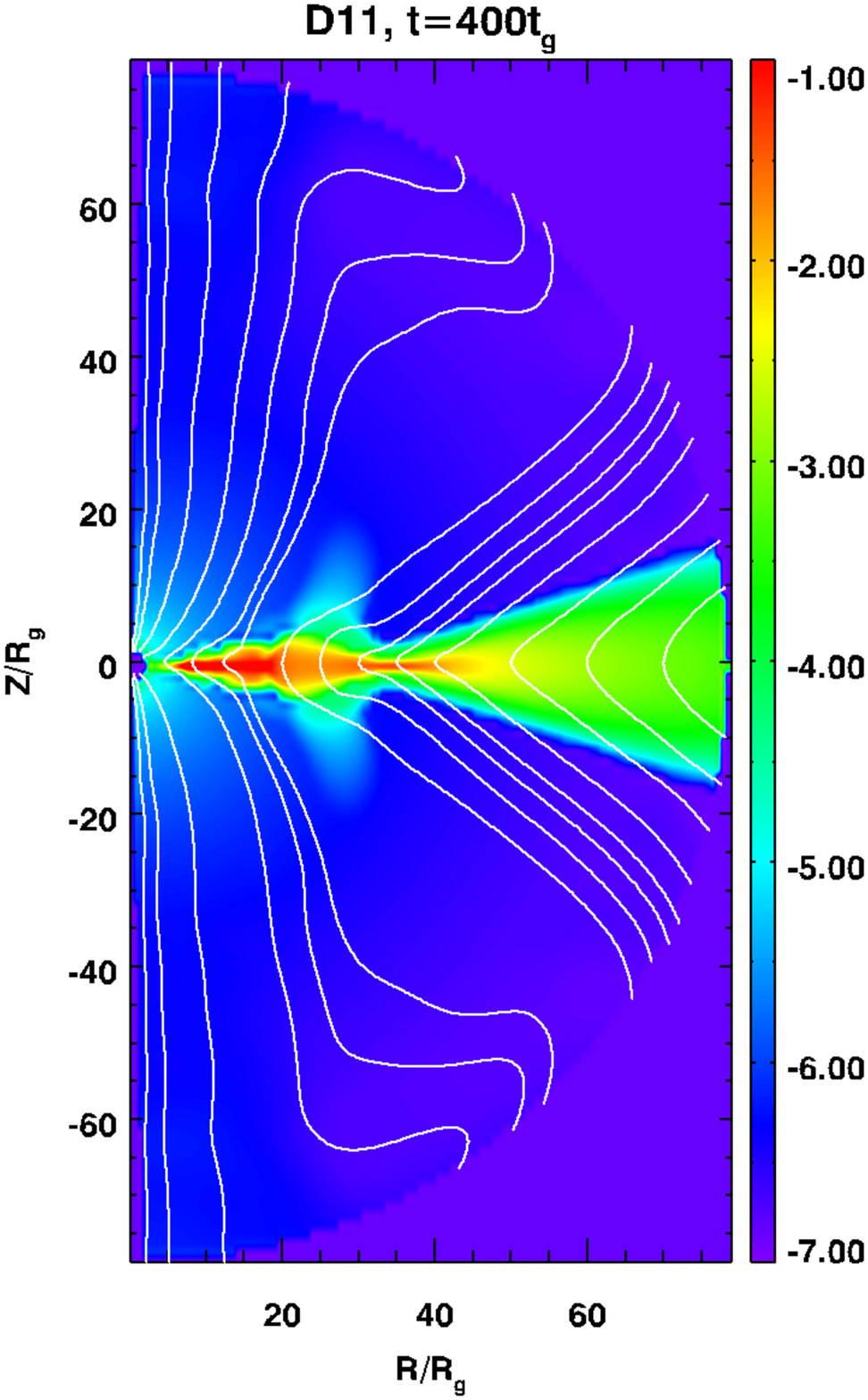}
\includegraphics[width=2.in]{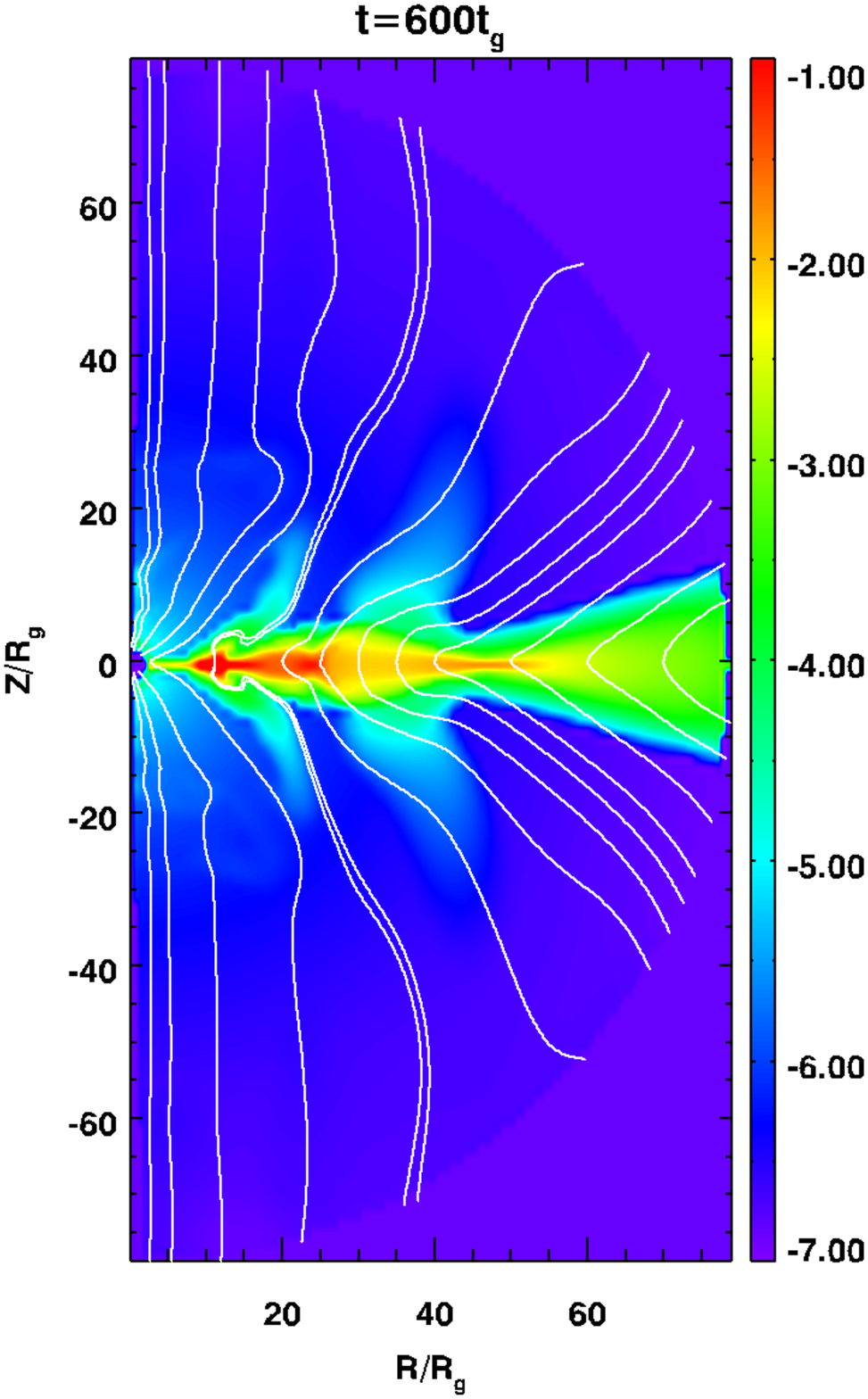}
\includegraphics[width=2.in]{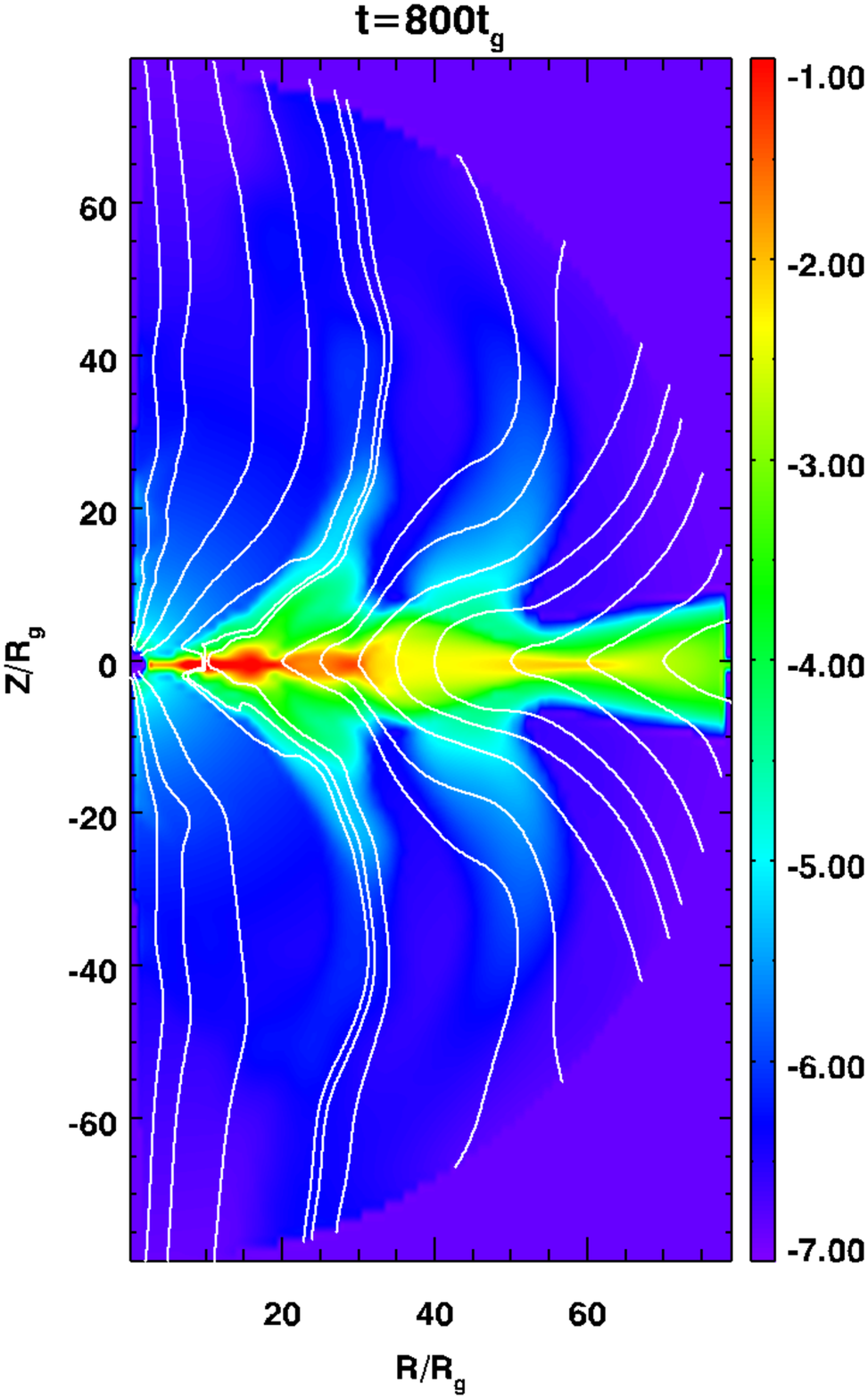}
\includegraphics[width=2.in]{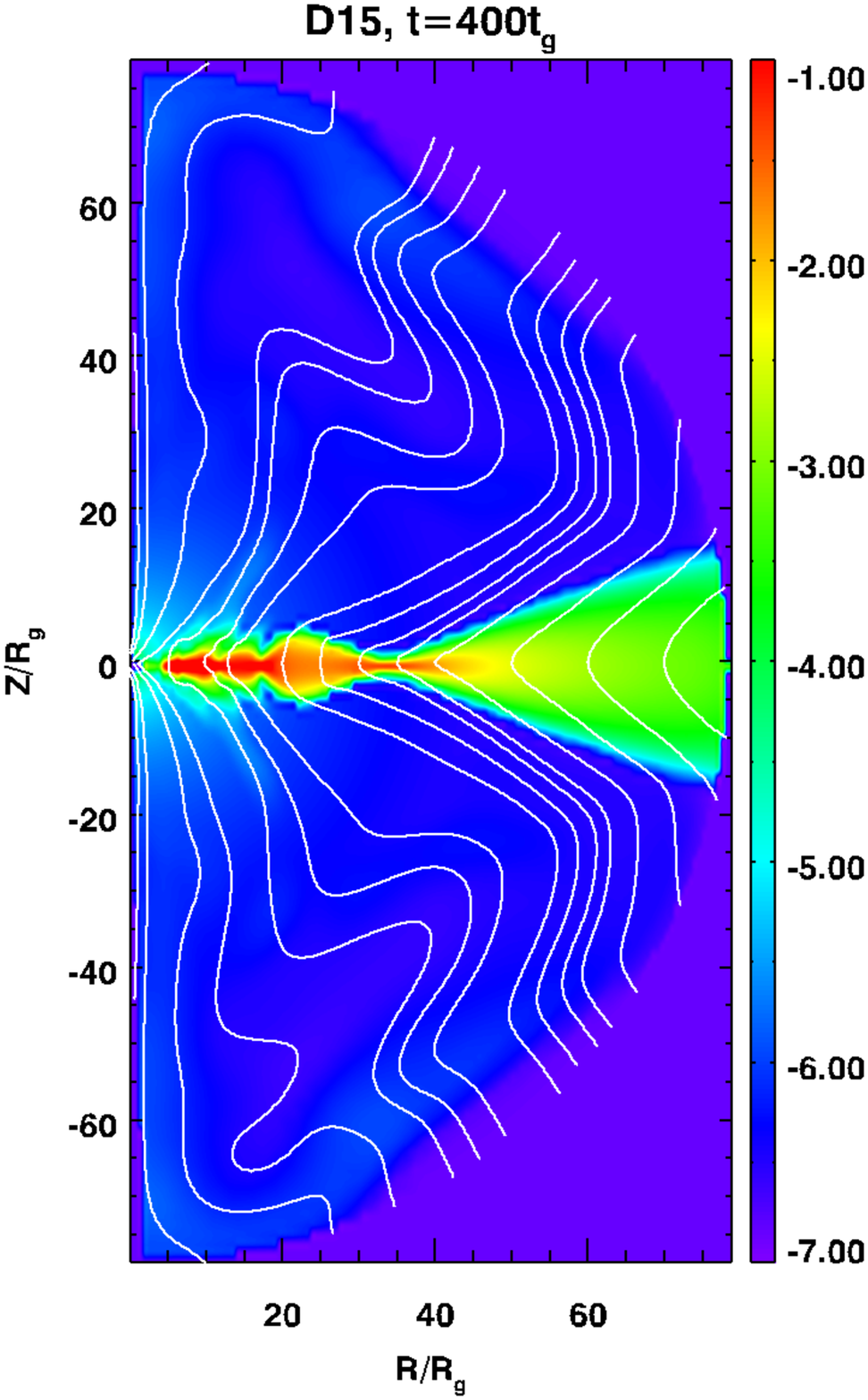}
\includegraphics[width=2.in]{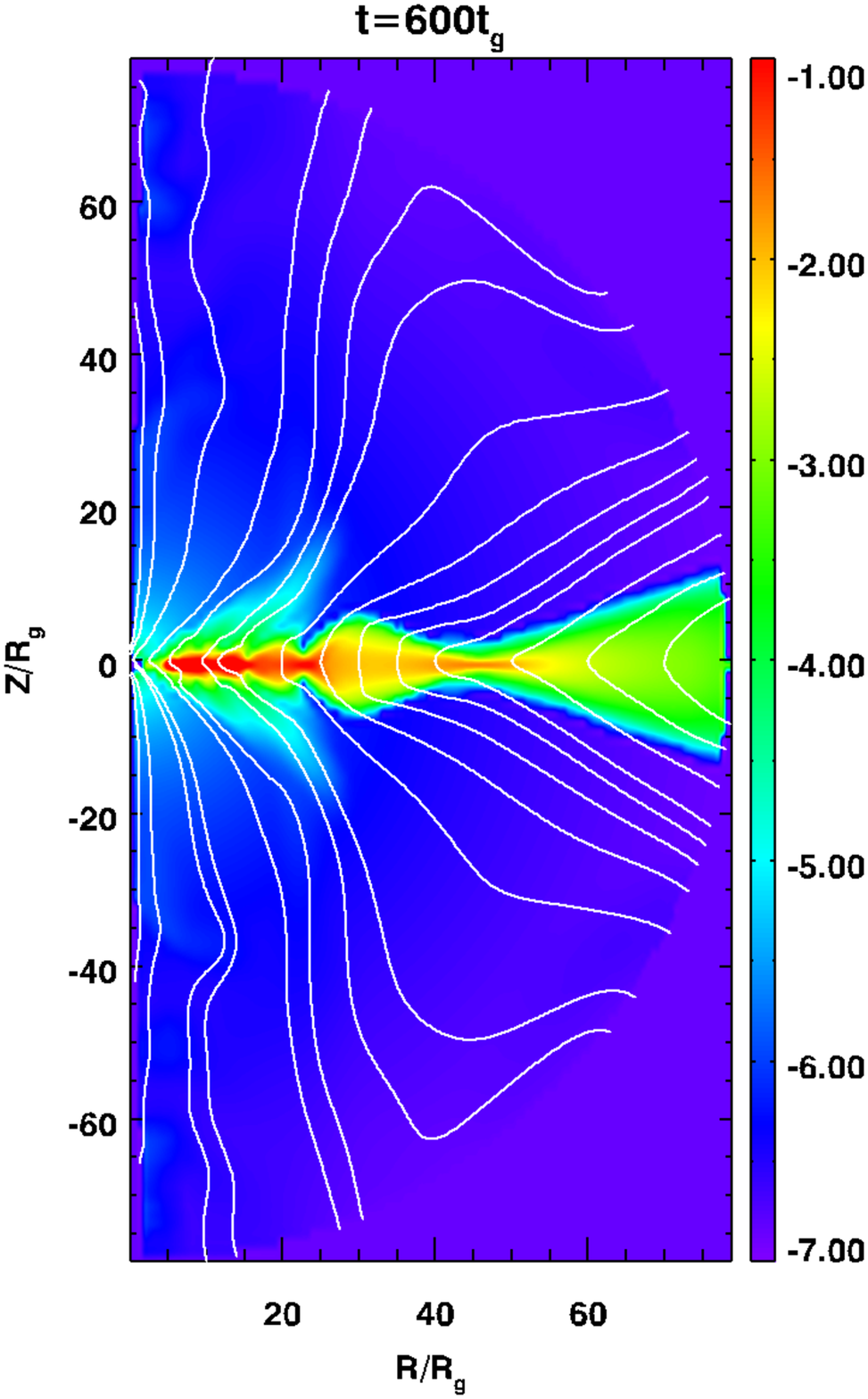}
\includegraphics[width=2.in]{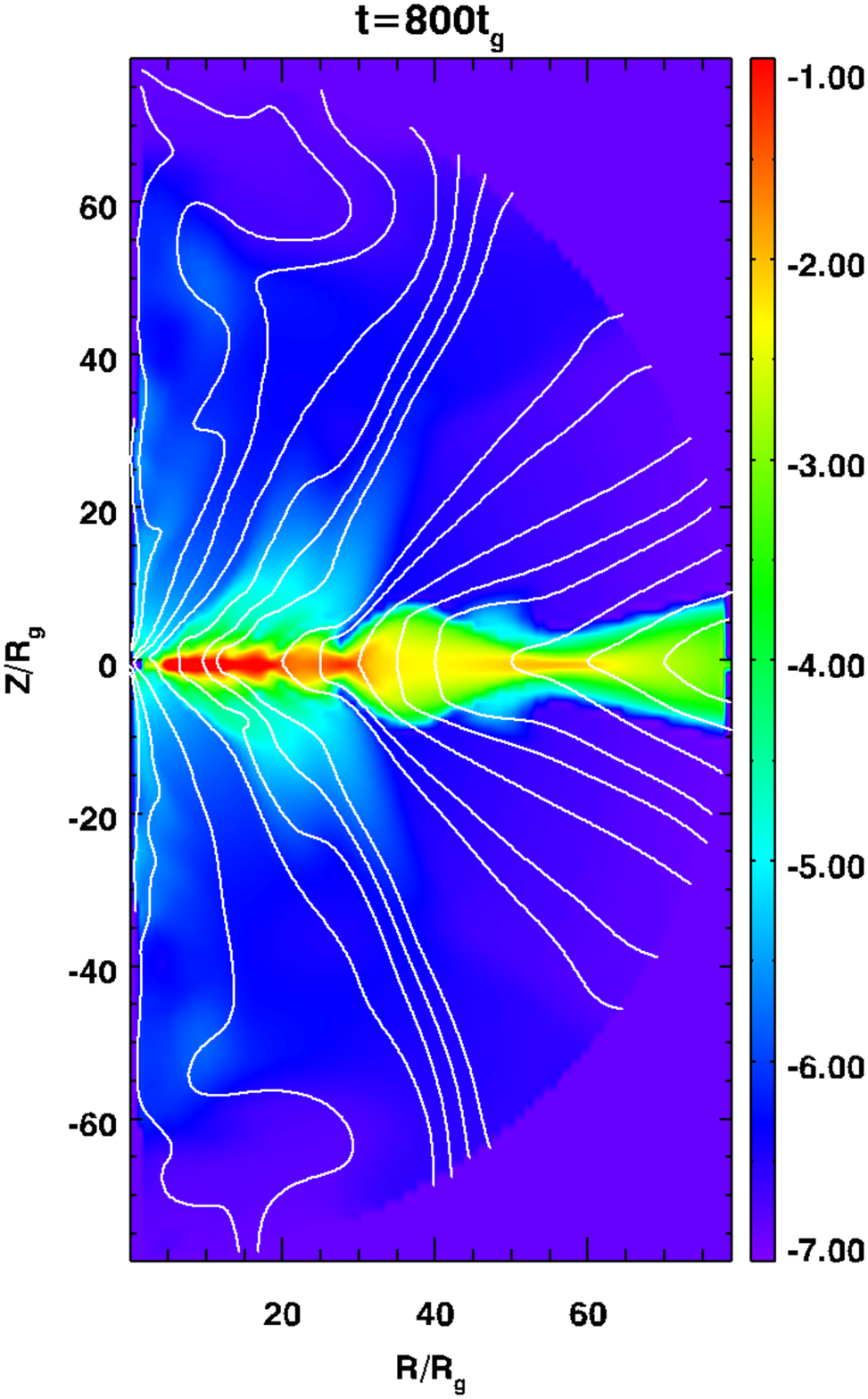}
\caption{Snapshots of density and poloidal magnetic field for simulations {\em D11} (top) and {\em D15} (bottom) at time
$t=400, 600, 800$. 
Note the difference in the disk wind morphology from simulation {\em D15} and {\em D11}. }
\label{astro_dis_morph_D11D15_rho_img}
\end{figure*}

In Figure~\ref{astro_dis_morph_D11D15_rho_img} we show density snapshots for simulations {\em D11} (upper panels) 
and {\em D15} (lower panels) at times $t=400, 600, 800$.
Simulation {\em D11} shows the time evolution for a non-rotating black hole, different only from {\em D7} by the
profile of the magnetic diffusivity parameter $\chi$. 
Indeed, the disk wind evolution in simulation {\em D11} is very similar to that in simulation {\em D7} (see 
Figure~\ref{astro_dis_morph_D6_rho_img}). 

Simulation {\em D15} has the largest spin parameter $a=0.9375$. 
Clearly, we see the launching of disk winds as well. 
However, the morphology of the outflow is different from that in simulation {\em D11}.
In general, the launching of disk outflow in simulation {\em D15} takes more time than in simulation {\em D11}.
If we compare the snapshots at $t=800$, we find already two outflow streams leaving the disk surfaces in 
simulation {\em D11}, while there is only one in simulation {\em D15}.
We understand this as due to the general interrelation between accretion and ejection.
We find that accretion becomes less efficient for larger black hole spin (see discussion below).
Assuming a general interrelation between accreting and ejection - thus a specific ejection efficiency
for a disk with certain diffusivity and plasma-$\beta$ - the outflow will be weakened if the accretion 
decreases.

\begin{figure}
\centering
\includegraphics[width=6.cm]{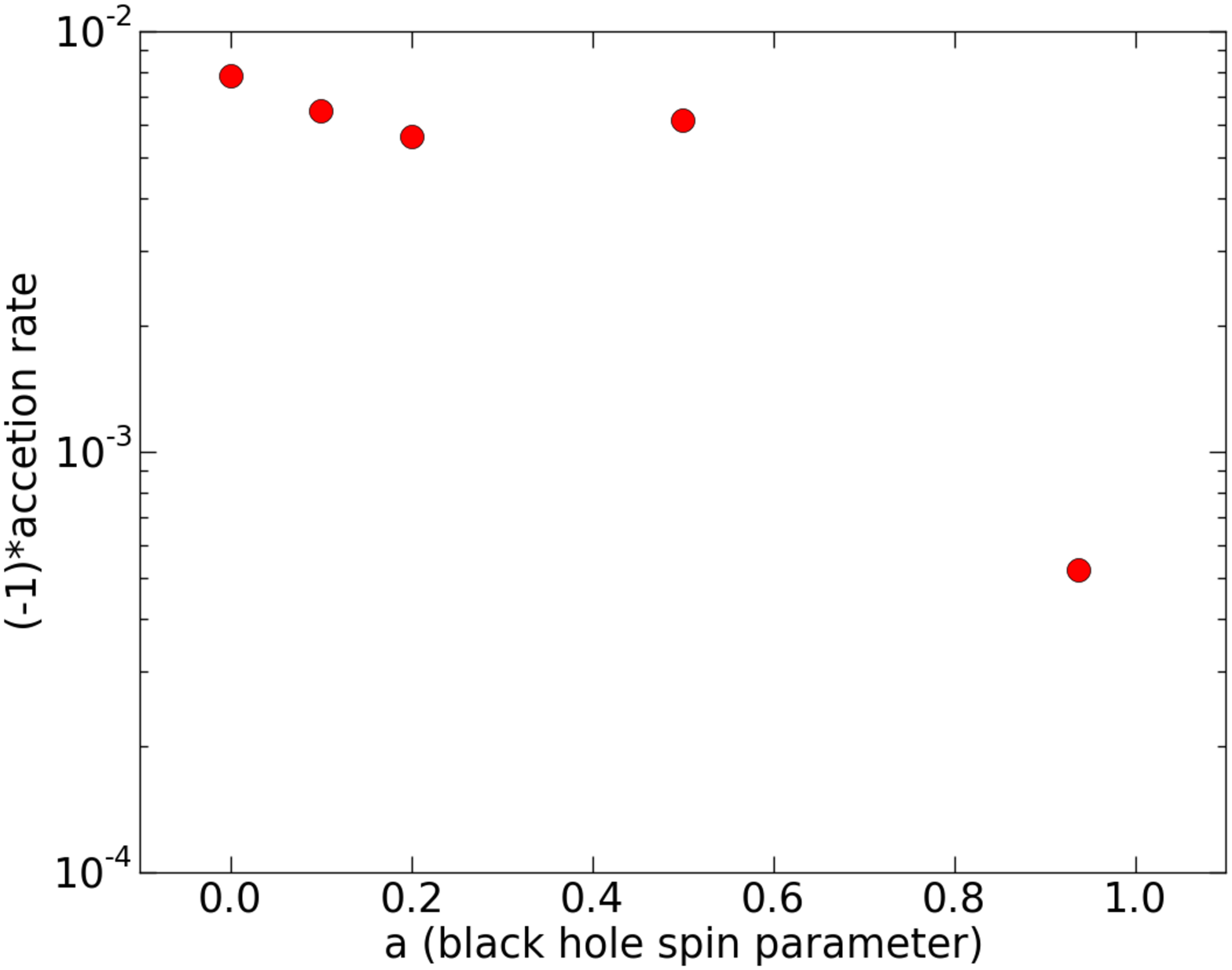}
\includegraphics[width=6.cm]{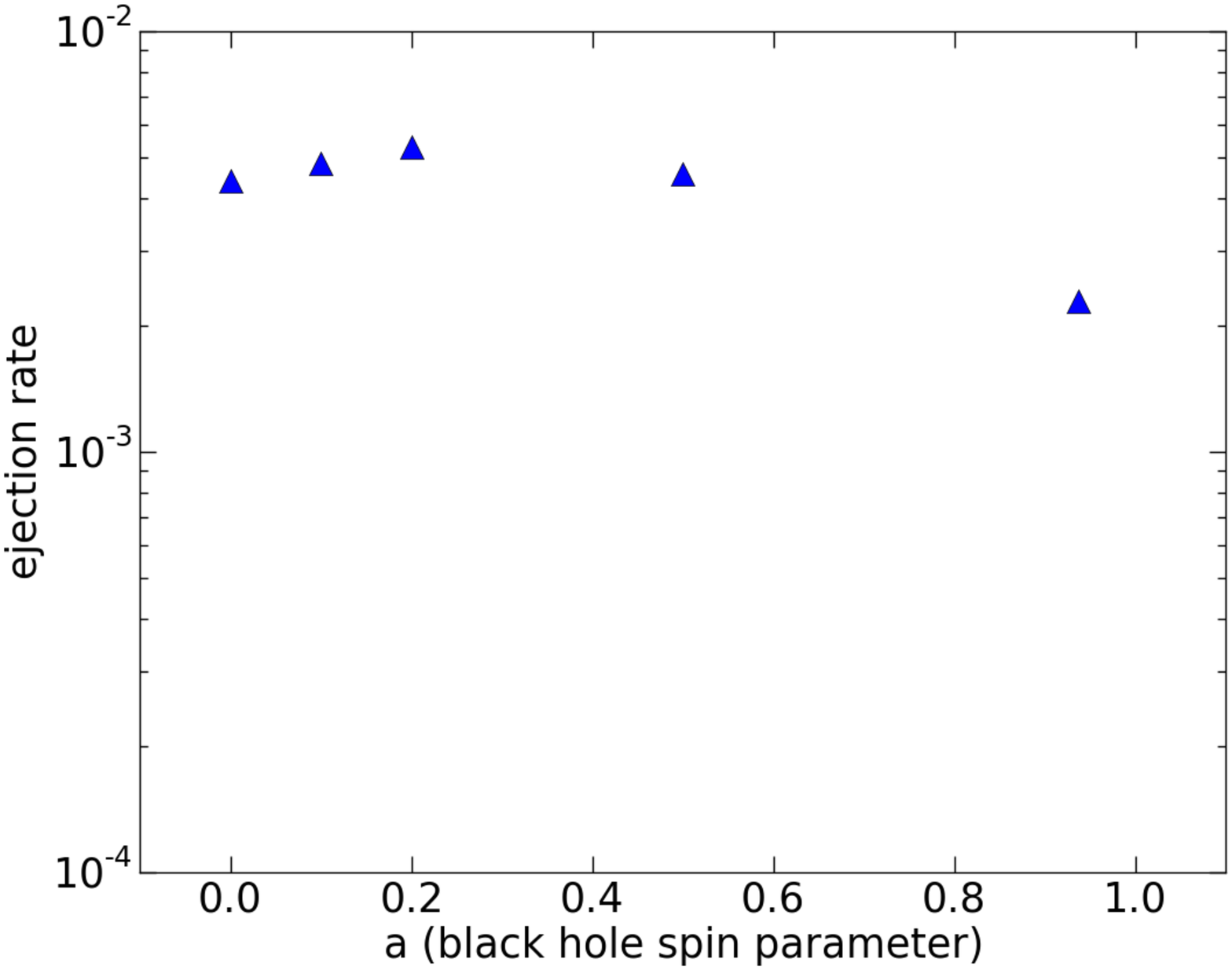}
\caption{The time averaged accretion rate (top) and ejection rate (bottom) for simulations {\em D11} - {\em D15} 
with the black hole spin parameter $a=0$, $0.1$, $0.2$, $0.5$ and $0.9375$, respectively. 
The averages are taken in the time interval from $t=500$ to $t=800$. 
Negative accretion rates were considered in the upper panel in order to plot them on a 
log scale.
We note that the result for {\em D14} that is offset the trend defined by the other data points.}
\label{astro_dis_acc-eje-a_relation_img}
\end{figure}

In Figure \ref{astro_dis_acc-eje-a_relation_img}, the time averaged accretion and ejection rates for simulations 
{\em D11} - {\em D15} are shown with respect to the black hole spin parameter.
As we see, simulations with higher spin parameters tend to return weaker accretion rates. 
The only exception is for simulation {\em D14} and is may be caused by the choice of the time interval for averaging; 
note that the averages are taken from $t=500$ to $t=800$ and the accretion system is not yet in steady state
at time $t=800$. 


\begin{figure}
\centering
\includegraphics[width=4.2cm]{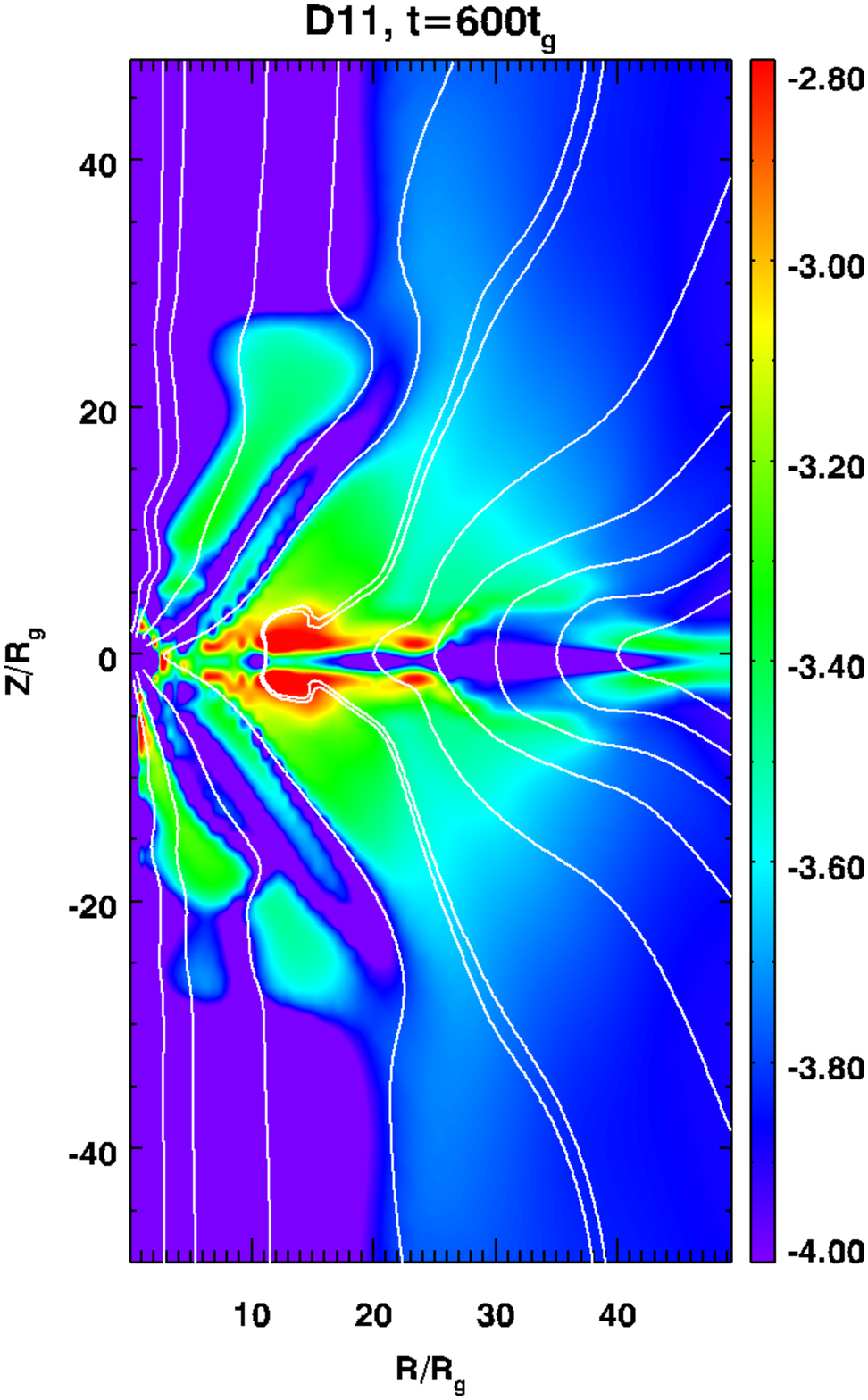}
\includegraphics[width=4.2cm]{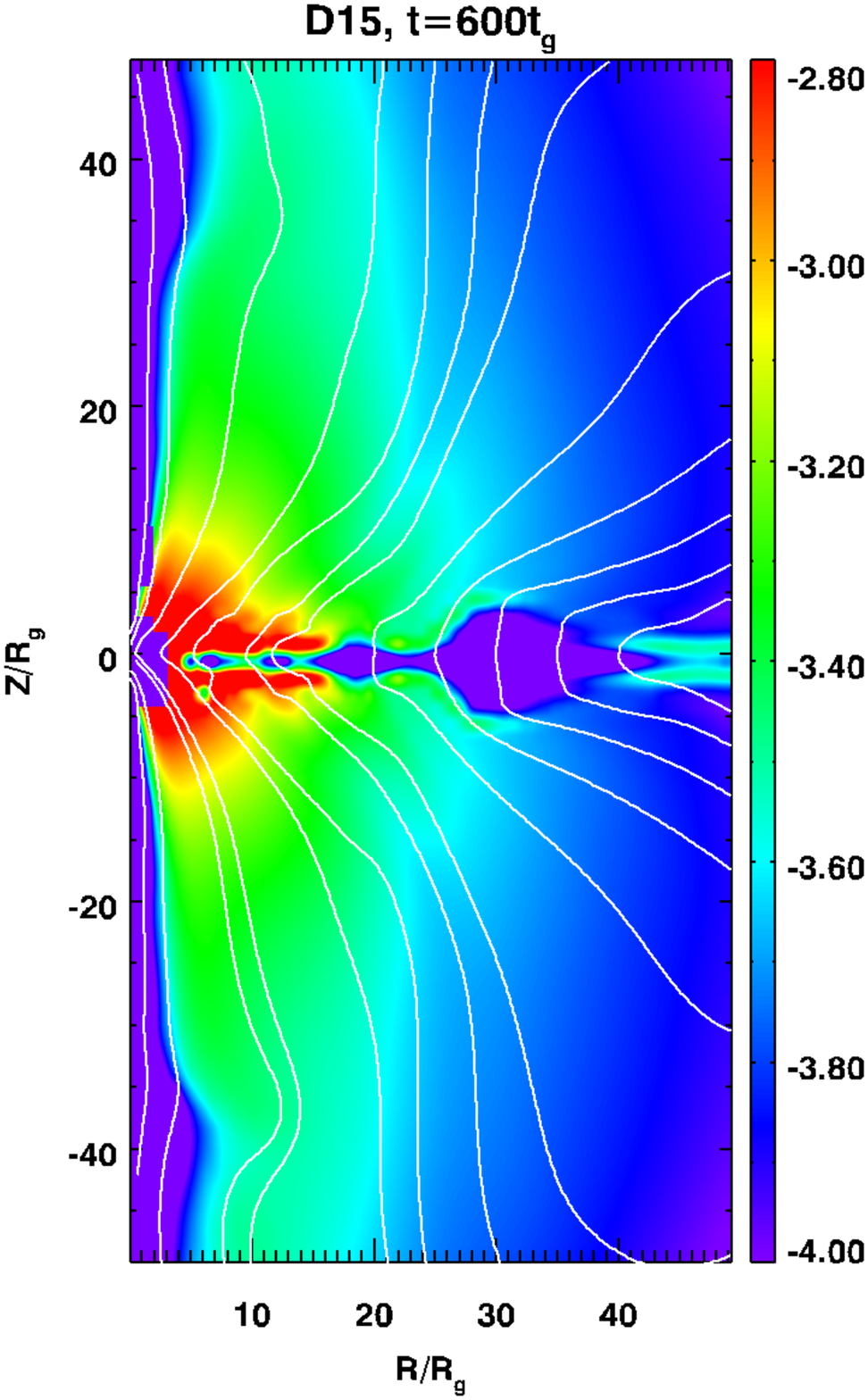}
\caption{The toroidal magnetic field strength (log scale) for simulation {\em D11} (left) and {\em D15} (right) 
at time $t=600$. 
}
\label{astro_dis_D9D16_logBphi_img}
\end{figure}

The influence of the black hole spin on the accretion rate may be explained by the behavior of the magnetic field 
lines that penetrating the ergosphere. 
The magnetic field lines that comes close to the rotating black hole will be tangled due to frame-dragging.
We thus expect a larger toroidal field near a rotating black hole than a non-rotating black hole. 
To confirm this point, we show the toroidal field strength for simulations {\em D11} and {\em D15} in Figure \ref{astro_dis_D9D16_logBphi_img}. 
In simulation {\em D11} the toroidal magnetic field is only induced by the rotation of the accretion disk,
and can be found mainly in the disk and disk outflow region.
In simulation {\em D15} the rapid rotation of the black hole induces also a strong toroidal field in the region 
above the black hole and also close to the horizon.
In comparison, no toroidal field is visible in the black hole vicinity in simulation {\em D11}. 
We see this as strong evidence of the black hole frame dragging mentioned above.

We believe that as a result of the additional toroidal magnetic field pressure the accretion flow is
decelerated, and, hence, accretion rate is suppressed. 


On the other hand, the ejection rates presented in Figure~\ref{astro_dis_acc-eje-a_relation_img} do not show 
a clear trend. 
This is understandable, since these disks are all similar, and thus deliver a similar outflow.
Naturally, with similar ejection rates but different accretion rates, the outflow efficiency 
varies.
We measure an outflow efficiency $\xi=0.56, 0.75, 0.94, 0.75$ and $4.34$, respectively, that are on average increasing
for an increasing black hole spin.

\subsection{Blandford-Znajek launching}
The power of the Blandford-Znajek process we measure by the electromagnetic flux across a surface close to 
the horizon.
We follow \citet{2004ApJ...611..977M} and define the total electromagnetic energy flux that goes through a surface
with radius $R$ as
\begin{eqnarray}
\dot{E}^{\rm (EM)}(R)=2\pi \int^{\pi}_{0} d\theta \sqrt{-g|_{r=R}}\,F^{\rm (EM)}(R,\theta),
\label{astro_dis_EM_Edot_def_eq}
\end{eqnarray}
where 
\begin{eqnarray}
F^{\rm (EM)}(r,\theta) &=& -{T^{r}_{t}}^{\rm (EM)}
\nonumber \\            
                       &=& -[ (b^{2}+e^{2})(u^{r}u_{t}+\frac{1}{2}g^{r}_{\,\,t}) -  b^{r}b_{t} - e^{r}e_{t} 
\nonumber \\ 
                       & &- u_{\lambda}e_{\beta}b_{\kappa}
                       ( u^{r}\epsilon_{t}^{\,\,\lambda \beta \kappa} + u_{t}\epsilon^{r \lambda \beta \kappa})]
\label{astro_dis_Fem_def_eq}
\end{eqnarray}

is the electromagnetic energy flux per solid angle. 

\begin{figure}
\centering
\includegraphics[width=6.cm]{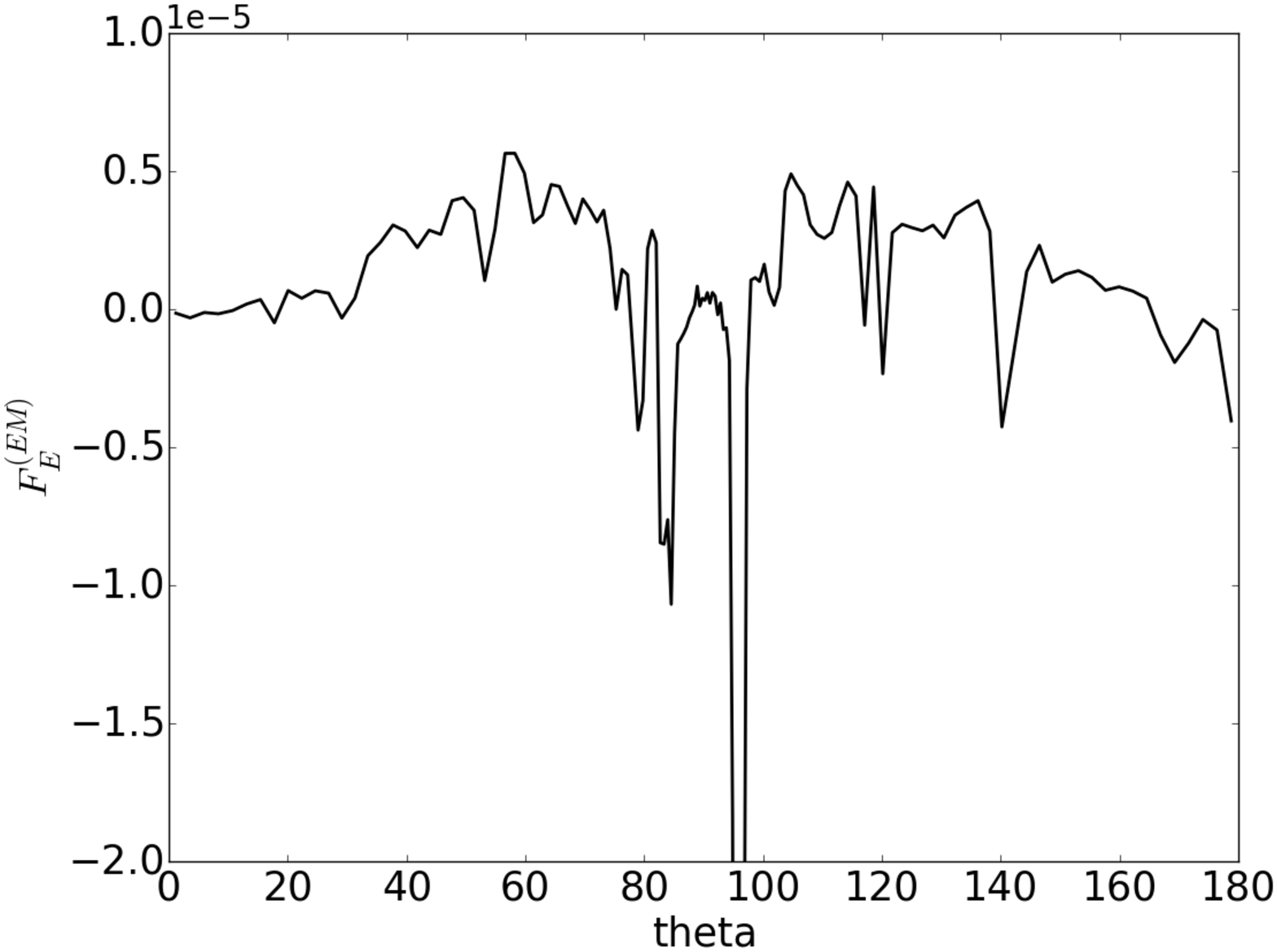}
\includegraphics[width=6.cm]{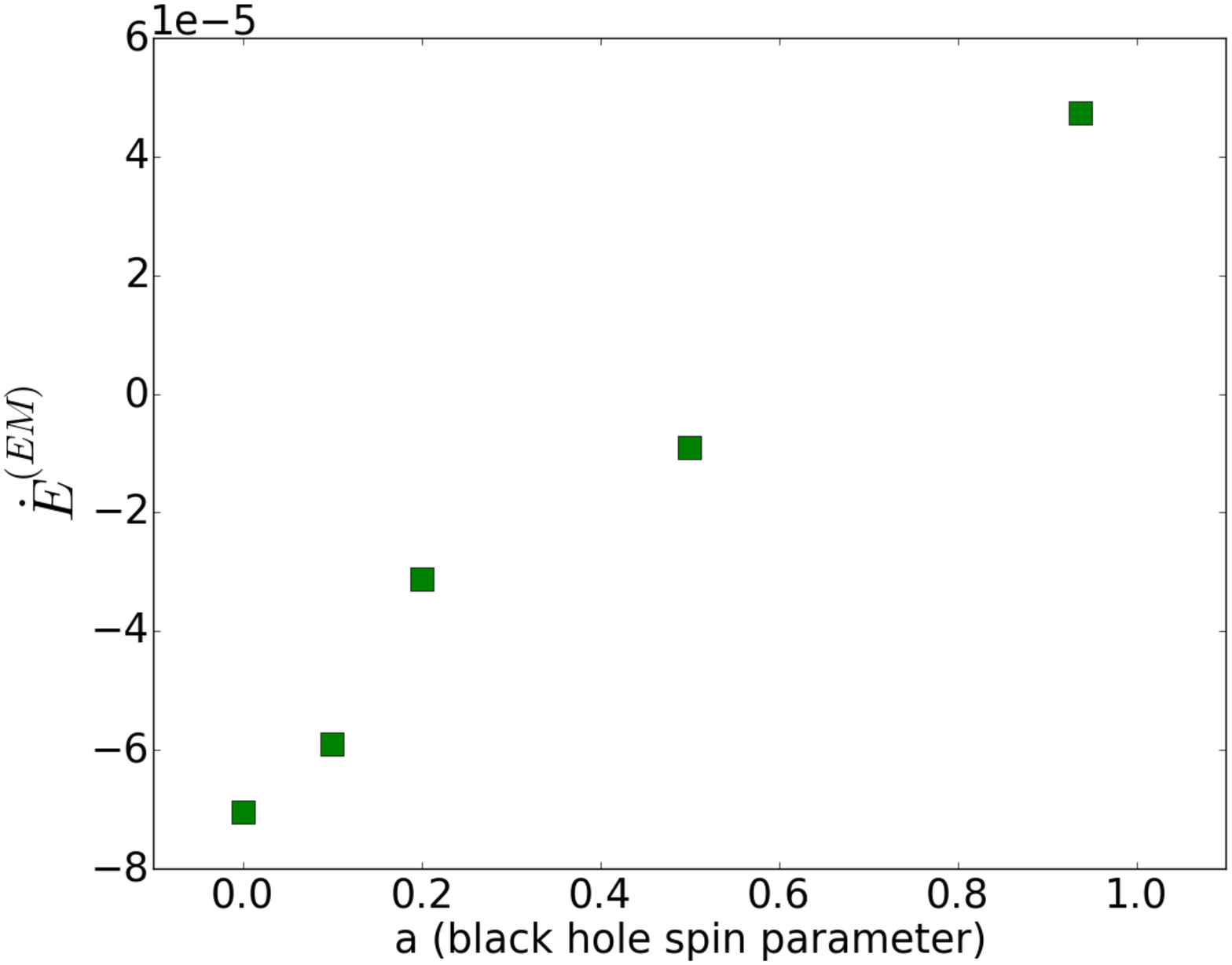}
\caption{
The time averaged profile of electromagnetic energy flux per solid angle $F^{\rm (EM)}$ 
along $\theta$ for simulation {\em D15} (top).
The values of $F^{\rm (EM)}(\theta) $ are calculated at $r=2$.
The time averaged electromagnetic energy flux measured at $r=2$ for simulation runs
{\em D11}-{\em D15} (bottom). 
The integration is done only for $\theta$ from $0^{\circ}$ to $75^{\circ}$ and from $105^{\circ}$ to $180^{\circ}$
in order to disentangle the outflow from the accretion flow. 
All averages are taken in the time interval from $t=500$ to $t=800$. 
}
\label{astro_dis_D15_Fem_vs_theta+Fem_at_r=2_img}
\end{figure}

The time averaged (from $t=500$ to $t=800$) flux $F^{\rm (EM)}$ for simulation {\em D15} at radius $r=2$ is
shown in the upper panel of Figure \ref{astro_dis_D15_Fem_vs_theta+Fem_at_r=2_img}.
The positive values for $F^{\rm (EM)}$ in the regions $\theta$ from $20^{\circ}$ to $70^{\circ}$ and from $110^{\circ}$ 
to $160^{\circ}$ specifically measure the energy output from the black hole through the Blandford-Znajek mechanism. 
The negative values that peak near the disk mid-plane indicate the electromagnetic flux advected with the accreting flow 
from the disk. 
The reason why there are several peaks is not really clear.
The reason for the low (absolute) value around $\theta =90^{\circ}$ results from the fact that the toroidal magnetic field
is almost vanishing at the equatorial plane (the Poynting flux is $\propto B_{\phi} B_{\rm p}$). 

We may compare this plot to similar, time averaged profiles from ideal GR-MHD simulation in
\citet{2004ApJ...611..977M} (see Fig. 9b in \citealt{2004ApJ...611..977M}). 
Their simulation initially employed a gas torus surrounding a rotating black hole with spin parameter $a=0.5$
and with a magnetic field initially confined in the torus.
In general, our figure shows a trend very similar to their's, showing also flux values $F^{\rm (EM)}$ influenced 
by the advection of magnetic flux due to mass accretion. 

We note that \citet{2004ApJ...611..977M} employ ideal GR-MHD, a substantially different field structure
(not net magnetic flux initially), and a lower coronal density.
Together all these differences lead to a stronger influence of the mass accretion from the disk on the
$F^{\rm (EM)}$, as is hardly a disk wind and the infalling material covers a much wider area.

\begin{figure}
\centering
\includegraphics[width=4.2cm]{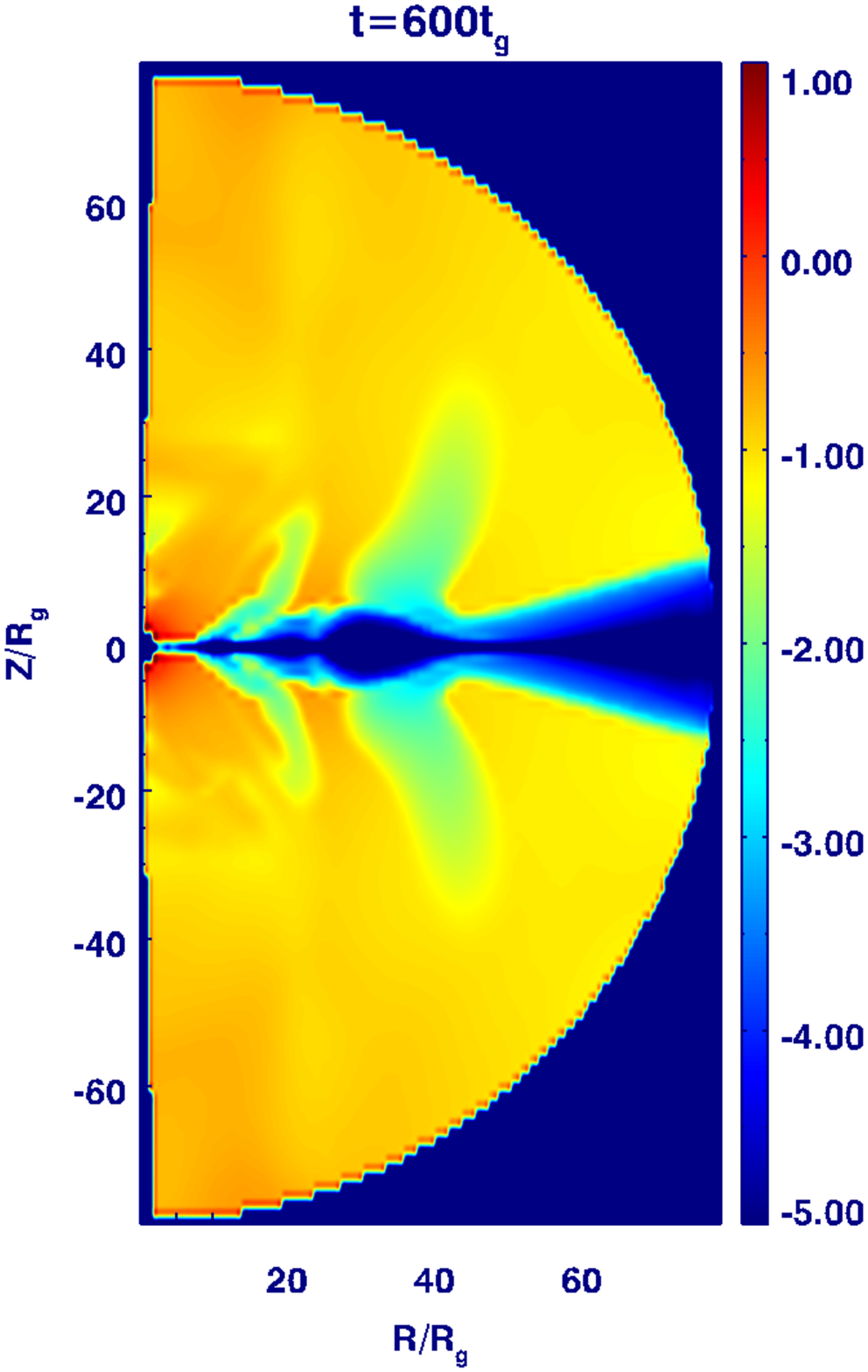}
\includegraphics[width=4.2cm]{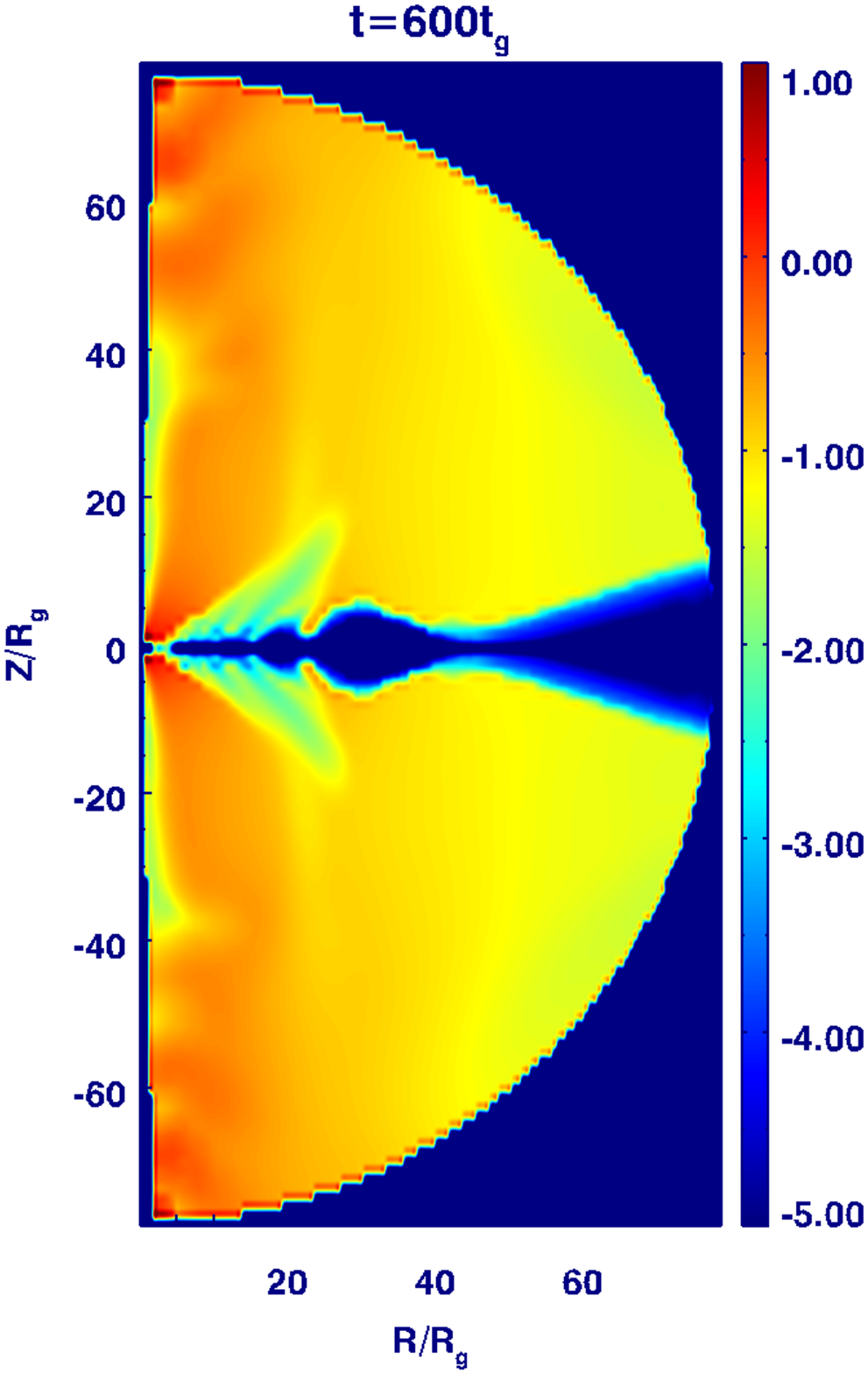}
\caption{Comparison of the magnetization $b^2/\rho$ for simulation {\em D11} (left) and {\em D15} (right) at time $t=600$.
}
\label{astro_dis_D11D15_magnetization_img}
\end{figure}

In Figure~\ref{astro_dis_D11D15_magnetization_img}, we compare the magnetization ($\sigma = b^{2}/\rho$) for 
simulation {\em D11} and {\em D15} at $t=600$. 
In the case of the rapidly spinning black hole ({\em D15}, a=0.9375), we clearly identify the regions with positive
electromagnetic flux $F^{\rm (EM)}>0$ shown in Figure~\ref{astro_dis_D15_Fem_vs_theta+Fem_at_r=2_img} with those 
areas in Figure~\ref{astro_dis_D11D15_magnetization_img} 
that are highly magnetized in the funnel region that is driven by the Blandford-Znajek mechanism. 
In contrary, for the case of a non-spinning black hole (simulation {\em D11} with $a=0$), the magnetization is low, 
consistent with an inefficient Blandford-Znajek mechanism (see also below).


Using Equation~(\ref{astro_dis_EM_Edot_def_eq}), we plot the time averaged electromagnetic energy flux measured at 
$r=2$ for simulations {\em D11} - {\em D15} in Figure \ref{astro_dis_D15_Fem_vs_theta+Fem_at_r=2_img} (lower panel). 
The simulations with $a \leqslant 0.5$ return negative electromagnetic energy fluxes - thus, an inefficient Blandford-Znajek (BZ)
driving - even if we do not consider the advection of electromagnetic flux by mass accreting from the disk.
We attribute this behavior of the energy flux to the time evolution of the corona density that we have initially set. 
Since the corona is not stable against the accretion at the horizon, the matter will fall into the black hole together 
with the electromagnetic energy they carry. 
This will cancel any positive energy extraction from the black hole. 
Thus the electromagnetic energy flux becomes negative for the case of low black hole rotation, where the Blandford-Znajek 
mechanism is inefficient.

Here it is interesting to discuss arguments suggested by \citet{2009MNRAS.397.1153K}, who investigated the activation
of the BZ-mechanism in collapsar stars.
\citet{2009MNRAS.397.1153K} suggest that for the BZ-mechanism to operate, the Alfv\'en speed in the ergosphere should
be larger than the free fall speed.
Applying a non-relativistic estimate, \citet{2009MNRAS.397.1153K} derive a critical condition 
$\beta_{\rho} \equiv 4\pi \rho c^2 / B^2 < 1$, thus a plasma energy density smaller than the magnetic field energy 
density.
When we calculate $\beta_{\rho}$ for our simulations we find indeed values below unity.
Note that we apply the same (original) floor model of HARM that has been used for BZ mechanism simulations in the
literature \citep{2006ApJ...641..626N}.
Thus, following the arguments of \citet{2009MNRAS.397.1153K} we find the BZ mechanism likely to be
triggered\footnote{There is actually a numerical difficulty 
here such that the BZ mechanism is supposed to be favored for low $\beta_{\rho} <1$, while on the other hand the 
code (as typical for relativistic MHD codes) has convergence issues for low $\rho/B$. We note also that disk jet launching 
works most efficient for strong field s and low densities.}.
We further observe an increase of axial Poynting flux (a jet) for an increasing Kerr parameter, just es expected 
for the BZ mechanism.

In our simulations we need to support the disk corona with a substantial density, higher than in the case of simulations with an
initial gas torus (as in the literature). 
As the black hole cannot support a steady-state corona, infall of this (rather heavy) coronal gas will advect Poynting flux.
We believe that for low $a$ the advection of electromagnetic flux is larger than the Poynting flux that can be launched
by the Blandford-Znajek mechanism, and thus the overall electromagnetic energy flux of the black hole will be negative. 
We note that although this effect is a by-product of our model setup, it can indeed be astrophysically realistic in 
all cases where a free falling corona of substantial density must be considered.
Only when (numerical) floor values with arbitrarily low density are considered, the Blandford-Znajek mechanism
will dominate the energy output from the black hole.
In any case, taking this into account, the energy output from the black hole clearly increases with the black hole spin parameter. 

As a consequence of the effects discussed just above, the energy flux attributed to the Blandford-Znajek mechanism 
does not show an obvious non-linear growth with increasing 
$a$ in Figure \ref{astro_dis_D15_Fem_vs_theta+Fem_at_r=2_img} as for example in the results in
\citet{2004ApJ...611..977M, 2010ApJ...711...50T}. 
Furthermore, in the time interval from $t=500$ to $t=800$, the evolution of the accretion-ejection system is not yet 
steady.
Thus, mass accretion from the disk and the corresponding advection of magnetic flux will disturb the ``pure" 
Blandford-Znajek mechanism that we would be observed in a numerical experiment treating a disk-less black hole
\citep{2005MNRAS.359..801K}.


\begin{figure}
\centering
\includegraphics[width=6.cm]{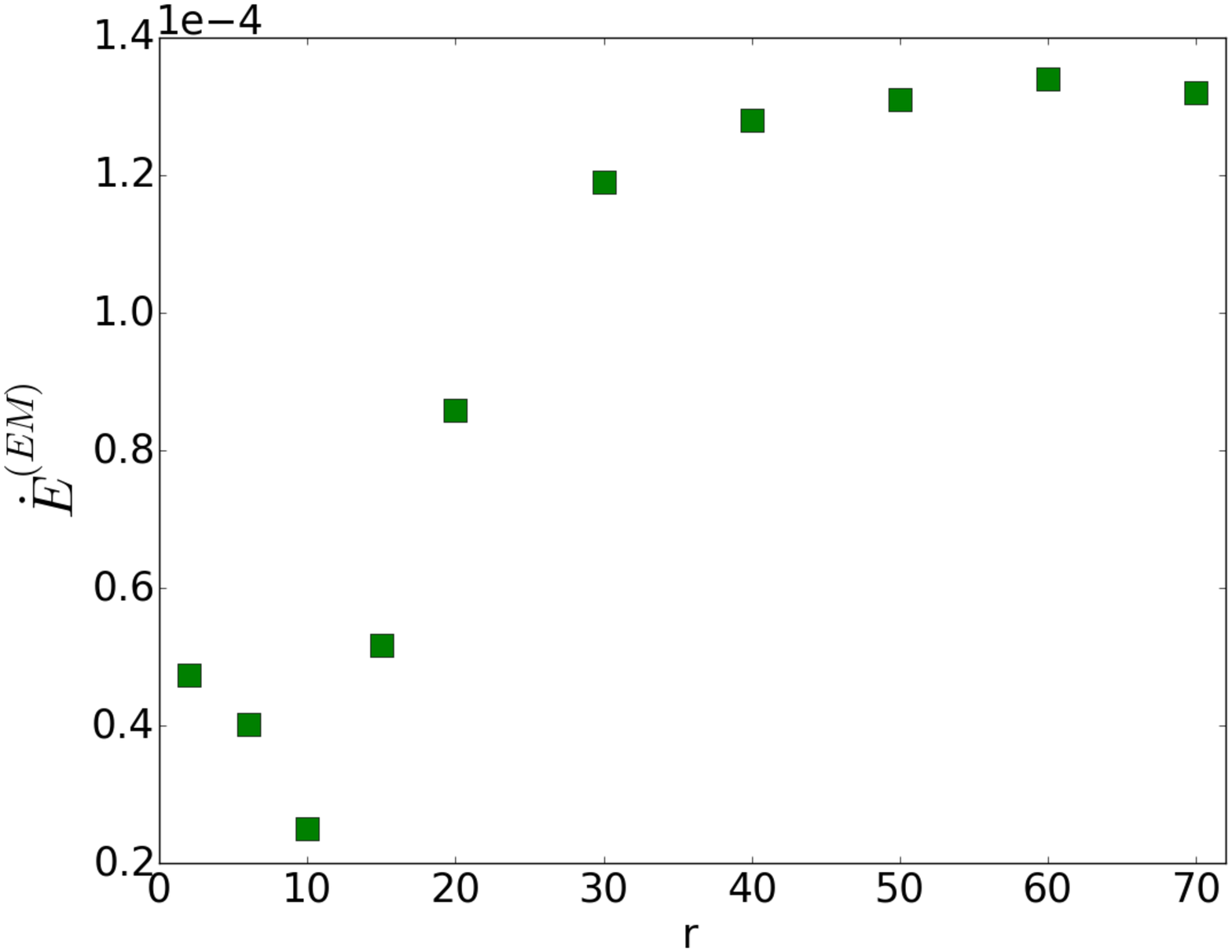}
\includegraphics[width=6.cm]{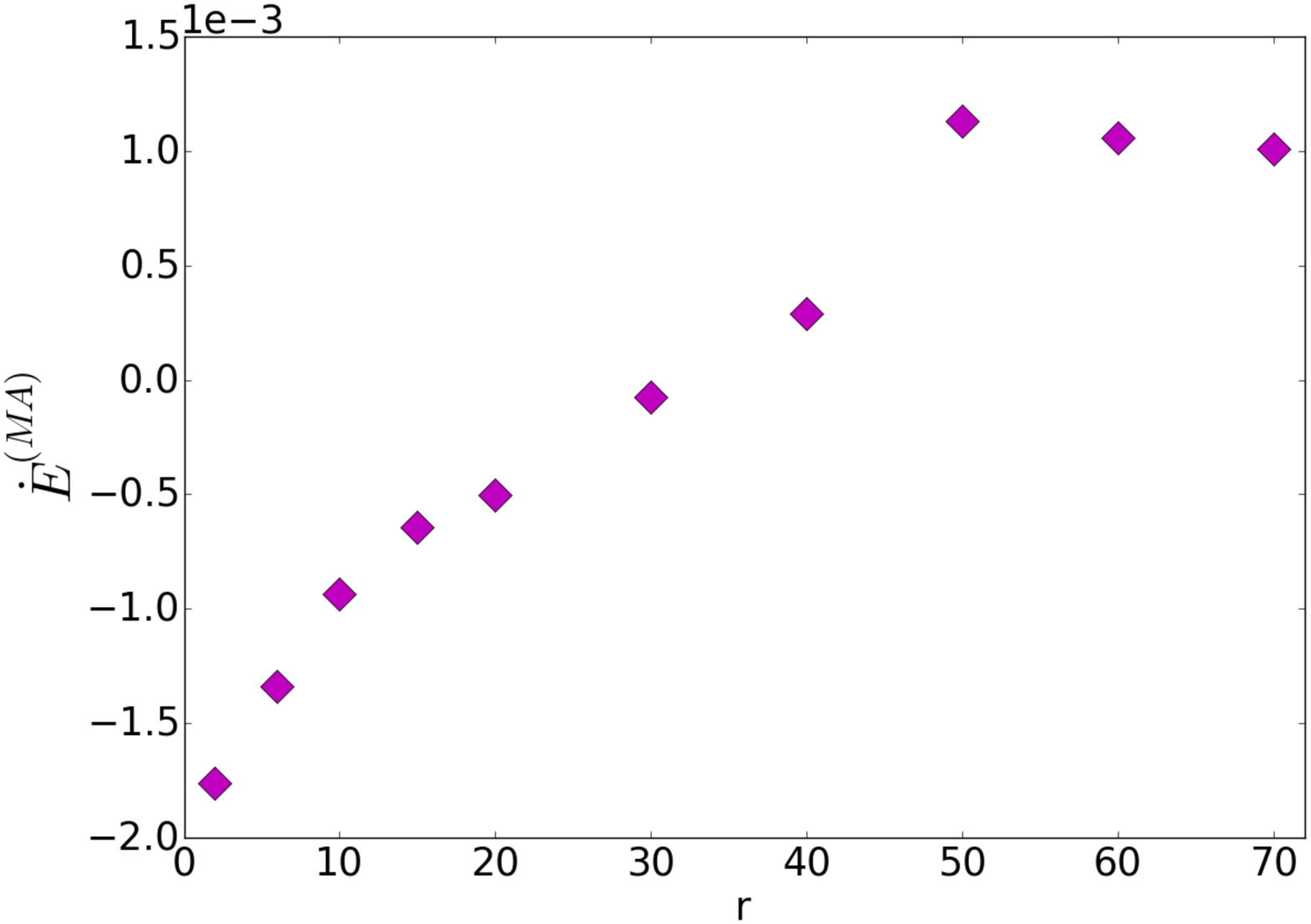}
\caption{
Time averaged electromagnetic energy flux (top) and time averaged matter energy flux (bottom)
for simulation {\em D15} measured at different radii as indicated. 
Averages are taken in the time interval from $t=500$ to $t=800$. 
The integration is done only for $\theta$ from $0^{\circ}$ to $75^{\circ}$ and from $105^{\circ}$ to $180^{\circ}$
in order to disentangle the outflow from the accretion flow.
The energy fluxes increase with radius clearly indicating that the disk wind increasingly dominates the
electromagnetic and matter energy flux in the overall outflow.
}
\label{astro_dis_D16_Fem_Fma_vs_r_img}
\end{figure}

\subsection{Disk wind vs. ergospheric driving}
We finally investigate which of the two jet launching mechanisms is more powerful, the disk wind/outflow or the
jet launched by the rotating black hole - Blandford-Payne or Blandford-Znajek?
In the following, we obviously focus on simulation {\em D15}, which has the largest spin parameter. 
In order to compare both effects quantitatively, we measure both the electromagnetic energy flux $\dot{E}^{(EM)}$ and the matter energy flux $\dot{E}^{(MA)}$ in radial direction.
Similar to Equation~(\ref{astro_dis_EM_Edot_def_eq}), the matter energy flux through a sphere
with radius $R$, $\dot{E}^{(MA)}(R)$ is defined by

\begin{equation}
\dot{E}^{(\rm MA)}(R) = 2\pi \int^{\pi}_{0} d\theta \,\sqrt{-g(R,\theta)} \,F^{(\rm MA)}(R,\theta),
\label{astro_dis_MA_Edot_def_eq_cf}
\end{equation}
where 
\begin{eqnarray}
F^{(\rm MA)}(r,\theta) &=& -{T^{r}_{t}}^{(\rm MA)}
\nonumber \\            
                       &=& -[ (\rho+u)u^{r}u_{t} + p(u^{r}u_{t}+\frac{1}{2}g^{r}_{\,\,t})]
\label{astro_dis_Fma_def_eq}
\end{eqnarray}
is the matter energy flux per solid angle. 

With this definition, we calculate the time averaged electromagnetic energy flux and matter energy flux for simulation 
{\em D15} trough spheres of different radii $R$.
Figure \ref{astro_dis_D16_Fem_Fma_vs_r_img} shows on the top the electromagnetic component of the energy flux.
We observe a clear increase of flux from $r=10$ to $r=40$.
This obviously demonstrates that the disk wind substantially contributes to the radial component of the electromagnetic 
energy flux.
We note that the geometric effect is involved. 
Disk outflow originating from larger radii have a larger cross section.
Even if the energy density in the disk outflow decreases with radius, the volume increases, and, thus, makes the
disk wind contribution to the overall energy budget substantial.
The energy flux levels off for $r > 50$ since the launching area is not yet established so far at these radii
at $t \leqslant 800$ (see Figure \ref{astro_dis_morph_D11D15_rho_img}).

In the bottom panel, the matter component of the energy flux is presented. 
Similar to the electromagnetic energy flux, the matter energy flux also grows with radius.
Within $r \lesssim 30$, the accretion flow dominates the matter energy flux (negative flux),
while for $r > 30$, outflow begins to dominate which leads to positive matter energy flux
(as visible in Figure \ref{astro_dis_D6_ur_img}, some regions above the disk surface have negative radial velocity).
Beyond $r = 50$, the growth of matter energy flux ceases, just because of the same reason as for the 
electromagnetic energy flux growth.

On one hand, according to the lower panel in Figure \ref{astro_dis_D15_Fem_vs_theta+Fem_at_r=2_img}, the pure energy 
output from the black hole in simulation {\em D15} is $\dot{E}^{(EM)} \sim 5 \times 10^{-5}$. 
This value adds to $\sim 1.4 \times 10^{-4}$ if we take the Poynting flux from disk wind into account (see 
Figure~\ref{astro_dis_D16_Fem_Fma_vs_r_img}). 
On the other hand, the energy output from the disk wind is $\dot{E}^{(MA)} \sim  1.2 \times 10^{-3}$. 
This about 10 times larger than the total $\dot{E}^{(EM)}$ and 20-30 times larger than the pure energy output 
from the black hole.
We explicitly note that here we have considered the total matter energy flux, including the rest mass flux.
We follow e.g. \citet{2004ApJ...611..977M} who define the efficiency of the BZ-mechanism by comparing the 
in-falling matter energy flux to the out-going electromagnetic energy flux, and arriving at values
$ \dot{E}^{(EM)}/\dot{E}^{(MA)}  \simeq 2-10\%$. 
Note that the latter is for a completely different initial setup (a zero-net-magnetic flux hydrostatic torus).

Compared to our case, were we find accretion matter fluxes and disk wind matter fluxes of the same order 
(within a factor 3), we would derive a similar efficiency between the matter fluxes and the BZ electromagnetic 
flux.

We now compare the BZ electromagnetic flux $\dot{E}^{(EM, BZ)}$ and the kinetic energy flux, that is the (total) 
matter energy flux, rest mass flux-subtracted.
This is the energy flux that could eventually be converted to e.g. heat or radiation, similar to the electromagnetic 
energy flux.
For the rest mass-subtracted matter flux at large radii we find 
$\dot{E}^{(MA, kin)} = 7.9\times 10^{-5}, 9.6\times 10^{-5}, 1.1\times 10^{-4}, 5.5\times 10^{-5}$ 
for radii $r=70,60,50,40$, respectively.
This has to be compared to the electromagnetic energy flux as measured close to the black hole 
for which we find $\dot{E}^{(EM, BZ} \simeq 5\times 10^{-5}$ (see Figure~20 for {\em D15}), 
a value that is somewhat lower than the kinetic energy flux from the disk.

Thus, the conclusion in this preliminary study is that the disk wind substantially contributes in the total energy
production within this specific evolution period of simulation {\em D15},
while the kinetic energy flux of the disk wind is only somewhat larger than the BZ electromagnetic energy flux.

In \citet{1977MNRAS.179..433B}, the ideal MHD condition was assumed in their derivation of the Blandford-Znajek effect. 
They argued that ``the magnetic flux will be frozen into the accreting material and so the field close to the horizon 
can become quite large-much larger than the field at infinity." 
This condition is, nevertheless, not quite satisfied in the simulations in this section because the disks are 
magnetically diffusive. 
On the other hand, the ideal MHD condition above the disk required by disk-driven wind is well satisfied since the 
diffusivity value drops exponentially in the direction away from the disk mid-plane 
(see Equation~\ref{astro_dis_eta_init_eq}). 
This contrast of the diffusive level may be reflected, to some extent, by the dominance of the energy output from the 
disk wind. 

\section{Summary}
In this paper, we have applied the newly developed resistive GR-MHD code {\HAR} \citep{2017ApJ...834...29Q} to the 
astrophysical context of jet launching from thin accretion disks surrounding a black hole.
In our model the disk is threaded by inclined open poloidal field lines.
In general our simulation results demonstrate how the magnetic field strength, the disk magnetic diffusivity, and 
the black hole spin influence the MHD launching of disk winds and the Blandford-Znajek jet from the black hole.
Essentially we are able to compare the strength and power of both jet components.
In the following we summarize our results.

As a test simulation we have applied the code for a setup considering a very low magnetic diffusivity ($\eta < 10^{-12}$) 
together with a very weakly magnetized (plasma-$\beta=10^{8}$) thin disk around a non-rotating black hole.
No outflow is observed during the evolution. 

We then ran simulations with the same setup, but with a strong disk magnetic fields (plasma-$\beta = 10$) and a variation
in the strength of magnetic diffusivity. 
In these simulations, the launching of a disk wind is clearly observed.
This is consistent with classical non-relativistic accretion-ejection
MHD simulations that require a strong magnetic field for jet launching.

We then investigate a possible relation between accretion and ejection rates and the different levels of magnetic diffusivity.
We find that magnetic diffusivity can affect the mass fluxes in several ways.
It primarely decreases the coupling between matter and magnetic field, and so may decrease the efficiency of magnetic driving
(lower mass fluxes, lower jet velocities).
We find that an increasing diffusive level suppresses the
angular momentum transport through magnetic torques inside the disk and, as
a consequence, lowers the accretion rate. 
Quantitatively, for our setup we find a 100 times higher accretion rate for a magnetic diffusivity $\eta_0 <10^{-4}$ than 
for $\eta_0 < 10^{-2}$.
The lower accretion rate is always accompanied by a lower disk wind ejection rate.
Despite of the generally weaker mass fluxes for higher diffusivity, the ejection efficiency - the ratio of ejection rate to
accretion rate is about 10 time larger, a fact that we attribute to the more efficient mass loading in diffusive MHD. 
As probably expected, a higher level of diffusivity results in a smoother magnetic field distribution above the disk,
thus a less turbulent field structure.

From our analysis, we attribute the predominant driving force of the disk wind launching in the simulations with 
a strong initial magnetic field to the toroidal magnetic field pressure gradient, producing so called 
''tower jets" \citep{1995ApJ...439L..39U,1996MNRAS.279..389L}. 
We show that the contributions from the poloidal field pressure gradient and thermal pressure gradient are minor for wind 
launching.
This statement holds for the investigated time scales and magnetic field strength, plasma-$\beta = 10$.

Having found persistent outflows launched by magnetically diffusive disks we are able to compare the energy output carried
by the disk winds and the axial jets driven by rotation black holes (Blandford-Znajek mechanism).

Concerning the disk accretion we find lower accretion rates for increasing black hole spin parameters $a$ which we 
interpret as due to the magnetic pressure of the tangled toroidal field close to the horizon that is induced by frame 
dragging of the rotating black hole.
Quantitatively, the accretion rate for the simulation with black hole spin $a=0.9375$ is $10$ times smaller than that for 
the simulation with a non-rotating black hole. 
Furthermore, the magnetic pressure from the tangled field also supports the ejection of outflows.

For the energy output from rotating black holes we find that the Blandford-Znajek mechanism is seemingly less efficient 
than is presented in the literature.
We do not observe a noticeable non-linear growth in our simulations with increasing $a$ as shown in 
\citet{2004ApJ...611..977M, 2010ApJ...711...50T}.
We believe that this difference emerges because of advection of magnetic flux with the infalling disk corona
that is heavier in our disk simulations than for the torus simulations.

Comparing the overall energy output from the black hole-disk system,
we find that the matter energy flux dominates over the Poynting flux by a factor of ten.
Considering that only half of the the electromagnetic energy flux comes from the rotating black hole ($a=0.9375$), the disk 
contribution to the energy output is about 20 times larger than what we gain from the black hole rotational 
energy extraction.
Since the disk wind velocity is only mildly relativistic, the {\em kinetic} energy flux is 
relatively small and comparable to the Blandford-Znajek-driven electromagnetic flux.

\acknowledgements
We thank again Scott Noble and Matteo Bugli for guidance when implementing magnetic diffusivity in the HARM code. 
We thank an unknown referee for many constructive comments, in particular concerning the feasibility of the 
magneto-rotational instability in our disks.
We thank Amelia Hankla for preparing shearing-box test simulations of the MRI. 
Q.Q. thanks the Max Planck Institute for Astronomy and Hans-Walter Rix for the financial support of the 
thesis project.

\appendix

\section{Resolution study}
\label{astro_dis_high_resolution_sec}
As discussed in Section \ref{astro_dis_acc_sec}, our code allows to determine the numerical diffusivity of our setup.
By comparing the evolution of our simulations for lower and lower physical magnetic diffusivity $\eta_0$ we find a 
numerical diffusivity of $\eta_{\rm num} \lesssim 10^{-4}$ beyond which the system evolves following the numerical
diffusivity and not the physical diffusivity $\eta_0 <\eta_{\rm num}$ (see also our test simulations in
\citealt{2017ApJ...834...29Q}).
 
Simulation {\em D7} with our typical grid resolution $128 \times 128$ shows accretion and ejection rates that are clearly 
influenced by the magnetic diffusivity (see Figure \ref{astro_dis_acc-eje-eta_relation_img}).
In order to prove the reliability of these conclusions, we have run a comparison simulation {\em D8} with the same 
initial conditions, but with the double grid resolution, thus $256 \times 256$. 

\begin{figure*}
\centering
\includegraphics[width=2.in]{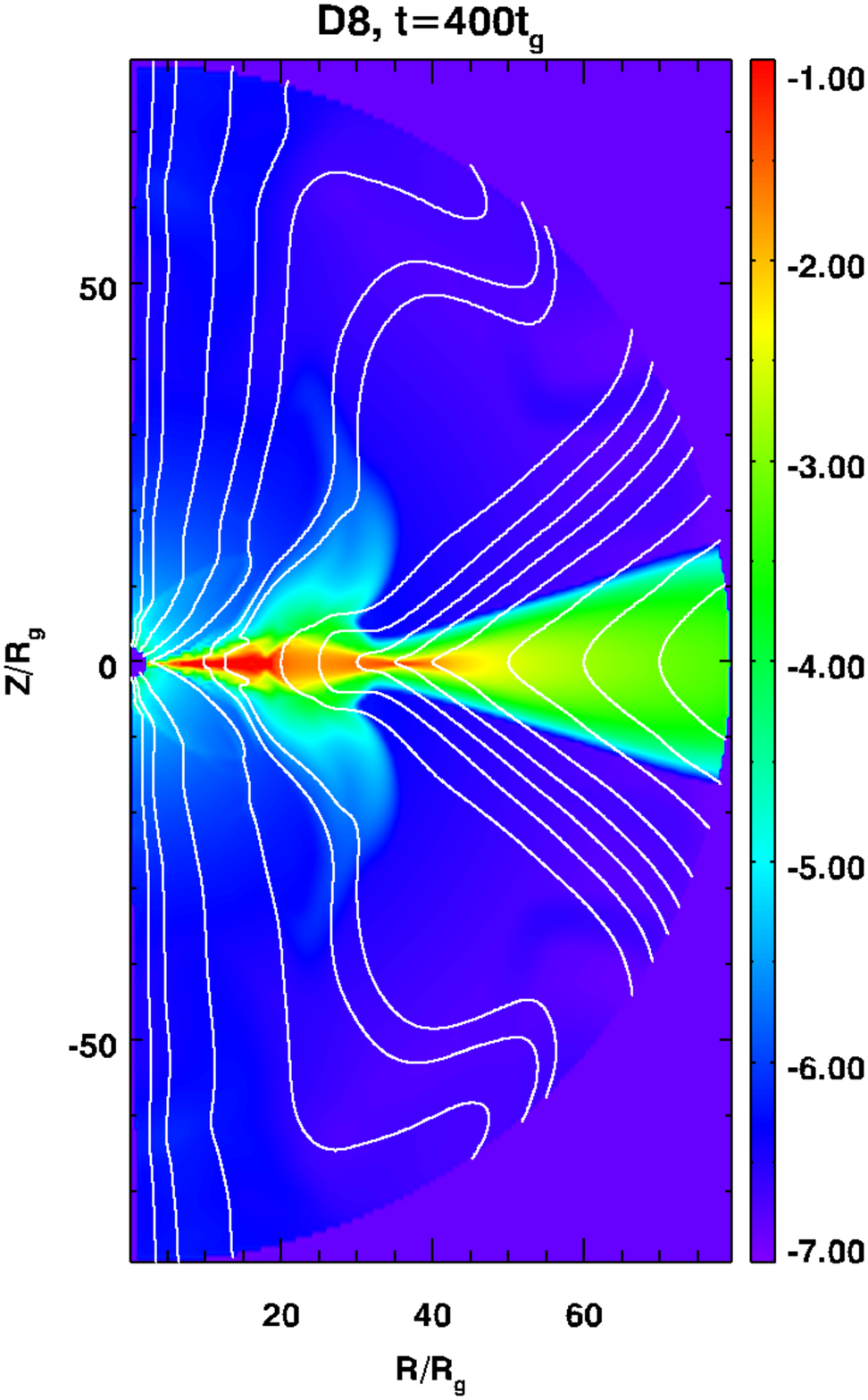}
\includegraphics[width=2.in]{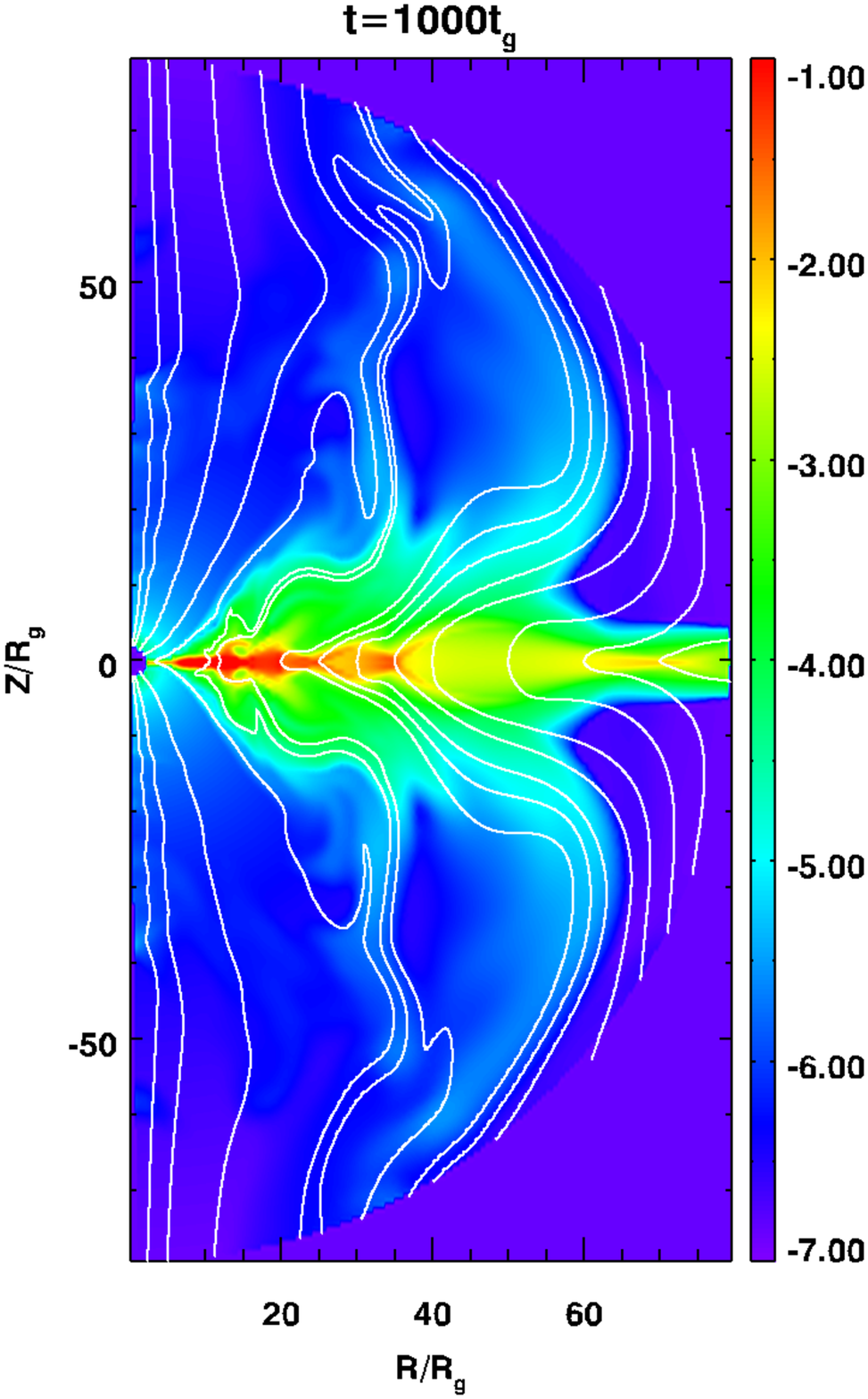}
\includegraphics[width=2.in]{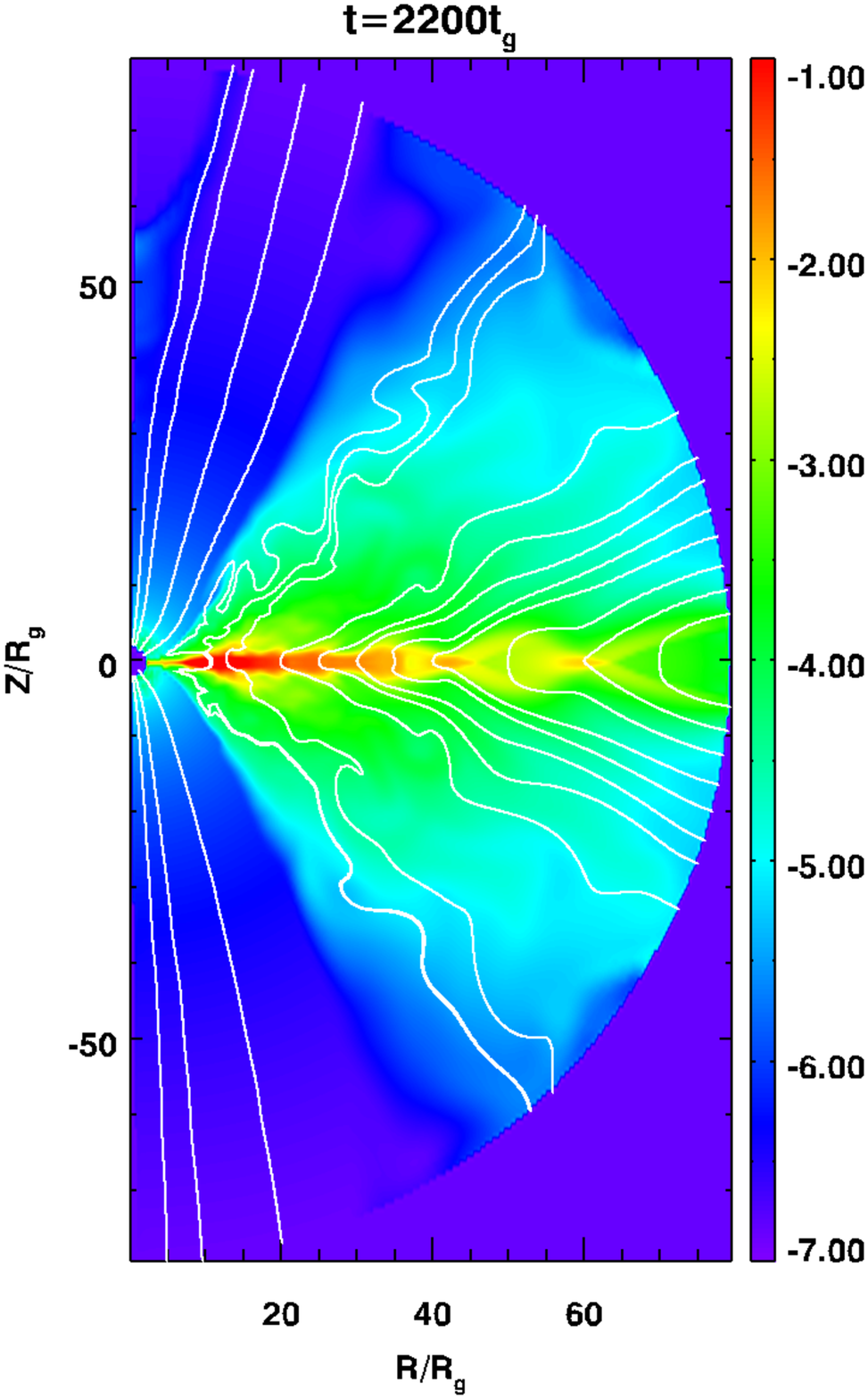}
\caption{Density and poloidal magnetic field lines of simulation {\em D8} at $t=400$ (left), $t=1000$ (middle),
and $t=3000$ (right). 
The morphology of the disk wind evolution show only slight differences when compared to simulation {\em D7} in Figure \ref{astro_dis_morph_D6_rho_img}, which has the same initial condition but a lower resolution. }
\label{astro_dis_morph_D7_rho_img}
\end{figure*}

In Figure \ref{astro_dis_morph_D7_rho_img}, we show the density and magnetic field distribution for simulation {\em D8} 
at times $t=400, 1000, 2200$. 
Comparing this to the corresponding figures for simulation {\em D7} (see Figure~\ref{astro_dis_morph_D6_rho_img}), we
may identify just the same overall disk wind launching processes.
Nevertheless, the launching processes in both simulations show some subtle differences that we attribute to the 
numerical resolution.
The disk structure and field distribution in simulation {\em D8} looks ``more evolved" that those for 
simulation {\em D7}. 
Certain features such as the field lines are more tangled in the higher resolution simulation.
However, the overall structure is indeed the same, in particular also the magnitude of the magnetic field and the density.

We may then compare the time averaged accretion rate for simulations {\em D7} and {\em D8}, which are 
$-2.0\times 10^{-3}$ and $-5.8\times 10^{-3}$, respectively.
The time averaged ejection rates are $7.0 \times 10^{-3}$ and $7.6 \times 10^{-3}$, respectively. 
The averages are taken from $t=2000$ to $t=3000$. 
The accretion rate for simulation {\em D8} triples that for simulation {\em D7}, while the ejection rate for 
simulation {\em D8} is only slightly larger than that for simulation {\em D7}. 
This is a strong indication for the fact that the numerical diffusivity in simulation {\em D7} weakens 
the angular momentum transport process through the magnetic torque (see Section \ref{astro_accretion_discussion_subsec}).
In this case, we expect a less evolved field line structure in the accretion disk.

\begin{figure*}
\centering
\includegraphics[width=2.6in]{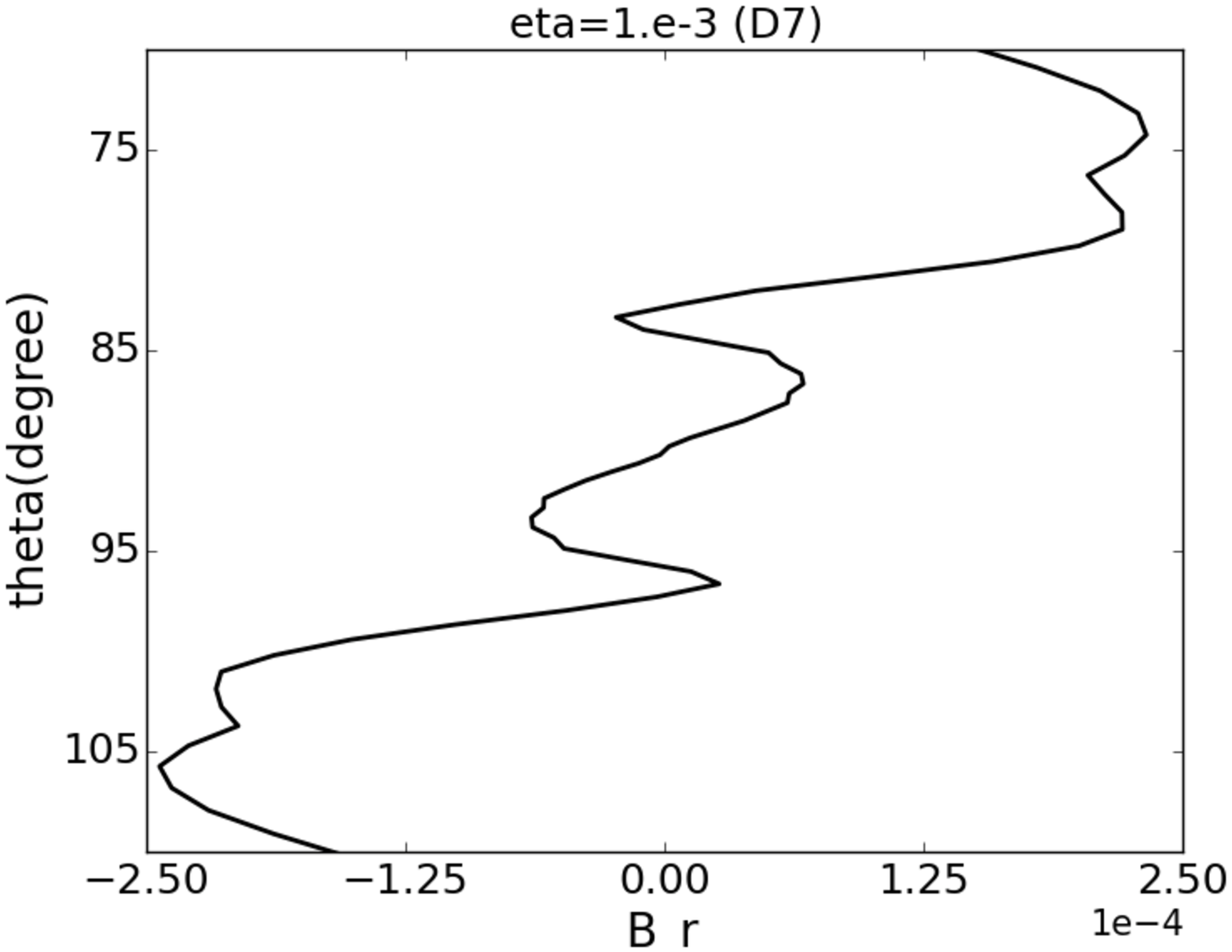}
\includegraphics[width=2.6in]{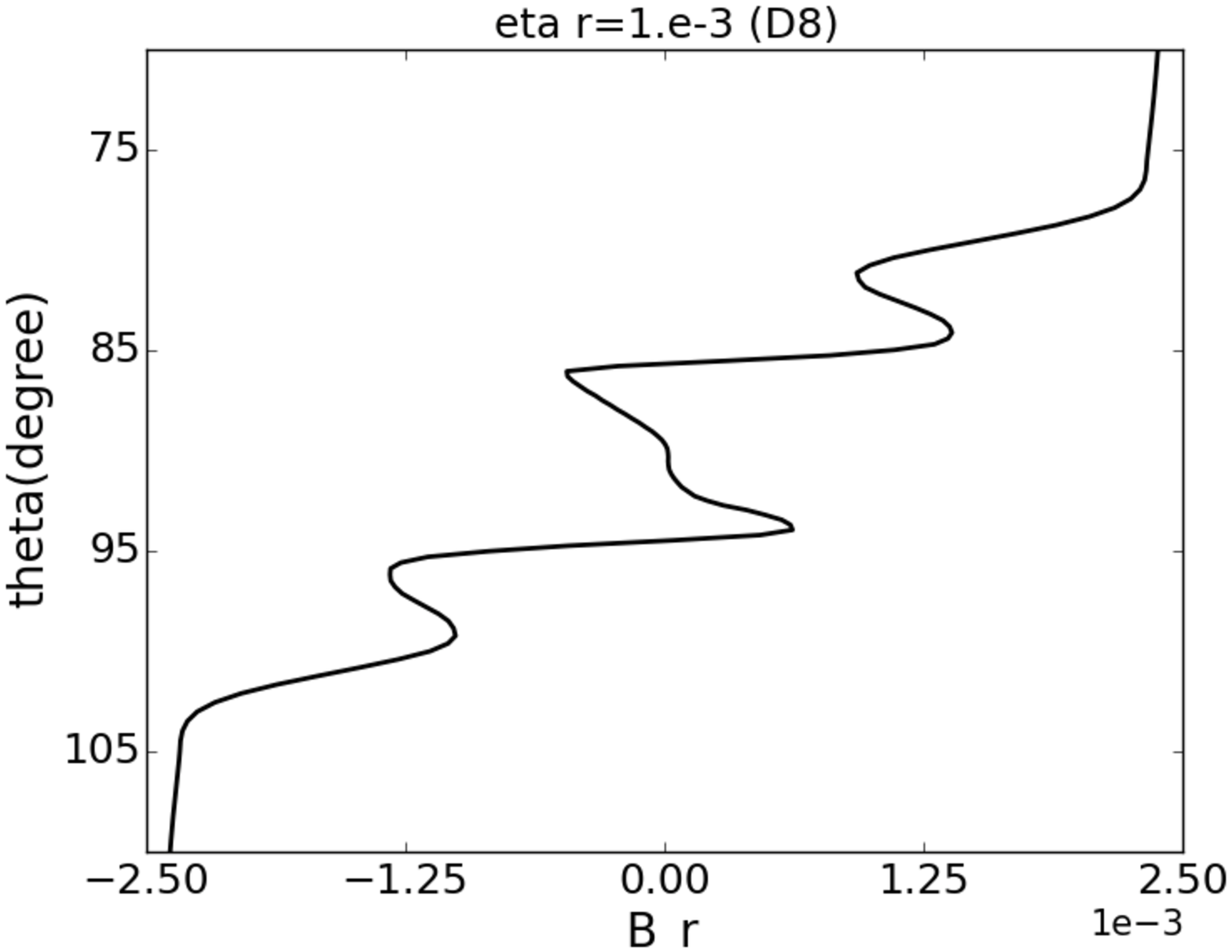}
\caption{Radial component of the magnetic field for simulations 
{\em D7} (left; resolution $ 128\times 128 $) and 
{\em D8} (right; resolution $ 256 \times 256 $) measured 
at $r=12.6$ and $t=1000$. 
Note that the Keplerian rotation period at radius $r=12.6$ corresponds to about 260 time units.
}
\label{astro_dis_resolution_MRI_img}
\end{figure*}

To confirm this, we compare the radial component of the magnetic field vector inside the disk at radius $r=12.6$ 
at $t=1000$ for both simulations. 
In Figure~\ref{astro_dis_resolution_MRI_img} we clearly see that the growth of radial magnetic field in simulation
{\em D7} has been much slower (note the magnitude of the field amplitude) than that in simulation {\em D8}.
This is leading to a less tangled magnetic field in simulation {\em D7}. 
In the end, we measure an ejection efficiency is higher in simulation {\em D7} 
than in simulation {\em D8} (see Section \ref{astro_dis_outflow_eff_subsec}).

In summary, the simulation with higher resolution shows only subtle differences in the overall disk and outflow structure.
Nevertheless, the integral physical may be different and depend somewhat on resolution.
Due to CPU time constraints, however,  we have restricted ourselves to a resolution of $128\times128$ for our serial 
version of rHARM (one simulation run such as {\em D7} takes about a month).
We believe that by comparing our simulations with similar resolution, the derived physical variables such as mass and energy
fluxes are indeed comparable.
The next obvious step is to repeat these simulations with a parallel code.
However, such code is not yet available.

\section{Influence of the outer boundary condition}
\begin{figure*}
\centering
\includegraphics[width=2.in]{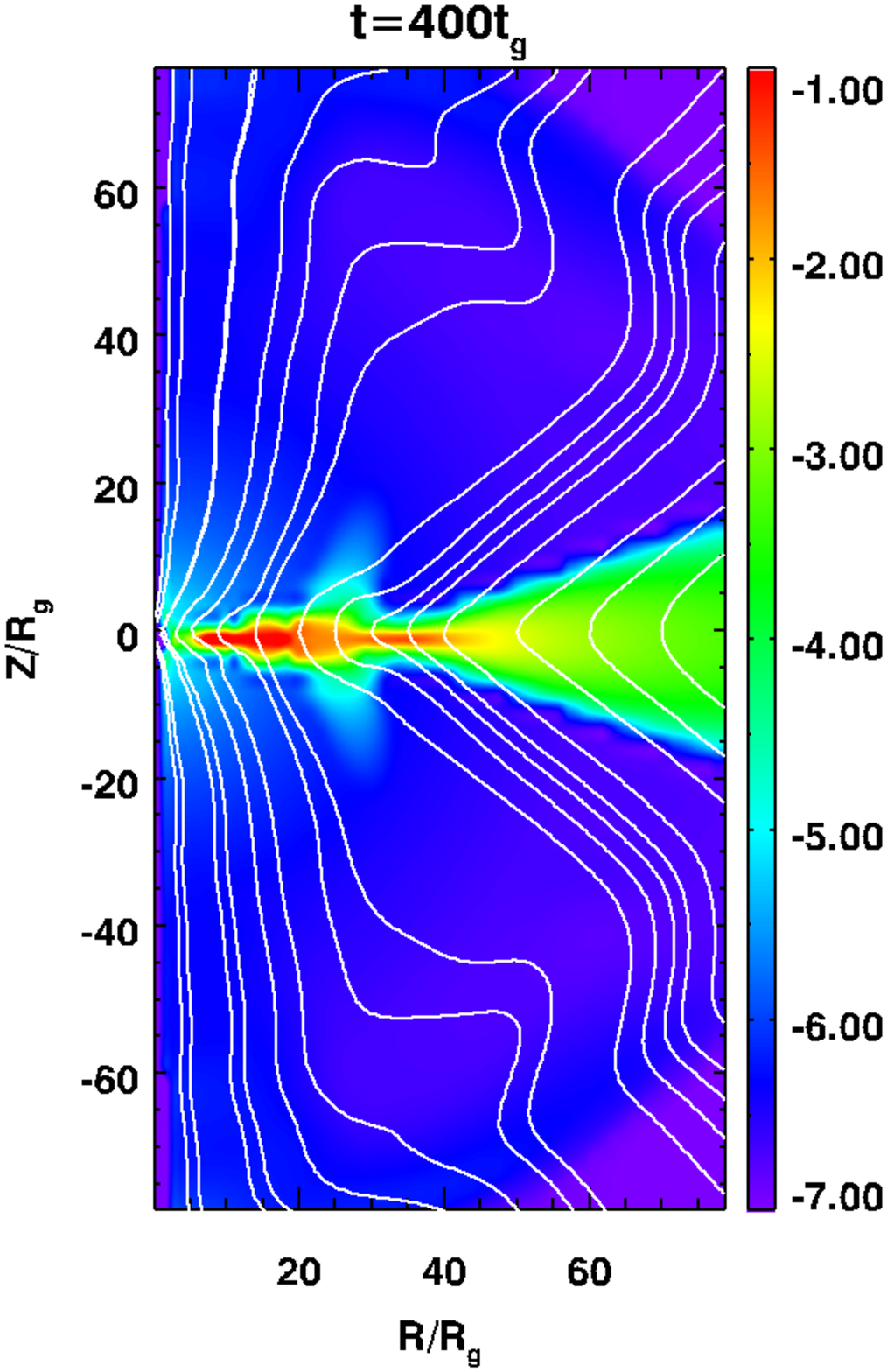}
\includegraphics[width=2.in]{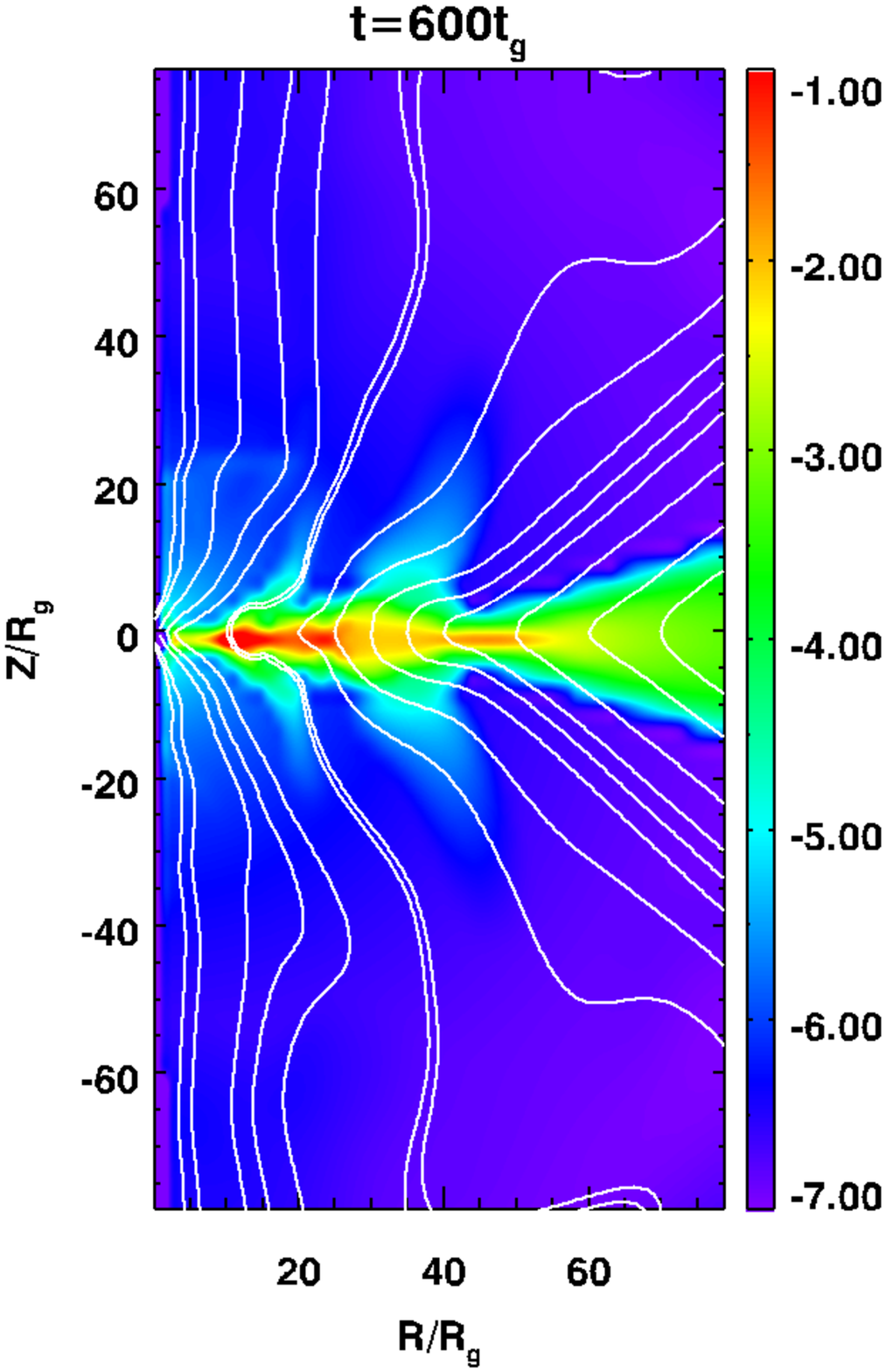}
\includegraphics[width=2.in]{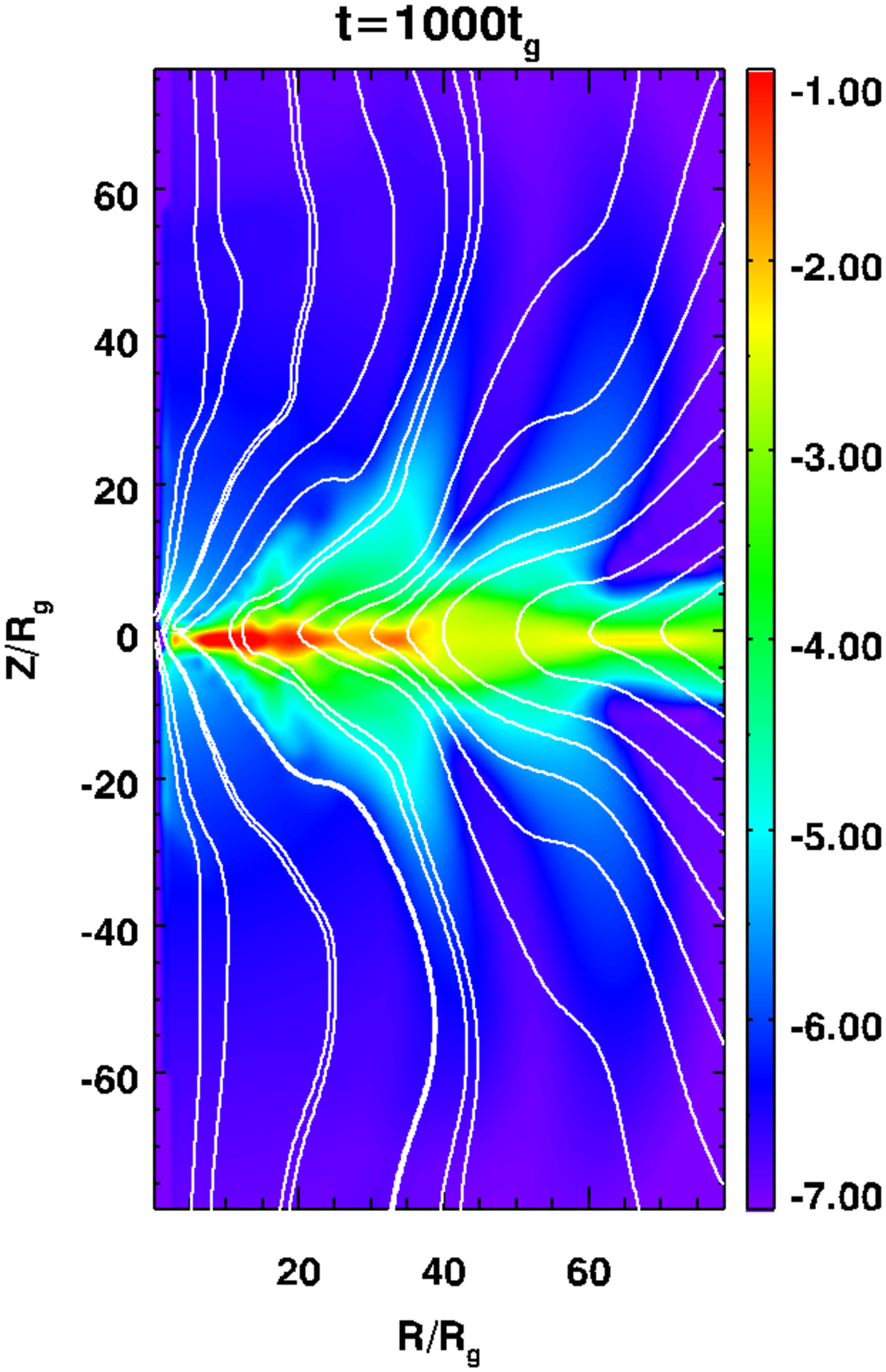}

\includegraphics[width=2.in]{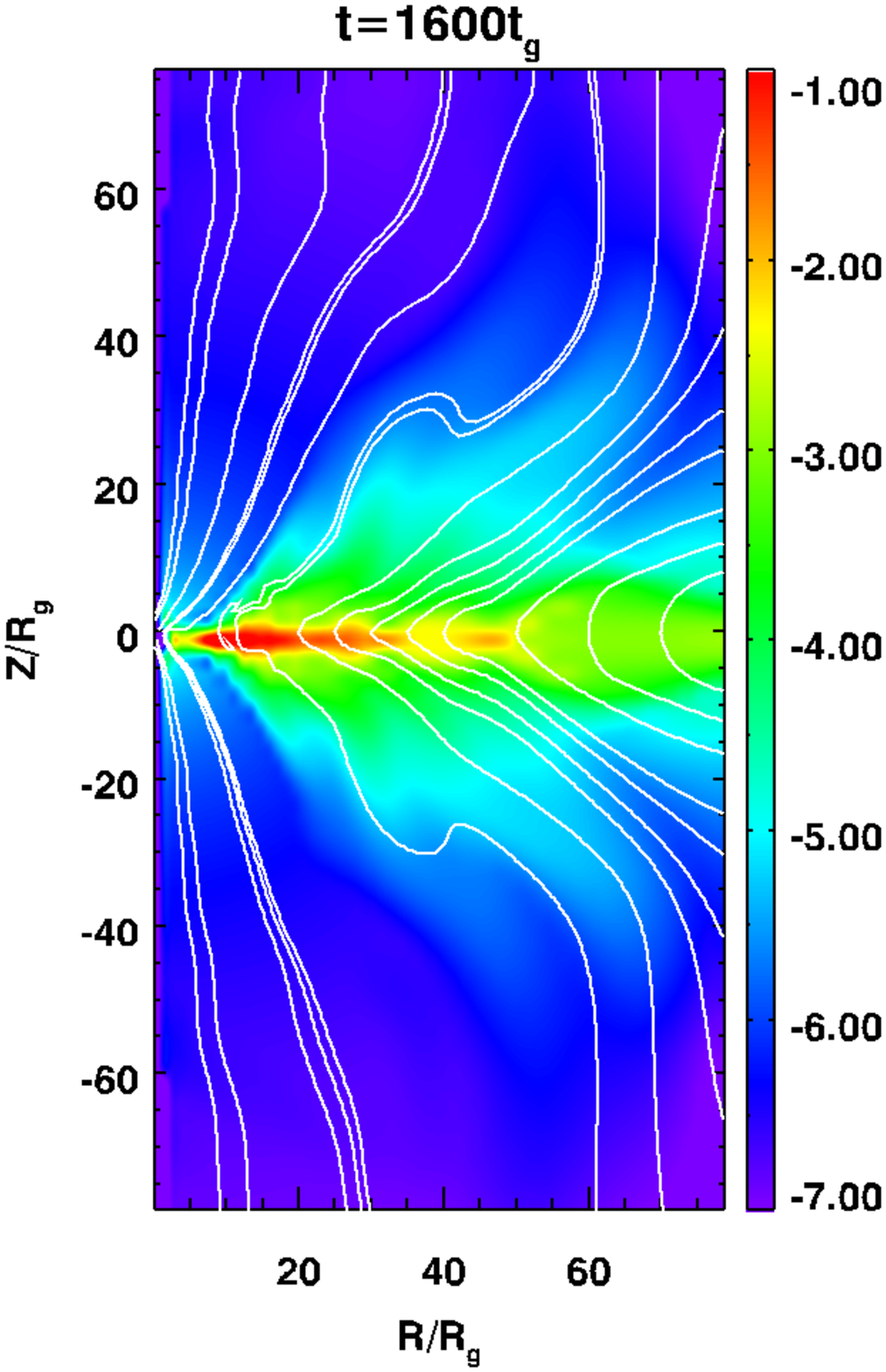}
\includegraphics[width=2.in]{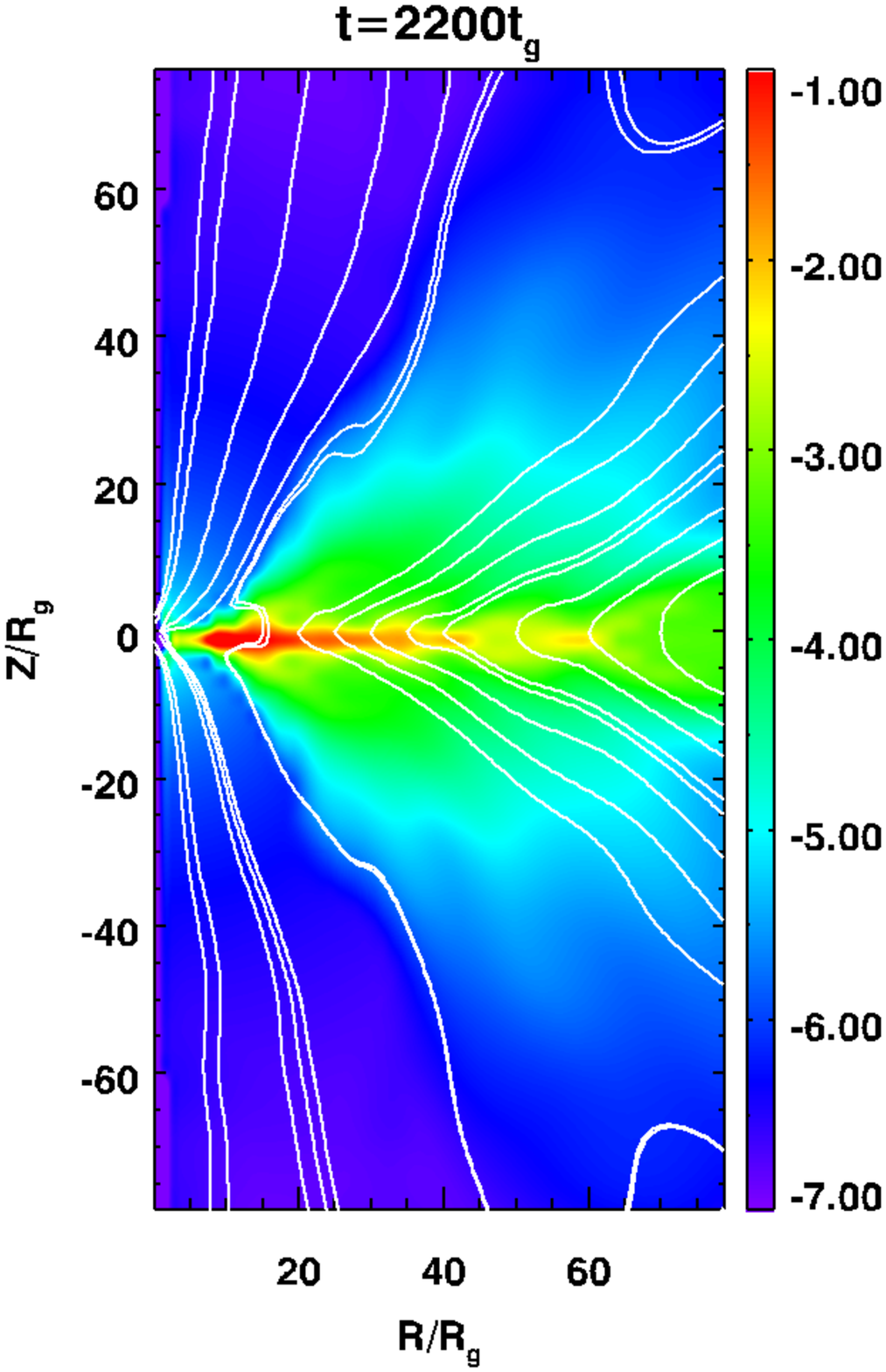}
\includegraphics[width=2.in]{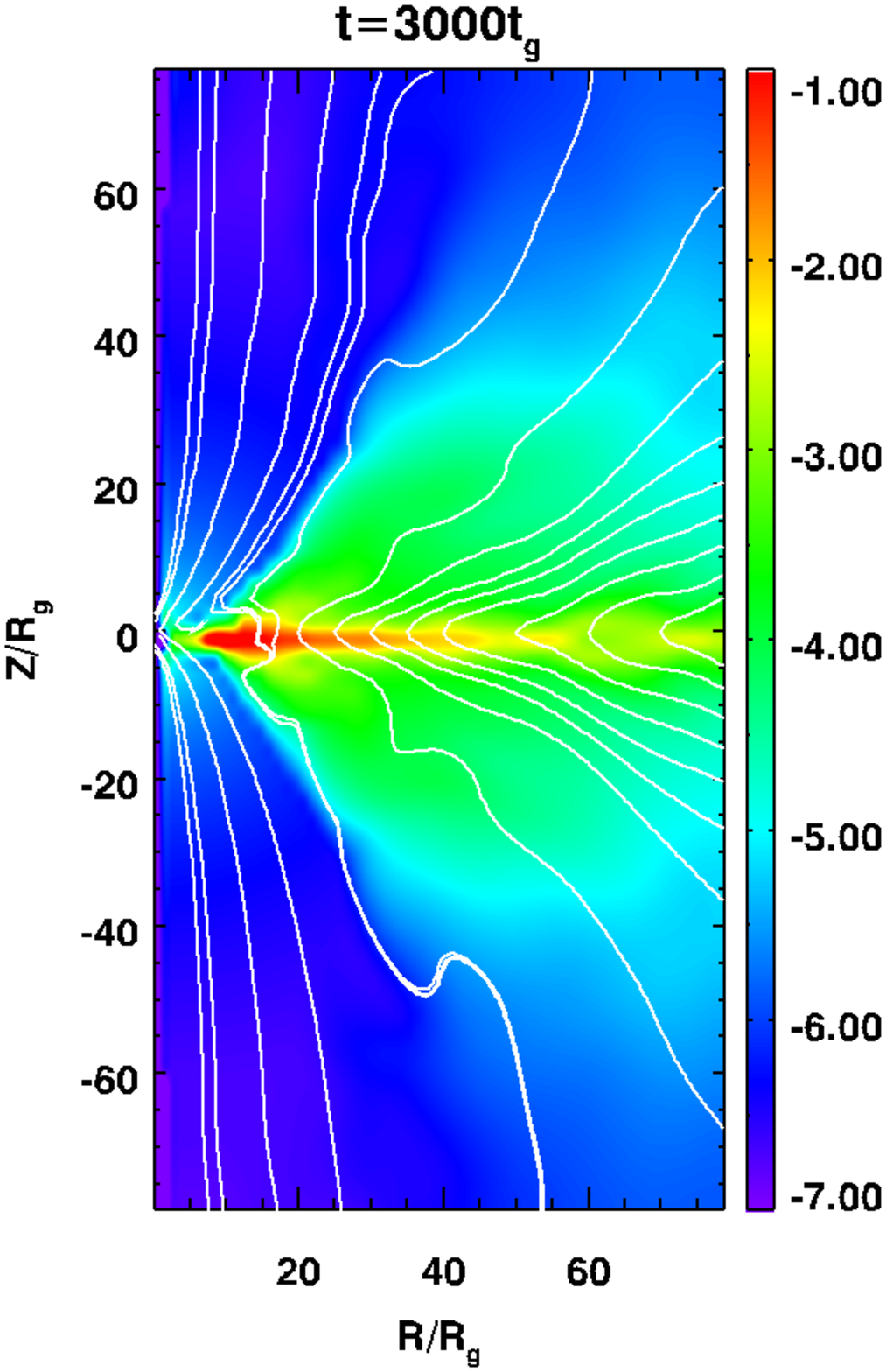}

\caption{Shown is the time evolution of density for the simulation described in the text. To compare the plots to 
those in Figure \ref{astro_dis_morph_D6_rho_img}, we clipped out the region $r \epsilon [0,80]$ and $z \epsilon [-80,80]$. 
This region is resolved by 110 grid cells in redial direction, which is very similar to simulation {\em D7}.
}
\label{astro_dis_morph_Rout160_rho_img}
\end{figure*}

In our simulations, we apply the standard outflow boundary condition of HARM that has been used in the
literature extensively so far.
In difference to the literature we investigate the formation of disk winds and their potential
evolution into disk jets.
What we typically observe for the outflow evolution far from the disk is an appearent over-collimation or
re-collimation towards the axis (see Figure~\ref{astro_dis_morph_D6_rho_img}, upper right and the lower left panels).
In order to prove whether this collimation is a physical effect or an artifact of the boundary condition, ran a simulation 
that has the same set up as simulation {\em D7} except the physical grid size of $r_{out}=160$ instead of $80$. 
In this simulation, the outer region of the area within  $r=80$ is resolved radially by 110 cells, thus very similar to that in simulation
{\em D7}. 
The time evolution of density for this simulation is shown in Figure \ref{astro_dis_morph_Rout160_rho_img}. 
We see that the corresponding plots here do not show the same ``collimation" effect as for {\em D7} shown in Figure~\ref{astro_dis_morph_D6_rho_img}. 
At $r=80$ the material is still moving out in an almost radial direction, only slightly collimated.
Thus, we interpret the collimation of outflow material in simulation {\em D7} potentially as a boundary effect.
It has been stated that sub-Alfv\'enic flows may became reflected at boundaries obeying a classical zero-gradient condition
(see e.g. \citealt{1999ApJ...516..221U}).
However, our outflow is super-Alfv\'enic and is also evolved on a spherical grid.
Nevertheless, for forthcoming papers dealing with the asymptotic acceleration and collimation of disk jets it would be essential
to revise the outflow boundary condition such that artificial collimation effects are excluded.
This has been already been successfully applied by \citet{2010ApJ...709.1100P} who applied ''zero-current" outflow conditions for
their relativistic MHD simulations of jet formation.
Such an attempt is, however, beyond the topic of the present paper in which we are more concerned about the 
launching conditions for jets from the disk and from the ergosphere.  

\bibliographystyle{apj}


\end{document}